\documentclass[pdftex,nonacm,sigconf]{acmart}

\usepackage{ifthen}

\newcommand{\version}{full}

\usepackage{algorithm}
\usepackage[noend]{algpseudocode}

\algrenewcommand{\algorithmiccomment}[1]{\textcolor{gray}{\# #1}}\algrenewcommand\algorithmicdo{\textbf{\texttt{:}}}\algrenewcommand\algorithmicthen{\textbf{\texttt{:}}}\algrenewcommand\algorithmicfunction{\textbf{\texttt{def}}}\algrenewcommand\algorithmicindent{1.5ex}

\usepackage{varwidth}
\usepackage{wrapfig}

\usepackage{amsmath}

\usepackage{amssymb}
\usepackage{array}

\usepackage{stmaryrd}

\usepackage{caption}
\usepackage{subcaption}
\usepackage{listings}
\usepackage{mathpartir} 

\usepackage{xcolor}

\usepackage{xspace} \usepackage{paralist} 

\usepackage{tikz}
\usepackage{pgf}
\usepackage{pgfplots}

\usepackage{multirow}

\usepackage{circledsteps}
\tikzset{/csteps/outer color=blue}

\usetikzlibrary{positioning}
\usetikzlibrary{trees}
\usetikzlibrary{calc}
\usetikzlibrary{tikzmark}
\usetikzlibrary{matrix}
\usepgfplotslibrary{statistics}

\ifthenelse{\equal{\version}{conf}}{
\usepackage[appendix=strip,bibliography=common]{apxproof}
}{
\usepackage[appendix=append,bibliography=common]{apxproof}
}

\usepackage{xpatch}
\usepackage{textcase}
\makeatletter
\xpatchcmd{\@sect}{\uppercase}{\MakeTextUppercase}{}{}
\xpatchcmd{\@sect}{\uppercase}{\MakeTextUppercase}{}{}
\makeatother

\theoremstyle{plain}
\newtheoremrep{theorem}{Theorem}[section]
\newtheoremrep{lemma}[theorem]{Lemma}
\newtheoremrep{proposition}[theorem]{Proposition}
\newtheoremrep{claim}[theorem]{Claim}
\newtheoremrep{corollary}[theorem]{Corollary}
\newtheoremrep{observation}[theorem]{Observation}
\newtheoremrep{remark}[theorem]{Remark}

 \ifthenelse{\equal{\version}{conf}}{
  \newcommand{\inConfVersion}[1]{#1} \newcommand{\inFullVersion}[1]{}   }{
  \ifthenelse{\equal{\version}{full}}{
    \newcommand{\inConfVersion}[1]{} \newcommand{\inFullVersion}[1]{#1} }{
\newcommand{\inConfVersion}[1]{#1} \newcommand{\inFullVersion}[1]{#1} }
}

\newcommand{\rev}[1]{\textcolor{black}{#1}}

\newcommand{\join}{\Join}
\newcommand{\semijoin}{\ensuremath{\ltimes}}

\newcommand{\setof}[2]{\{#1\mid#2\}}

\DeclareMathOperator*{\argmin}{arg\,min}

\newcommand{\multileft}{\ensuremath{\{\!\!\{}}
\newcommand{\multiright}{\ensuremath{\}\!\!\}}}
\newcommand{\bag}[1]{\multileft #1 \multiright}
\newcommand{\bagof}[2]{\bag{#1 \mid #2}}

\newcommand{\defeq}{\stackrel{\text{def}}{=}}

\newcommand{\card}[1]{\ensuremath{|#1|}}
\DeclareMathOperator{\dom}{\textit{dom}}
\DeclareMathOperator{\atoms}{\textit{atoms}}
\DeclareMathOperator{\supp}{\textit{supp}}
\DeclareMathOperator{\attr}{\textit{attr}}

\newcommand{\jtree}{\ensuremath{\mathcal{T}}}

\newcommand{\concat}{\mathbin{{+}\!\!{+}}}

\newcommand{\seq}[1]{\ensuremath{\overline{#1}}}
\newcommand{\nats}{\ensuremath{\mathbb{N}}}

\newcommand{\restr}[2]{\ensuremath{{#1}[{#2}]}}

\newcommand{\w}{\ensuremath{\seq w}}
\newcommand{\x}{\ensuremath{\seq x}}
\newcommand{\y}{\ensuremath{\seq y}}
\newcommand{\z}{\ensuremath{\seq z}}

\newcommand{\db}{\ensuremath{\textit{db}}}

\newcommand{\insize}{\ensuremath{\textsc{in}}\xspace}
\newcommand{\outsize}{\ensuremath{\textsc{out}}\xspace}

\newcommand{\bigo}{\ensuremath{\mathcal{O}}}

\newcommand{\lande}{L\&E\xspace}
\newcommand{\ya}{\textsf{YA}\xspace}
\newcommand{\nra}{\textsf{NRA}\xspace}
\newcommand{\nsa}{\textsf{NSA}\xspace}
\newcommand{\twonsa}{\textsf{2NSA}\xspace}
\newcommand{\ra}{\textsf{RA}\xspace}
\newcommand{\algname}{\textsf{SYA}\xspace}
\newcommand{\algnamefull}{Shredded Yannakakis\xspace}
\newcommand{\binplan}{\textsf{DF-Bin}\xspace}
\newcommand{\duckdbplan}{\textsf{DuckDB-Bin}\xspace}
\newcommand{\umbra}{\textsf{Umbra-Bin}\xspace}
\newcommand{\umbrale}{\textsf{Umbra-L\&E}\xspace}

\newcommand{\threeq}{\ensuremath{Q_3}}

\newcommand{\threej}{\ensuremath{J_3}}

\newcommand{\lookupsymbol}{\tikz[baseline=-2.5pt]{\node[inner sep=0pt,draw,circle] (a) at (0,0) {\tiny \ensuremath{\to}};}}
\newcommand{\expandsymbol}{\tikz[baseline=-2.5pt]{\node[inner sep=0.5pt,draw,circle] (a) at (0,0) {\footnotesize \textnormal{e}};}}
\DeclareMathOperator{\lookup}{\mathbin{\lookupsymbol}}
\DeclareMathOperator{\expand}{\mathbin{\expandsymbol}}

\newcommand{\attrs}{\ensuremath{\mathcal{A}}} \newcommand{\hf}[1]{\ensuremath{HF(#1)}}

\newcommand{\dto}{\ensuremath{\rightsquigarrow}} \newcommand{\dscheme}[2]{\ensuremath{#1\!\!\dto\!\! #2}}

\DeclareMathOperator{\nest}{\nu}
\DeclareMathOperator{\unnest}{\mu}
\newcommand{\flatten}{\ensuremath{\mu^*}}
\DeclareMathOperator{\group}{\gamma}
\newcommand{\groupby}[2]{\ensuremath{\group_{#1}}}
\newcommand{\nsemijoinsymb}{\tikz[baseline=-2pt]{\draw (0,-0.6ex)--(0.5ex,0)--(0,0.6ex)--(0,-0.6ex);\draw[->, >=stealth,draw] (0.5ex,0)--(1.75ex,0);}}

\DeclareMathOperator{\nsemijoin}{\nsemijoinsymb}

\DeclareMathOperator{\flatsub}{\attrs}
\DeclareMathOperator{\sub}{\textit{sub}}

\DeclareMathOperator{\shred}{\textit{shred}}
\DeclareMathOperator{\ishred}{\textit{ishred}}
\DeclareMathOperator{\weightval}{\textit{weight}}
\newcommand{\hol}[1]{\textsf{hd\_}#1}
\newcommand{\holshort}[1]{\textsf{hd}#1}
\newcommand{\weight}[1]{\textsf{w\_}#1}

\newcommand{\nxt}{\textsf{nxt}}

\newcommand{\len}[1]{\ensuremath{\card{#1}}}

\newcommand{\allsel}[1]{\texttt{all}\ensuremath{_{\phys{#1}}}}
\DeclareMathOperator{\size}{\textit{size}}

\newcommand{\phys}[1]{\ensuremath{\mathtt{#1}}}
\newcommand{\rep}[1]{\ensuremath{\mathcal{#1}}}
\newcommand{\store}[1]{\ensuremath{\Sigma_{#1}}}

\newcommand{\sel}[1]{\texttt{#1}}

\newcommand{\newphys}{\texttt{new\_physical\_relation}}
\newcommand{\take}{\texttt{take}\xspace}

\newcommand{\opgroup}{\texttt{groupby}\xspace}
\newcommand{\opsemijoin}{\texttt{semijoin}\xspace}
\newcommand{\opunnest}{\texttt{unnest}\xspace}
\newcommand{\opflatten}{\texttt{flatten}\xspace}
\newcommand{\opflatteninner}{\texttt{rflatten}\xspace}

\newcommand{\opgenerate}{\texttt{generate}\xspace}

\newcommand{\posvec}{\texttt{pos}\xspace}
\newcommand{\repvec}{\texttt{rep}\xspace}

\newcommand{\rootnode}{A} \newcommand{\depthone}{B} \newcommand{\depthtwo}{C} \newcommand{\depththree}{E} \newcommand{\depthfour}{F} \newcommand{\depthfive}{G}

\newcommand{\cost}[2]{\ensuremath{\mathcal{C}\llbracket #1 \rrbracket_{#2}}}
\DeclareMathOperator{\cbuild}{\textit{build}}
\DeclareMathOperator{\cprobe}{\textit{probe}}
\DeclareMathOperator{\cgen}{\textit{take}}
\newcommand{\posreals}{\ensuremath{\mathbb{R}_{\geq 0}}}

\newcommand{\numcols}[1]{\ensuremath{\# #1}}

\DeclareMathOperator{\lleaf}{\ensuremath{LL}}
\DeclareMathOperator{\la}{\ensuremath{LA}}
\DeclareMathOperator{\ja}{\ensuremath{JA}}
\newcommand{\jvar}{\ensuremath{\ja}}

\newcommand{\tonsemijoin}[1]{\ensuremath{#1^{\nu}}}
\newcommand{\totwonsa}[1]{\ensuremath{\flatten(\tonsemijoin{#1})}}

\newcommand{\nrofqueries}{1,849}

\title[]{Instance-Optimal Acyclic Join Processing Without Regret: Engineering the Yannakakis Algorithm in Column Stores}

\author{Liese Bekkers}
\orcid{0000-0002-2725-2131}
\affiliation{
  \institution{UHasselt, Data Science Institute}
  \country{}
}

\author{Frank Neven}
\orcid{0000-0002-7143-1903}
\affiliation{
  \institution{UHasselt, Data Science Institute}
  \country{}
}

\author{Stijn Vansummeren}
\orcid{0000-0001-7793-9049}
\affiliation{
  \institution{UHasselt, Data Science Institute}
  \country{}
}

\author{Yisu Remy Wang}
\orcid{0000-0002-6887-9395}
\affiliation{
  \institution{University of California, Los Angeles}
  \country{}
}

\begin{document}

\begin{abstract}
  Acyclic join queries can be evaluated instance-optimally using Yannakakis'
  algorithm, which avoids needlessly large intermediate results through
  semi-join passes. Recent work proposes to address the significant hidden
  constant factors arising from a naive implementation of Yannakakis by
  decomposing the hash join operator into two suboperators, called Lookup and
  Expand.  In this paper, we present a novel method for integrating Lookup and
  Expand plans in interpreted environments, like column stores, formalizing them
  using Nested Semijoin Algebra (NSA) and implementing them through a shredding
  approach. We characterize the class of NSA expressions that can be evaluated
  instance-optimally as those that are 2-phase: no `shrinking' operator is
  applied after an unnest (i.e., expand).  We introduce \algnamefull (\algname),
  an evaluation algorithm for acyclic joins that, starting from a binary join
  plan, transforms it into a 2-phase NSA plan, and then evaluates it through the
  shredding technique. We show that \algname is provably robust (i.e., never
  produces large intermediate results) and without regret (i.e., is never worse
  than the binary join plan under a suitable cost model) on the class of
  well-behaved binary join plans. Our experiments on a suite of \nrofqueries\
  queries show that \algname improves performance for \rev{$85.3$}\% of the queries with
  speedups up to \rev{62.5}x, while remaining competitive on the other queries.  We
  hope this approach offers a fresh perspective on Yannakakis' algorithm,
  helping system engineers better understand its practical benefits and
  facilitating its adoption into a broader spectrum of query engines.
\end{abstract}

 \maketitle

\section{Introduction}
\label{sec:intro}

\inFullVersion{Computing joins efficiently has been a fundamental challenge in query processing
since the inception of the relational model. Thanks to decades of research and
engineering, contemporary query engines excel on common benchmark such as TPC-H
featuring foreign-key joins of a limited number of relations. However, queries
with up to a thousand of relations featuring many-to-many joins are not uncommon
anymore in modern data analysis scenarios~\cite{10.1145/3183713.3183733,
  10.1145/3514221.3517871, DBLP:journals/pvldb/ChenHWSS22}. Unfortunately, for
such queries, consistently finding a good join order is very difficult. At the
same time, a poorly chosen join order will bring even state-of-the-art systems
to their knees~\cite{DBLP:conf/pods/000124}. In recent
work~\cite{robust-diamond-hardened-joins}, Birler, Kemper, and Neumann
(henceforth BKN) have dubbed the problem underlying this phenomenon the
\emph{diamond problem}: a poor query plan will compute subresults that are
orders of magnitude larger than the output, even if these subresults are
unnecessary to produce this final output---thereby wasting significant
processing time.}

\inConfVersion{
Computing joins efficiently has been a fundamental challenge in query processing
since the inception of the relational model. Unfortunately, consistently finding a good join order remains very difficult. This is especially true for the increasingly common queries
with up to a thousand relations featuring many-to-many joins~\cite{10.1145/3183713.3183733,
  10.1145/3514221.3517871, DBLP:journals/pvldb/ChenHWSS22}. 
 At the
same time, a poorly chosen join order will bring even state-of-the-art systems
to their knees~\cite{DBLP:conf/pods/000124}. In recent
work~\cite{robust-diamond-hardened-joins}, Birler, Kemper, and Neumann
(henceforth BKN) have dubbed the problem underlying this phenomenon the
\emph{diamond problem}: a poor query plan will compute subresults that are
orders of magnitude larger than the output, even if these subresults are
unnecessary to produce this final output---thereby wasting significant
time.}

Avoiding the diamond problem is intrinsically linked to query engine \emph{robustness}:
by limiting the sizes of intermediate results, the engine's runtime becomes
bounded and predictable.  How to avoid the diamond problem has in fact been a
major topic in database theory for decades. From the concept of
acyclicity~\cite{DBLP:conf/stoc/BeeriFMMUY81,DBLP:journals/jacm/Fagin83} and
Yannakakis' seminal algorithm (\ya) for optimally processing acyclic
queries~\cite{DBLP:conf/vldb/Yannakakis81}, over various notions of query width
and query decompositions~\cite{DBLP:conf/pods/GottlobGLS16}, to the more recent
worst-case-optimal
(WCO)~\cite{DBLP:conf/icdt/Veldhuizen14,DBLP:journals/jacm/NgoPRR18,DBLP:conf/pods/000118}
and factorized~\cite{DBLP:journals/tods/OlteanuZ15,DBLP:conf/pods/KhamisNR16}
processing algorithms: much research has been done to identify and exploit
structural properties of join queries that can either completely eliminate or
bound the size of intermediate results. Although many of these techniques have been
known for decades, they have not yet found wide-spread adoption in practical
query engines. Indeed, most contemporary
systems~\cite{DBLP:journals/tods/AbergerLTNOR17,DBLP:journals/pvldb/FreitagBSKN20,DBLP:journals/pvldb/MhedhbiS19,
  DBLP:conf/sigmod/RaasveldtM19, DBLP:conf/sigmod/LambSHCKHS24,
  DBLP:conf/cidr/NeumannF20} continue to use non-robust binary join plans for
most queries, possibly resorting to WCO joins in certain cases---in particular
for cyclic queries. The reason for this lack of adoption is that the
above-mentioned research focuses on \emph{asymptotic} complexity and optimizes
for the worst-case input instance in avoiding the diamond problem. In fact, when
implemented in a concrete system, these techniques can be significantly
slower than traditional techniques on common-case instances and
queries~\cite{robust-diamond-hardened-joins,DBLP:conf/pods/000124}. From an
engineering viewpoint we are hence in search for provably robust query
processing algorithms \emph{without regret}: competitive with traditional join
algorithms while avoiding the diamond problem.

Towards this goal, BKN suggest to move to a larger space of query
plans~\cite{robust-diamond-hardened-joins}.  Concretely, they propose to
decompose the traditional hash join operator into two suboperators called Lookup
and Expand (or \lande for short). Lookup (denoted $\lookup$) finds the first match
of a given tuple in a hash table, while expand ($\expand$) iterates over the rest of the matches.
By considering query plans where these two
suboperators can be freely combined and reordered, 
dangling tuples (i.e., tuples that do not
contribute to the output) can be eliminated as early as possible, hence avoiding
the diamond problem. It is shown that \lande plans can be used to optimally process acyclic
joins as well as effectively process certain cyclic joins when an additional operator is added. However, their approach to create \lande plans 
does not formally guarantee to always avoid the diamond problem (see point (4) below for more detail). 

While BKN successfully implement \lande plans inside
Umbra~\cite{DBLP:conf/cidr/NeumannF20}, a compiled query engine, it is unclear
how to effectively implement \lande plans inside \emph{interpreted} query
engines.
Indeed, Umbra generates code from \lande plans 
using the produce-consume interface~\cite{10.14778/2002938.2002940} favored in compiled engines, and then rely on compiler optimizations to remove inefficiencies.
Obtaining the same behavior in an interpreted engine poses two
challenges. First, in the typical architecture of an interpreted engine, 
(physical) operators adopt a \emph{uniform} (physical) data model. In column stores, this data model is simply a relation, implemented as set of
column segments.  While BKN state that an \lande plan is also meant to produce a
relation, they also impose several constraints. For instance, after performing
$R \lookup S$, one cannot access the non-join attributes of $S$ without first
applying an expand operation. This suggests that the output of $\lookup$ is not
a standard relation, making it unclear what exactly the (physical) data model is one should implement for \lande plans.

The second challenge, particular to column stores, is that the common wisdom in
column stores is to let operators process \emph{a column-at-a-time}.  This is
empirically a (large) constant factor faster than element at-a-time processing
since it allows using vectorization when applicable as well as amortize
function-call overhead.  Yet, the code generated for \lande plans by BKN
proceeds tuple-at-a-time.

In this paper, we build further upon the ideas in BKN by investigating the
implementation of L\&E-based query processing inside the more common
\emph{interpreted} query engines, in particular column stores.  We address the
challenges above and obtain an evaluation algorithm for acyclic joins, named
\algnamefull (\algname), that is \emph{provably robust without regret} for a
subclass of queries. We summarize our contributions next and highlight the
differences with as well as improvements over BKN.

(1) In contrast to BKN who describe lookup and expand in terms of
their effect on some intermediate state during execution, we provide a formal
semantics to \lande plans based on the nested relational
model~\cite{DBLP:journals/acr/ThomasF86,DBLP:journals/tcs/BunemanNTW95} which is
an extension of the relational model where individual records may themselves
contain entire relations.  In particular, we design a set of nested relational
operators that we call the \emph{Nested Semijoin Algebra} (\nsa).  Here, lookup
can be expressed as a form of nesting while expand is a form of unnesting. By
formalizing these operations algebraically, we explicitly define the logical
data model, allowing us to extend beyond lookup/expand and joins, and generalize
to all standard relational operators.

(2) We use \nsa  to implement \lande plans inside conventional
\emph{interpreted} query engines, in particular column stores. Our
implementation is based on \emph{query shredding} techniques for
simulating nested relational algebra with standard relational
algebra~\cite{DBLP:journals/tcs/Bussche01,DBLP:conf/sigmod/CheneyLW14,DBLP:journals/pvldb/SmithBNS20,DBLP:conf/pods/Wong93}.   We take special care to provide an efficient column-oriented
implementation for completely unnesting deeply nested relations.

(3) BKN observe that \lande plans consisting of two distinct phases---where the first phase exclusively performs Lookups and the second phase exclusively performs Expands---execute in time $\bigo(\insize + \outsize)$ for all inputs. In other words, such 2-phase plans are \emph{instance-optimal}. 
We extend this result to include all \nsa operators, not only \lande, by defining 2-phase \nsa expressions as those in which no `shrinking' operator is applied after an unnest (i.e., expand) operation has been performed. We show that a join query can be evaluated by means of a 2-phase
NSA join plan if and only if it is acyclic.
This result, therefore, generalizes the instance-optimality of \ya
to \nsa plans and provides an additional characterization for the class of acyclic joins.

(4) The aforementioned formal guarantees focus on asymptotic complexity, which
often overlooks crucial constant
factors. To address this, we perform a finer-grained analysis of NSA plans in terms of a
cost model that takes such constant factors into account (more specifically, the
cost of building and probing hash maps as well as generating single column
vectors). We identify a class of traditional binary join plans---referred to as
\emph{well-behaved}---that can be transformed into equivalent 2-phase \nsa
plans that are guaranteed to \emph{always} have a cost that is no worse than the
binary join plan. For binary join plans that are not well-behaved, we offer a
heuristic to transform them into equivalent well-behaved plans, while
minimizing additional cost.

\algnamefull (\algname) refers to the algorithm that takes a binary join plan as
input, transforms it into a well-behaved plan if needed, and then evaluates the resulting 2-phase \nsa plan using shredding.
Importantly, \algname
can be seamlessly integrated with an existing query optimizer that
generates traditional binary join plans, providing a provably robust solution
that \emph{consistently} avoids the diamond problem. Additionally, \algname is
guaranteed to be robust \emph{without regret} on the class of well-behaved
binary join plans.

In comparison, while BKN observe that 2-phase \lande plans can achieve
instance-optimality, they adopt a cost-optimisation-based approach to generating
\lande plans that does not require, nor guarantee these plans to be 2-phase. As
a result, the generated plans are not guaranteed to be instance-optimal. Thus,
as with binary join algorithms, the robustness of the system still depends on
the quality of the cost estimation and the optimizer. In contrast, the rewriting
we propose in this paper is always provably robust, and without regret on a
clear subclass.

(5) We implement \algname inside Apache
Datafusion~\cite{DBLP:conf/sigmod/LambSHCKHS24}, a high-performance
main-memory-based columnar interpreted query engine written in Rust.  Our experimental
set-up comprises multiple established benchmarks and includes \nrofqueries\
queries evaluated over real-world data. We show that the performance of \algname is \emph{always} competitive with that
of binary join plans, and often much better---improving performance for \rev{$85.3$}\%
of the queries with speedups up to \rev{62.5}x---while at the same time guaranteeing
robustness.  \rev{Additionally, our implementation of \algname shows relative  speedups that are comparable, and sometimes exceed, those observed by BKN  in their compiled query engine.  The  idea of \lande decomposition  can hence be successfully formulated and implemented in interpreted query engines, in particular  column stores.}

In summary, we show how to process acyclic joins instance-optimally and without
regret. We hope that this perspective can help system engineers to better
understand \ya, and pave the way for its adoption into existing systems.

This paper is organized as follows. We introduce background in
Section~\ref{sec:prelim}, \nsa in Section~\ref{sec:nsa}, and 
shredding in Section~\ref{sec:representation-and-processing}. We discuss
asymptotic complexity and instance-optimality of 2-phase \nsa in
Section~\ref{sec:nsa-instance-optimal}, and cost-based complexity and \algname
in Section~\ref{sec:left-deep}. We discuss experiments in
Section~\ref{sec:experiments}, and conclude in
Section~\ref{sec:discussion}. Related work is discussed throughout the paper.
\inConfVersion{Interested reviewers may find the proofs of formal statements in
  the full paper version~\cite{anon-full-version}.}  \inFullVersion{Full proofs
  of formal statements are given in the Appendix.}

\section{Preliminaries and Background}
\label{sec:prelim}

For a natural number $n > 0$ we denote the
set $\{1,\dots,n\}$ by $[n]$.  
We are concerned with the evaluation of \emph{natural join queries}, a.k.a.
full conjunctive queries, which are queries of the form:
\begin{equation}
  \label{eq:1}
  Q = R_1(\x_1) \Join \dots \Join R_k(\x_k).
\end{equation}
Here, $k \geq 1$; each $R_i$ is a relation symbol; and each $\x_i$ is a tuple of
pairwise distinct attributes  that denotes the schema of
$R_i$, for $i \in [k]$.  Expressions of the form $R_i(\x_i)$ are called
\emph{atoms}.

Following the SQL-standard, we adopt bag semantics for join queries.  Each input
relation $R_i(\x_i)$ is assumed to be a bag (i.e., multiset) of input tuples
over $\x_i$, and $Q$ computes a bag of tuples over $\x_1 \cup \dots \cup
\x_k$. Tuple $t$ occurs in the result of $Q$ if for every $i \in [k]$ the tuple
$t[\x_i]$ (i.e., $t$ projected on $\x_i$), occurs with multiplicity $m_i > 0$ in
input relation $R_i$. The result multiplicity of $t$ is then
$m_1 \times \dots \times m_k$.  In what follows, we use doubly curly braces
$\bag{\dots}$ to denote bags as well as bag comprehension and denote by
$\supp(M)$ the \emph{set}, without duplicates, of all elements present in a bag $M$.

\begin{example}
  \label{ex:queries}
  We 
  use $\threeq = R(x,y) \Join S(y,z) \Join T(z,u)$ as
  an example query throughout the paper. The query is over binary relations and can be seen to compute graph paths of length three.
\end{example}

\smallskip\noindent\textbf{Binary Join Plans.}
The standard approach to processing a join query $Q$ is to compute one binary join
at a time. A \emph{binary plan} (also known as a binary join order) is a rooted
binary tree where each internal node is a join operator $\Join$ and each leaf
node is one of the atoms $R_i(\x_i)$ of the query. To be correct under  bag semantics, it is required that each atom occurs
exactly as many times in the plan as it occurs in $Q$.  We will only
consider such valid plans in what follows. A binary plan is \emph{left-deep} if
the right child of every join node is a leaf; it is \emph{right-deep} if the
left child of every join node is a leaf; and it is \emph{bushy} otherwise.
\inFullVersion{For example, valid plans for 
$\threeq$ are
$(R \Join S) \Join T$, which is left-deep, and $R \Join (S \Join T)$ which is
right-deep. An example of a bushy plan is 
$(R \Join S) \Join ((T \Join U) \Join V)$. }

We interpret binary plans as physical query plans where all the joins are
evaluated by means of hash-joins. We focus on hash-joins as they are the most
common type of joins in database systems. Concretely, every join node in a
binary plan indicates a hash-join where the left child is the probe side and the
right child is the build side. Leaf nodes indicate input relations.  

\begin{example}
  \label{ex:threeq-binaryjoin}
  Consider the binary plan $P = (R \Join S) \Join T$ for $Q_3$.
  Figure~\ref{fig:threeq-inputs} illustrates two input databases. In database
  $\db_1$, every relation has $N$ tuples and every tuple joins with exactly one
  tuple of the other relations. On this database $\threeq$ hence returns $N$
  output tuples. Processing $\threeq$ on $\db_1$ by means of left-deep plan $P$
  involves building a hash table on $S$ and $T$; $\card{R}$ probes of
  $R$-tuples in the hash table on $S$; and $\card{R \Join S} = N$ probes into
  the hash table on $T$, hence doing $\bigo(N)$ work in total, which is optimal.

  The second database $\db_2$ has $N+1$ tuples in $R$ and $T$, and $2N$ tuples
  in $S$. \rev{Specifically, the relations are defined as follows:
\begin{align*}
  R &= \{(x_1, y_1)\} \cup \setof{(x_{i+1}, y_{N+1})}{i \in [N]} \\
  S &= \setof{(y_i, z_1)}{i \in [N]} \cup \setof{(y_{N+1},z_{i+1})}{i\in [N]} \\
  T &= \setof{(z_1, u_{i})}{i \in [N]} \cup \{(z_{N+1},u_{N+1})\}
\end{align*}}While there are only $2N$ output tuples to be produced, plan $P$ is
  $\Omega(N^2)$ since it will do at least $\card{R \join S} = N^2 + 1$ probes
  into $T$.  It hence wastes time computing tuples in $R \Join S$ which in the
  end do not contribute to the output.
\end{example}

\begin{figure}[tbp]
 \begin{minipage}[b]{0.13\columnwidth}
  \centering
  \subcaptionbox{\threej \label{fig:join-tree-q3}\label{fig:join-trees}}{\begin{tikzpicture}[scale=0.8, every node/.style={transform shape}]
\node (R) at (0,0) {$R(x,y)$};
    \node (S) at (0,-1) {$S(y,z)$};
    \node (T) at (0,-2) {$T(z,u)$};

\draw[-] (R) -- (S);
    \draw[-] (S) -- (T);
\end{tikzpicture}
 } \end{minipage}
  \centering
  \begin{minipage}[b]{0.40\columnwidth} \centering
      \subcaptionbox{Database $\db_1$ \label{fig:threeq-inputs:good}}{\begin{tikzpicture}[node distance=0.9cm and 0.8cm, auto, transform shape, scale=0.9]

\node (x1) {$x_1$};
    \node (y1) [right of=x1] {$y_1$};
    \node (z1) [right of=y1] {$z_1$};
    \node (u1) [right of=z1] {$u_1$};
    
    \node (x2) [below of=x1, node distance=0.5cm] {$x_2$};  \node (y2) [right of=x2] {$y_2$};
    \node (z2) [right of=y2] {$z_2$};
    \node (u2) [right of=z2] {$u_2$};
    
    \node (xdots) [below of=x2, node distance=0.3cm] {$\vdots$};  \node (ydots) [right of=xdots] {$\vdots$};
    \node (zdots) [right of=ydots] {$\vdots$};
    \node (udots) [right of=zdots] {$\vdots$};
    
    \node (xn) [below of=xdots, node distance=0.5cm] {$x_N$};  \node (yn) [right of=xn] {$y_N$};
    \node (zn) [right of=yn] {$z_N$};
    \node (un) [right of=zn] {$u_N$};
    
\draw[-] (x1) -- (y1) node[midway, above, yshift=3pt] {$R$};
    \draw[-] (y1) -- (z1) node[midway, above, yshift=3pt] {$S$};
    \draw[-] (z1) -- (u1) node[midway, above, yshift=3pt] {$T$};
    
    \draw[-] (x2) -- (y2);
    \draw[-] (y2) -- (z2);
    \draw[-] (z2) -- (u2);
    
    \draw[-] (xn) -- (yn);
    \draw[-] (yn) -- (zn);
    \draw[-] (zn) -- (un);

\end{tikzpicture}       }
  \end{minipage}
\begin{minipage}[b]{0.40\columnwidth} \centering
      \subcaptionbox{Database $\db_2$ \label{fig:threeq-inputs:bad}}{\begin{tikzpicture}[node distance=1cm and 0.8cm, auto, transform shape, scale=0.9]

\node (x1) {$x_1$};
    \node (y1) [right of=x1] {$y_1$};
    \node (z1) [right of=y1] {$z_1$};
    \node (u1) [right of=z1] {$u_1$};
    
    \node (y2) [below of=y1, node distance=0.5cm,color=gray] {$y_2$};
    \node (u2) [below of=u1, node distance=0.5cm] {$u_2$};

    \node (ydots) [below of=y2, node distance=0.3cm,color=gray] {$\vdots$};
    \node (udots) [below of=u2, node distance=0.3cm] {$\vdots$};

    \node (yN) [below of=ydots, node distance=0.5cm,color=gray] {$y_N$};
    \node (uN) [below of=udots, node distance=0.5cm] {$u_N$};

\draw[-] (x1) -- (y1) node[midway, above, yshift=3pt] {$R$};
    \draw[-] (y1) -- (z1) node[midway, above, yshift=3pt] {$S$};
    \draw[-] (z1) -- (u1) node[midway, above, yshift=3pt] {$T$};
    
    \draw[-,color=gray] (y2) -- (z1);
    \draw[-,color=gray] (yN) -- (z1);
    \draw[-] (z1) -- (u2);
    \draw[-] (z1) -- (uN);

\node (yN1) [below of=yN, node distance=0.5cm] {$y_{N+1}$};
    \node (x2) [left of=y2] {$x_2$}; \node (z2) [right of=y2,color=gray] {$z_2$}; 

    \node (x3) [below of=x2, node distance=0.5cm] {$x_3$};
    \node (z3) [below of=z2, node distance=0.5cm,color=gray] {$z_3$};

    \node (xdots) [below of=x3, node distance=0.3cm] {$\vdots$};
    \node (zdots) [below of=z3, node distance=0.3cm,color=gray] {$\vdots$};

    \node (xN1) [below of=xdots, node distance=0.5cm] {$x_{N+1}$};
    \node (zN1) [below of=zdots, node distance=0.5cm] {$z_{N+1}$};
    \node (uN1) [right of=zN1] {$u_{N+1}$};

\draw[-] (x2) -- (yN1);
    \draw[-,color=gray] (yN1) -- (z2);
    \draw[-] (x3) -- (yN1);
    \draw[-,color=gray] (yN1) -- (z3);
    \draw[-] (xN1) -- (yN1);
    \draw[-] (yN1) -- (zN1);
    \draw[-] (zN1) -- (uN1);

\end{tikzpicture}       }
  \end{minipage}

  \caption{Join tree $\threej$ for the three-path query $\threeq$, and two input
    databases.  Tuples in $\db_2$ not contributing to the final output are in
    gray.}
  \label{fig:threeq-inputs}
\end{figure}  

While we may be tempted to think that we were just unlucky in choosing an
suboptimal binary plan to process $\db_2$ in the previous example, this is not
the case: it is straightforward to verify that \emph{any} binary join plan for
$\threeq$ will produce a quadratic subresult. As such, binary join plans are
highly effective on certain inputs but cannot efficiently process joins on
\emph{all} inputs, even if the query is acyclic---a concept that we introduce next.

\smallskip\noindent\textbf{Acyclicity and Yannakakis' Algorithm.}  A join query
$Q$ is
\emph{acyclic} if
it admits a join tree~\cite{DBLP:conf/stoc/BeeriFMMUY81,DBLP:journals/jacm/Fagin83}. A \emph{join tree for $Q$} is a rooted undirected tree
$J$ in which each node is an atom of $Q$. To be correct under bag semantics, it
is required that each atom in $Q$ appears exactly as as many times in $J$ as it
does in $Q$.  Join trees are required to satisfy the \emph{connectedness
  property}: for every attribute $x$, all the nodes containing $x$ form a
connected subtree of $J$.  To illustrate, Figure~\ref{fig:join-trees} shows a
join tree for $\threeq$.

Checking whether a query is acyclic and constructing a join tree if it exists
can be done in linear time w.r.t. the size of the query by means of the GYO
algorithm~\cite{DBLP:conf/compsac/YuO79,graham-gyo,DBLP:journals/siamcomp/TarjanY84}. A
seminal result by Yannakakis~\cite{DBLP:conf/vldb/Yannakakis81} states that
acyclic join queries can be processed \emph{instance-optimally} under data complexity, i.e., in time
that is asymptotically linear in the size of the input plus the output. Yannakakis' Algorithm (\ya) does so by fixing a
join tree and computing in three passes. Define the \emph{semijoin
  $R \semijoin S$} of bag $R$ by $S$ to be the bag  containing all $R$-tuples
for which a joining tuple in $S$ exists. If a tuple $t$ appears in
$R \semijoin S$ it has the same multiplicity as in $R$.
\begin{compactenum}[1.]
\item The first pass operates bottom-up over the join tree. For the leaves there
  is nothing to do. When we reach an internal node $R$ with children
  $S_1, \dots, S_k$ \ya will replace $R$ by the semijoin of $R$ and all of its
  children, i.e., we set
  $R := (\dots ((R \semijoin S_1) \semijoin S_2) \dots S_k)$. 
\item The second pass operates top-down over the join tree. There is nothing to
  do for the root. For all other nodes $R$ with parent $P$, $R$ is replaced by the semijoin of $R$ and its parent,
  $ R:= R \semijoin P$. 
\item The final pass uses standard binary joins to join the relations resulting
  from the second pass.  While \ya is typically described to again work
  bottom-up over the join tree, any binary join plan $P$ for $Q$ that avoids
  needless Cartesian products\footnote{Meaning that if in a subplan
    $P' = P_1 \join P_2$ of $P$ no attributes are shared between $P_1$ and
    $P_2$, then the same must hold for all ancestors of $P'$.} can be used in
  this step.
\end{compactenum}

\smallskip
\noindent The first two passes are known as a \emph{full semijoin reduction}
and remove so-called \emph{dangling tuples} from the input: input tuples that
cannot be joined to form a complete join result. Once dangling tuples are
removed, standard binary joins can be used to compute the actual join result. At
that point any intermediate result tuple produced is guaranteed to participate
in at least one output tuple.

\begin{example}
  \label{ex:yann-q3-reduction}
  Reconsider $\threeq$ and the input database $\db_2$ from
  Example~\ref{ex:threeq-binaryjoin}. Assume we execute \ya using the join tree
  $\threej$ for $\threeq$ shown in Figure~\ref{fig:join-trees}.  
  \rev{The bottom-up pass of the algorithm
  first replaces $S$ with $S\semijoin T$,
  removing tuples over $z_2, \ldots, z_N$ from $S$.
  Note that this retains the full range of $y$-values in $S$.
  Next, $R$ is semijoined with the reduced $S$.
  However this does not remove any tuples from $R$, 
  since all $y$-values in $R$ are also in $S$.
  The top-down pass then replaces $S$ with $S \semijoin R$,
  removing tuples over $y_2, \ldots, y_N$.
  The final semijoin $T\semijoin S$ again removes nothing.
  At this point,} all gray-colored tuples in
  Figure~\ref{fig:threeq-inputs:bad} are removed, leaving only the black-colored tuples. 
  On this reduced database, any binary join plan without Cartesian product runs
  instance-optimally.  Note that the removal of dangling tuples is essential, as
  we know from Example~\ref{ex:threeq-binaryjoin} and the subsequent discussion
  that on the original input $\db_2$ every binary join plan will require
   $\Omega(N^2)$ time.
\end{example}

A straightforward way to implement \ya in a database engine is to record the
sequence of joins and semijoins that \ya does in a physical query
plan~\cite{DBLP:journals/corr/abs-2303-02723}. These kinds of query plans, which
we will refer to as \emph{semijoin plans}, are 
binary join plans where leaf nodes are replaced by trees that compute semijoins
on input relations. For example, the right of Figure~\ref{fig:semijoin_plans} shows a
semijoin plan for $\threeq$, corresponding to executing \ya using the join tree
$J_3$ of Figure~\ref{fig:join-tree-q3} and using the left-deep join order
$(R \join S) \join T$ in the last phase.

\begin{figure}[tbp]
  \begin{tikzpicture}

    \begin{scope}[xshift=-4cm,
        grow=down, level distance=0.7cm, sibling distance=1.3cm, level 2/.style={sibling distance=1.8cm}, level 3/.style={sibling distance=1cm},
        scale=0.75, every node/.style={transform shape}
        ]

        \node {$\join$ }
          child { node {$\join$}
            child { node {$\semijoin$}
              child {node{$R(x,y)$}}
              child { node {$\semijoin$}
                child {node {$S(y,z)$}}
                child {node {$T(z,u)$}}}}         
            child { node {$\semijoin$}
              child {node {$S(y,z)$}}
              child {node {$T(z,u)$}}}
          } 
          child { node {$T(z,u)$}};
          
      \end{scope}

\begin{scope}[xshift=-1.5cm,
    grow=down, level distance=0.7cm, sibling distance=2.5cm, level 2/.style={sibling distance=2cm}, level 3/.style={sibling distance=1cm}, 
    scale=0.75, every node/.style={transform shape}
    ]

    \small
    \node {$\join$ }
child [yshift=-1.5cm] { node {$\semijoin$}
            child {node{$R(x,y)$}}
            child [xshift=-1cm] {node {$\semijoin$}
                    child [xshift=-0.4cm] {node {$S(y,z)$}}
                    child {node {$T(z,u)$}}
                    }
            }
child { node [yshift=0.3cm]{$\join$}
                child {node [yshift=0.3cm]{$\semijoin$}
                    child {node{$S(y,z)$}}
                    child {node {$\semijoin$}
                        child {node{$R(x,y)$}}
                        child {node {$\semijoin$}
                                child [xshift=-0.4cm] {node {$S(y,z)$}}
                                child [xshift=-0.2cm] {node {$T(z,u)$}}
                        }
                    }
                }
child {node [yshift=0.3cm]{$\semijoin$}
                    child {node {$T(z,u)$}}
                    child {node{$\semijoin$}
                        child {node{$S(y,z)$}}
                        child [xshift=0cm] {node {$\semijoin$}
                            child {node{$R(x,y)$}}
                            child {node {$\semijoin$}
                                    child {node {$S(y,z)$}}
                                    child {node {$T(z,u)$}}
                            }
                        }
                    }
                }
        };
\end{scope}

\end{tikzpicture}   \caption{Semijoin plans induced by \ya on join tree \threej  (Fig.\ref{fig:join-trees}). Left: pass two and three combined. Right: all three passes.  \label{fig:semijoin_plans}
  }
\end{figure}

Unfortunately, this straightforward implementation of \ya creates significant
overhead when the input database contains no, or only few dangling
tuples. Indeed, for $\threeq$ observe that every relation now participates in at
least one join and at least one semijoin, while some relations, like $S$,
participate in five semijoins. Semijoins are also executed by means of hashing and therefore also incur build and probing costs even if they do not remove
any tuples in the concrete input database that we execute on.  This commonly
happens: BKN note that on the Join Order
Benchmark~\cite{DBLP:journals/vldb/LeisRGMBKN18}, this way of implementing \ya
by adding full semijoin reductions yields a 5-fold slowdown compared to binary
join plans.

One way to overcome this limitation is to adopt a cost-based approach and
selectively add semijoin operators only when they are deemed
useful~\cite{DBLP:conf/icde/StockerKBK01}. However, this no longer guarantees
instance-optimality. Another possibility, which preserves instance-optimality,
is to observe that instead of doing the full three passes of classical \ya, the
second and third pass can actually be
combined~\cite{DBLP:conf/csl/BaganDG07,DBLP:journals/vldb/IdrisUVVL20,DBLP:conf/sigmod/IdrisUV17}. It
then suffices to do only the first pass of semijoin-reductions. This
modification of \ya leads to somewhat simpler plans as illustrated in the left
of Figure~\ref{fig:semijoin_plans} for our running example $\threeq$ and join
tree $\threej$. Note, however, that while this reduces the overhead, it does not
completely eliminate it since $T$ and $S$ continue to participate in multiple
(semi)joins. Recent so-called \emph{enumeration-based} join evaluation
algorithms go one step further: they compute only the semijoin
$R \semijoin (S \semijoin T)$ and reuse the hash tables created during the
semijoin to \emph{enumerate} the join result $R \join S \join T$ using a
specialized
algorithm~\cite{DBLP:conf/csl/BaganDG07,DBLP:journals/vldb/IdrisUVVL20,DBLP:conf/sigmod/IdrisUV17}. While
such enumeration algorithms have previously been difficult to cast as operators
in a physical query plan algebra, and have to date been limited to specialized
research prototypes, \lande/\nsa plans will provide exactly this functionality.

\smallskip\noindent\textbf{In conclusion.} Binary join plans suffer from the diamond problem. By contrast, semijoin plans induced by running \ya (in full, or with the latter two phases combined) are instance-optimal and hence avoid the diamond problem, but on common inputs they may suffer from a constant-factor slowdown  compared to binary join plans.
Our objective in this paper, therefore, is to  engineer
the instance-optimality of \ya in a database engine without regret.

 \section{Nested Semijoin Algebra}
\label{sec:nsa}

In this section, we provide a formal syntax and semantics for \lande plans,
including how
they interact with other relational algebra (\ra) operators, in terms of a set of
nested relational operators that we call the \emph{Nested Semijoin Algebra}
(\nsa). 
Having specified the data model and nested
operators required to support \lande plans, we subsequently use this
formalisation in Section~\ref{sec:representation-and-processing} to derive an
implementation strategy of \lande plans in interpreted query engines.

The nested relational model is an extension of the standard relational model.
In a \emph{nested relation}, a tuple may
consist not only of scalar data values but also of entire relations in turn. The
\emph{nested relational algebra} (NRA) for querying nested relations is obtained
by generalizing the  operators of  relational algebra (selection,
projection, join, \dots) to work on nested relations, and by adding two extra
operators: \emph{nesting} and
\emph{unnesting}~\cite{DBLP:journals/acr/ThomasF86}. Many variants of the nested
relational model have been proposed, including extensions that allow for mixed
collection types such as sets, bags, lists,
arrays~\cite{DBLP:journals/tcs/BunemanNTW95} as well as
dictionaries~\cite{DBLP:conf/vldb/DeutschPT99}. In this paper, we consider a
variant where each (nested) relation is bag-based, and where we also have
dictionaries. To make the connection with \lande plans, we depart from the
standard set of operators of NRA, and instead introduce a set of operators that
we call the Nested Semijoin Algebra (\nsa).

\inFullVersion{
\smallskip\noindent\textbf{Schemes and Nested Relations.} We refer to the
attributes that appear in the schema of classical flat relations as \emph{flat}
attributes. Let $\attrs$ denote the set of all flat attributes. The set
$\hf{\attrs}$ of \emph{hereditarily finite sets over} $\attrs$ is the smallest
set containing $\attrs$, such that if $X_1,\dots, X_n \in \hf{\attrs}$ then also
$\{X_1,\dots,X_n\} \in \hf{\attrs}$. A \emph{scheme} is an element
$X \in \hf{\attrs} \setminus \attrs$ in which no flat attribute occurs more than
once. Here, an element $x$ is said to occur in $X$ if $x \in X$ or $x$ occurs
recursively in some set $Y \in X$. We write $\flatsub(X)$ for the set of all
flat attributes occurring in $X$, and $\sub(X)$ for the set of all schemes
occurring in $X$. Schemes are also called \emph{nested} attributes.  Note that a
flat attribute is not a scheme.  We range over flat attributes by lowercase
letters ($x$, $y$, \dots) and over nested attributes by uppercase letters ($X$,
$Y$, \dots), both from the end of the alphabet. A finite set of flat attributes
is denoted by $\seq{x}$. 

Fix a scheme $X$. A \emph{relation} over a $X$ is a finite bag of tuples over $X$. Here, a \emph{tuple over} $X$ is a mapping $t$ on $X$ such that $t(x)$ is a  scalar data value (of appropriate type) for each flat attribute $x \in X \cap \attrs$, and $t(Y)$ is a non-empty relation over $Y$ for each nested attribute $Y \in X \setminus \attrs$. Note that if $X$ is \emph{flat}, i.e., if
$X \subseteq \attrs$, then this definition of a relation over $X$ coincides with the usual one. We call $R$ a \emph{flat relation} in that case. We restrict inner nested relations to be non-empty as in this paper we always start from flat relations and the operators that we consider will never introduce empty inner nested relations.
We write $R\colon X$ and $t\colon X$ to denote that $R$ is a relation (resp. $t$
is a tuple) over scheme $X$. We write $\card{R}$ denote the \emph{cardinality} of
$R$, i.e., the total number of tuples in $R$. Note that $\card{R}$ only refers
to the number of tuples in the outer-most bag of $R$, and does not say anything
about the cardinality of the inner-nested relations appearing in those tuples.
}

\inConfVersion{

\smallskip\noindent\textbf{Schemes and Nested Relations.} \rev{Just like flat relations have flat schemes, nested relations have nested schemes. }
We refer to the
attributes that appear in the scheme of classical flat relations as \emph{flat}
attributes. Let $\attrs$ denote the set of all flat attributes. \rev{
 A \emph{(nested) scheme} is a finite set $X$, like $\{x, \{y\}, \{u,\{v\}\}\}$, that consists of flat attributes ($x$ in this case) and other schemes (i.e., $\{y\}$ and $\{u,\{v\}\}$). No flat attribute is allowed to occur  more than 
once, so 
$\{x, \{y\}, \{u,\{x\}\}\}$ is not a valid scheme. Schemes are also called \emph{nested}
attributes. }  We range over flat
attributes by lowercase letters $(x,y,\dots)$;  over schemes by uppercase letters ($X$, $Y$, \dots), both from the end of the
alphabet; and over finite sets of flat attributes by $\x$. \rev{We write $\flatsub(X)$ for the set of all flat attributes occurring
somewhere in $X$ (either directly or in some inner nested scheme); and $\sub(X)$
for the set of all schemes occurring in $X$ (again, either directly or in some
inner nested scheme). So for $X = \{x, \{y\}, \{u,\{v\}\}\}$,
$\flatsub(X) = \{x,y,u,v\}$ and $\sub(X) = \{ \{y\}, \{u,\{v\}\}, \{v\} \}$.}

Fix a scheme $X$. A \emph{relation} over a $X$ is a finite bag of tuples over $X$. Here, a \emph{tuple over} $X$ is a mapping $t$ on $X$ such that $t(x)$ is a  scalar data value (of appropriate type) for each flat attribute $x \in X$, and $t(Y)$ is a non-empty relation over $Y$ for each nested attribute $Y \in X$. Note that if $X$ is \emph{flat}, i.e., if
$X \subseteq \attrs$, then this definition of a relation over $X$ coincides with the usual one. We call $R$ a \emph{flat relation} in that case. We restrict inner nested relations to be non-empty as in this paper we always start from flat relations and the operators that we consider will never introduce empty inner nested relations.
We write $R\colon X$ and $t\colon X$ to denote that $R$ is a relation (resp. $t$
is a tuple) over scheme $X$. We write $\card{R}$ denote the \emph{cardinality} of
$R$, i.e., the total number of tuples in $R$. Note that $\card{R}$ only refers
to the number of tuples in the outer-most bag of $R$, and does not say anything
about the cardinality of the inner-nested relations appearing in those tuples.
Figure~\ref{fig:shredding} shows a nested relation with cardinality 2 and
  scheme $\{x, \{y\}, \{u, \{v\}\}\}$.
}

\begin{figure}[tbp]
  \begin{minipage}[t]{0.3\columnwidth}
  \centering
  \begin{tikzpicture}[
table/.style={
        matrix of nodes,
        nodes in empty cells,
row sep=-\pgflinewidth,
        column sep=-\pgflinewidth,
    },
    highlight/.style={
        draw=black,
        rounded corners,
    },
    scale=0.8, every node/.style={transform shape}
]

\newcommand{\mydist}{0.1cm}
\newcommand{\myddist}{0.2cm}
\newcommand{\mydddist}{0.3cm}

\node (root) {
     \begin{tikzpicture}
     \matrix (table1) [table] {
        $x$ & $\{y\}$            & $\{u,$ & $\{v\}\}$ \\
        \hline \\
        $a_1$ &|[name=b1]|$b_1$  &|[name=c1]|$c_1$  &|[name=d1]|$d_1$ \\
              &|[name=b2]|$b_2$  &        &|[name=d2]|$d_2$ \\[\mydist]
              &                   & $c_2$  &|[name=d1b]|$d_1$\\[\mydddist]
        $a_2$ &|[name=b1b]|$b_1$  &|[name=c3]|$c_3$ &|[name=d3]|$d_3$ \\
              &|[name=b2b]|$b_3$  &        &|[name=d4]|$d_4$ \\
     };
\draw[highlight] (b1.north west) rectangle (b2.south east);
     \draw[highlight] (d1.north west) rectangle (d2.south east);
     \draw[highlight] (d1b.north west) rectangle (d1b.south east);
     \draw[highlight] ($(c1.north west) + (0,\myddist)$) rectangle ($(d1b.south east) + (\mydist,-\mydist)$);
     \draw[highlight] (b1b.north west) rectangle (b2b.south east);
     \draw[highlight] ($(c3.north west) + (0,\myddist)$) rectangle ($(d4.south east) + (\mydist,-\mydist)$);
     \draw[highlight] (d3.north west) rectangle (d4.south east);
     \end{tikzpicture}
};

\end{tikzpicture}

   \end{minipage}
  \begin{minipage}[t]{0.65\columnwidth}
    \centering

\addtolength{\tabcolsep}{-2pt}
\small 

\begin{tikzpicture}[node distance=0.5cm]

\newcommand{\nodedistx}{0.2cm}
\newcommand{\nodedisty}{0.2cm}
\newcommand{\g}[1]{\textcolor{gray}{#1}}
\newcommand{\mybot}{0}

\tikzstyle{detailnode}=[inner sep=0pt, text centered]

  \node [detailnode] (R) { \scriptsize
\begin{tabular}{c|cc|cc}
    \multicolumn{5}{c}{$\phys{R}$} \\ \hline \hline
$x$ & \multicolumn{2}{c|}{$\{y\}$} & \multicolumn{2}{c}{$\{u,\{v\}\}$}\\
    & $\holshort{}$ & $w$ & $\holshort{}$ & $w$ \\
    \hline
    $a_1$  &  $2$& 2&$2$&3  \\
    $a_2$  &  $4$& 2&$3$&2 \\
\end{tabular}
};

\node [detailnode, anchor=north west] (uv)
      at ($ (R.north east) + (\nodedistx, 0em) $)  {\scriptsize
\begin{tabular}{c|c|cc|c}
    \multicolumn{5}{c}{$\store{R}(\{u,\{v\}\})$}\\ \hline \hline
&$u$ & \multicolumn{2}{c|}{$\{v\}$} &$\nxt$ \\
&  & $\holshort{}$ & $w$  \\ \hline  
    \g{$1$}  &  $c_1$ & $2$&2 &$\mybot$\tikzmark{s3}\\
    \g{$2$}  &  $c_2$ & $3$& 1&$\shortuparrow$\tikzmark{e3} \\
    \g{$3$}  &  $c_3$ & $5$& 2&$\mybot$\\    
\end{tabular}
};

\node [detailnode, anchor=north west] (v)
      at ($ (R.south west) + (0em, -\nodedisty) $)  {\scriptsize
\begin{tabular}{ccc}
    \multicolumn{3}{c}{$\store{R}(\{v\})$}\\ \hline \hline
       & $v$ & $\nxt$ \\ \hline
     \g{$1$}  &  $d_1$ & $\mybot$\tikzmark{s4}\\
     \g{$2$}  &  $d_2$ & $\shortuparrow$\tikzmark{e4} \\
     \g{$3$}  &  $d_1$ & $\mybot$ \\
     \g{$4$}  &  $d_3$ & $\mybot$\tikzmark{s5} \\
     \g{$5$}  &  $d_4$ & $\shortuparrow$\tikzmark{e5}\\
\end{tabular}
};

\node [detailnode, anchor=north west] (v)
      at ($ (uv.south west) + (0em, -\nodedisty) $)  {\scriptsize
\begin{tabular}{ccc}
    \multicolumn{3}{c}{$\store{R}(\{y\})$}\\ \hline \hline
       & $y$ & $\nxt$ \\ \hline
     \g{$1$}  &  $b_1$ & $\mybot$\tikzmark{s1}\\
     \g{$2$}  &  $b_2$ & $\shortuparrow$\tikzmark{e1} \\
     \g{$3$}  &  $b_1$ & $\mybot$\tikzmark{s2}\\
     \g{$4$}  &  $b_3$ & $\shortuparrow$\tikzmark{e2} \\
\end{tabular}
};
\end{tikzpicture}

     \end{minipage}
  \caption{\label{fig:shredding} (left) A nested relation $R$. (right) Its shredded representation $\rep{R}$. The gray numbers indicate tuple offsets; $\nxt$ points to the next tuple (via $\shortuparrow$) or is 0 when there is none. }
\end{figure}

\inFullVersion{
\begin{example}
  \label{ex:nested-rel}
  Figure~\ref{fig:shredding} shows a nested relation with cardinality 2 and
  scheme $\{x, \{y\}, \{u, \{v\}\}\}$. This scheme has two nested attributes,
  namely $\{y\}$ and $\{u, \{v\}\}$, and one flat attribute $x$.
\end{example}
}

We adopt the following notation on tuples. If $s\colon X$ and $t\colon Y$ are tuples over disjoint schemes then $s \circ t$ denotes their concatenation, which is a tuple over $X \cup Y$. 
Furthermore, if $Z \subseteq X$ then $\restr{t}{Z}$ denotes the restriction
(i.e., projection) of mapping $t$ to the attributes in $Z$.

\smallskip\noindent\textbf{Dictionaries.}  A \emph{dictionary scheme} is an
expression of the form $\dscheme{\y}{Z}$ \inConfVersion{\rev{such that no flat attribute in $\y$ occurs anywhere in $Z$.}} \inFullVersion{such that no flat attribute of $\y$ occurs anywhere in $Z$, i.e. $\y \cap \flatsub(Z) = \emptyset$.}
A \emph{dictionary over} $\dscheme{\y}{Z}$ is a finite mapping $D$ that maps
$\y$-tuples to non-empty relations over $Z$. The tuples in the domain $\dom(D)$ of $D$ are
called the \emph{keys} of $D$. The \emph{cardinality} of $D$, denoted $\card{D}$ is the number of keys. We
write $D\colon \dscheme{\y}{Z}$ to indicate that $D$ is a dictionary over
$\dscheme{\y}{Z}$. Conceptually, a dictionary is a special kind of nested
relation with scheme $\y \cup \{Z\}$; in contrast to a nested relation it also
allows to lookup keys.
\begin{figure}
  \begin{center}

\small

\addtolength{\tabcolsep}{-2pt} \begin{tikzpicture}[node distance=0.5cm, scale=0.8, every node/.style={transform shape}]
  \tikzstyle{detailnode}=[inner sep=0pt, text centered]

  \newcommand{\nodedistx}{0.2cm}
  \newcommand{\nodedisty}{0.2cm}
  
  \newcommand{\distop}{0.02cm}
  \newcommand{\ddistop}{0.04cm}
  \newcommand{\tdistop}{0.06cm}
  \newcommand{\nodedist}{0.5cm}
  \newcommand{\rowdist}{0.1cm}
  \newcommand{\halfrowdist}{0.05cm}
  \newcommand{\doublerowdist}{0.2cm}
  \newcommand{\triplerowdist}{0.3cm}
  \newcommand{\sizeshreddedtable}{\small}
  
  \newcommand{\selv}[1]{\texttt{[}#1\texttt{]}}
  \newcommand{\g}[1]{\textcolor{gray}{#1}}
  \newcommand{\mybot}{0}
  \renewcommand{\uparrow}{\shortuparrow}

  \tikzstyle{tabularnode}=[anchor=north,inner sep=3.6pt,text width=2.5cm, text centered]

  \tikzstyle{table}=[matrix of nodes,nodes in empty cells,inner sep=3pt,row sep=-\pgflinewidth, column sep=-\pgflinewidth]
  \tikzstyle{widertable}=[table,inner sep=4pt]
  \tikzstyle{smalltable}=[table,inner sep=2pt]

  \tikzstyle{nested}=[draw=black,rounded corners]

\node (root) {
    \begin{tikzpicture}
      \node (top) {\phantom{xx}$\unnest_{\{u\}}$};
      \matrix (table) [draw,smalltable,below=\ddistop of top,matrix anchor=table-1-2.north] {
         $x_1$ & $y_1$ & $z_1$ & $u_1$\\
         $x_1$ & $y_1$ & $z_1$ & $u_2$\\
         \rev{$x_2$} & \rev{$y_3$} & \rev{$z_3$} & \rev{$u_3$}\\
         $x_3$ & $y_3$ & $z_3$ & $u_3$\\
         };
    \end{tikzpicture}
  };

\node (unnest1) [below=\nodedist of root.south,xshift=-0.0cm] {
    \begin{tikzpicture}
      \node (top) {\phantom{xx}$\unnest_{\{z,\{u\}\}}$};
      \matrix (table) [draw,table,below=\ddistop of top,matrix anchor=table-1-2.north] {
          $x_1$ & $y_1$ & $z_1$ &|[name=u1]|$u_1$\\
                &       &       &|[name=u2]|$u_2$\\[\rowdist]
          $x_2$ & $y_3$ & $z_3$ &|[name=u3]| $u_3$\\[\rowdist]              
          $x_3$ & $y_3$ & $z_3$ &|[name=u3x3]| $u_3$\\        
        };
      \draw[nested] (u1.north west) rectangle (u2.south east);
      \draw[nested] (u3.north west) rectangle (u3.south east);
      \draw[nested] (u3x3.north west) rectangle (u3x3.south east);
    \end{tikzpicture}    
    };

\draw (root.south) -- (unnest1.north);

\node (semijoin2) [below=\nodedist of unnest1.south] {
  \begin{tikzpicture}
    \node (top) {\phantom{xx}$\nsemijoin$};
    \matrix (table) [draw,widertable,below=\tdistop of top,matrix anchor=table-1-2.north] {
        $x_1$ & $y_1$ &|[name=z1]| $z_1$ &|[name=u1]|$u_1$\\
              &       &       &|[name=u2]| $u_2$\\[\doublerowdist]
        $x_2$ & $y_3$ &|[name=z3]|$z_3$ &|[name=u3]| $u_3$\\[\doublerowdist]  
        $x_3$ & $y_3$ &|[name=z3x3]|$z_3$ &|[name=u3x3]| $u_3$\\        
      };
    \draw[nested] (u1.north west) rectangle (u2.south east);
    \draw[nested] ($(z1.north west) + (0,\halfrowdist)$) rectangle ($(u2.south east) + (\halfrowdist,-\halfrowdist)$);
    \draw[nested] (u3.north west) rectangle (u3.south east);
    \draw[nested] ($(z3.north west) + (0,\halfrowdist)$) rectangle ($(u3.south east) + (\halfrowdist,-\halfrowdist)$);
    \draw[nested] (u3x3.north west) rectangle (u3x3.south east);
    \draw[nested] ($(z3x3.north west) + (0,\halfrowdist)$) rectangle ($(u3x3.south east) + (\halfrowdist,-\halfrowdist)$);
  \end{tikzpicture}    
};

\draw (unnest1.south) -- (semijoin2.north);

\node (R) [below=\nodedist of semijoin2.south, xshift=-1.5cm] {
  \begin{tikzpicture}
    \node (top) {$R(x,y)$};
    \matrix (table) [draw,smalltable,below=\distop of top,matrix anchor=table-1-1.north,xshift=-0.2cm] {
        $x_1$ & $y_1$\\
        $x_2$ & $y_3$\\
        $x_3$ & $y_3$\\
      };
  \end{tikzpicture}    
};

\draw (semijoin2.south) -- (R.north);

\node (groupby3) [below=\nodedist of semijoin2.south, xshift=1cm] {
  \begin{tikzpicture}
    \node (top) {\phantom{xx}$\groupby{\{y\}}{\{z,u\}}$};
    \matrix (table) [draw,widertable,below=\tdistop of top,matrix anchor=table-1-2.north] {
       $y_1$ & $\rightarrow$ &|[name=z1]| $z_1$&|[name=u1]| $u_1$\\
                  &       &       &|[name=u2]| $u_2$\\[\doublerowdist]
      $y_2$ & $\rightarrow$ &|[name=z1y2]| $z_1$&|[name=u1y2]| $u_1$\\
                  &       &       &|[name=u2y2]| $u_2$\\[\doublerowdist]
            $y_3$ & $\rightarrow$ & |[name=z3]|$z_3$&|[name=u3]|$u_3$\\
      };
    \draw[nested] (u1.north west) rectangle (u2.south east);
    \draw[nested] ($(z1.north west) + (0,\halfrowdist)$) rectangle ($(u2.south east) + (\halfrowdist,-\halfrowdist)$);
    \draw[nested] (u1y2.north west) rectangle (u2y2.south east);
    \draw[nested] ($(z1y2.north west) + (0,\halfrowdist)$) rectangle ($(u2y2.south east) + (\halfrowdist,-\halfrowdist)$);
    \draw[nested] (u3.north west) rectangle (u3.south east);
    \draw[nested] ($(z3.north west) + (0,\halfrowdist)$) rectangle ($(u3.south east) + (\halfrowdist,-\halfrowdist)$);
  \end{tikzpicture}    
};

\draw (semijoin2.south) -- (groupby3.north);

\node (semijoin4) [below=\nodedist of groupby3.south] {
  \begin{tikzpicture}
    \node (top) {\phantom{x}$\nsemijoin$};
    \matrix (table) [draw,table,below=\distop of top,matrix anchor=table-1-2.north] {
       $y_1$ &$z_1$&|[name=u1]| $u_1$\\
             &       &|[name=u2]| $u_2$\\[\rowdist]
       $y_2$ &$z_1$&|[name=u1y2]| $u_1$\\
             &       &|[name=u2y2]| $u_2$\\[\rowdist]
       $y_3$ &$z_3$&|[name=u3]|$u_3$\\
      };
    \draw[nested] (u1.north west) rectangle (u2.south east);
    \draw[nested] (u1y2.north west) rectangle (u2y2.south east);
    \draw[nested] (u3.north west) rectangle (u3.south east);    
  \end{tikzpicture}    
};

\draw (groupby3.south) -- (semijoin4.north);

\node (S) [below=\nodedist of semijoin4.south, xshift=-1cm] {
  \begin{tikzpicture}
    \node (top) {$S(y,z)$};
    \matrix (table) [draw,smalltable,below=\distop of top,matrix anchor=table-1-1.north,xshift=-0.2cm] {
        $y_1$ & $z_1$\\
        $y_2$ & $z_1$\\
        $y_3$ & $z_2$\\
        $y_3$ & $z_3$\\
      };
  \end{tikzpicture}    
};

\draw (semijoin4.south) -- (S.north);

\node (groupby5) [below=\nodedist of semijoin4.south, xshift=1cm] {
  \begin{tikzpicture}
    \node (top) {$\groupby{\{z\}}{\{u\}}$};
    \matrix (table) [draw,table,below=\tdistop of top,matrix anchor=table-1-2.north] {
       $z1$ & $\rightarrow$ &|[name=u1]| $u_1$\\
            &               &|[name=u2]| $u_2$\\[\rowdist]
       $z_3$ & $\rightarrow$&|[name=u3]|$u_3$\\
      };
    \draw[nested] (u1.north west) rectangle (u2.south east);
    \draw[nested] (u3.north west) rectangle (u3.south east);    
  \end{tikzpicture}    
};

\draw (semijoin4.south) -- (groupby5.north);

\node (T) [below=\nodedist of groupby5.south] {
  \begin{tikzpicture}
    \node (top) {$T(z,u)$};
    \matrix (table) [draw,smalltable,below=\distop of top,matrix anchor=table-1-1.north,xshift=-0.2cm] {
        $z_1$ & $u_1$\\
        $z_1$ & $u_2$\\
        $z_3$ & $u_3$\\
      };
  \end{tikzpicture}    
};

\draw (groupby5.south) -- (T.north);

\node [right=-2.7cm of root,yshift=-0.2cm] {$(\rootnode)$}; \node [right=-3cm of unnest1,yshift=-0.1cm] {$(\depthone)$}; \node [right=-3.3cm of semijoin2,yshift=-0.1cm] {$(\depthtwo)$}; \node [right=-3.35cm of groupby3,yshift=-0.3cm] {$(\depththree)$}; \node [right=-2.5cm of semijoin4,yshift=-0.1cm] {$(\depthfour)$}; \node [right=-2.4cm of groupby5,yshift=-0.1cm] {$(\depthfive)$};

\node [detailnode, anchor=north west,right=0.3cm of groupby5,yshift=-0.1cm] (shreddict5store)
      {\sizeshreddedtable
\begin{tabular}{ccc}
    \multicolumn{3}{c}{$\store{\depthfive}(\{u\})$}\\ \hline \hline
       & $u$ & $\nxt$ \\ \hline
     \g{$1$}  &  $u_1$ & $\mybot$\\
     \g{$2$}  &  $u_2$ & $\uparrow$\\
     \g{$3$}  &  $u_3$ & $\mybot$ \\
\end{tabular}
};

\node [detailnode, anchor=north west,right=0.3cm of shreddict5store,yshift=-0.1cm] (shreddict5mapping)
      {\sizeshreddedtable
\begin{tabular}{l}
$\rep{\depthfive} =(\phys{h}_{\depthfive}, \store{\depthfive}) $
    \\
    \\
     $\phys{h}_{\depthfive}\colon z_1 \mapsto (2,2)$\\
     $\phantom{\phys{h}_{\depthfive}\colon} z_3 \mapsto (3,1)$\\
\end{tabular}
};

\node [detailnode,anchor=north west,right=0.3cm of semijoin4,yshift=-0.1cm] (shreddedreldepthfour) { \sizeshreddedtable
\begin{tabular}{ccc|cc}
    \multicolumn{5}{c}{$\phys{\depthfour}$} \\ \hline \hline
&$y$ & $z$ & \multicolumn{2}{c}{$\{u\}$}\\
   &    &   & $\holshort{}$ & $w$ \\
    \hline
    \g{1}&$y_1$  & $z_1$& $2$& 2\\
    \g{2}&$y_2$  & $z_1$& $2$& 2 \\
    \g{3}&$y_3$  & $z_2$& $0$& 0 \\
    \g{4}&$y_3$  & $z_3$& $3$& 1 \\    
\end{tabular}
};

\node [detailnode, anchor=north west,right=0.3cm of shreddedreldepthfour,yshift=0.1cm] (shreddedreldepthfourmapping)
      {\sizeshreddedtable
\begin{tabular}{l}
$\rep{\depthfour} =(\phys{\depthfour}, \store{\depthfive}, \sel{\MakeLowercase{\depthfour}}) $
    \\
    \\
$\sel{\MakeLowercase{\depthfour}} = \selv{1,2,4}$\\
\end{tabular}
};

\node [detailnode, anchor=north west,right=0.3cm of groupby3,yshift=-0.1cm] (shreddict3store)
      {\sizeshreddedtable
\begin{tabular}{ccc|cc}
    \multicolumn{5}{c}{$\store{\depththree}(\{z,\{u\}\})$}\\ \hline \hline
       & $z$ & $\nxt$ & \multicolumn{2}{c}{$\{u\}$}\\
       &     &        & $\holshort{}$ & $w$ \\ \hline
     \g{$1$}  &  $z_1$ & $\mybot$ &2&2\\
     \g{$2$}  &  $z_1$ & $\mybot$ &2&2\\
     \g{$3$}  &  $z_2$ & $\mybot$ &0&0\\
     \g{$4$}  &  $z_3$ & $\mybot$ &3&1\\
\end{tabular}
};

\node [detailnode, anchor=north west,right=0.3cm of shreddict3store,yshift=-0.1cm] (shreddict3mapping)
      {\sizeshreddedtable
\begin{tabular}{l}
$\rep{\depththree} =(\phys{h}_{\depththree}, \store{\depththree} \text{+} \store{\depthfive}) $
    \\
    \\
     $\phys{h}_{\depththree}\colon y_1 \mapsto (1,2)$\\
     $\phantom{\phys{h}_{\depththree}\colon} y_2 \mapsto (2,2)$\\
     $\phantom{\phys{h}_{\depththree}\colon} y_3 \mapsto (4,1)$\\
\end{tabular}
};

\node [detailnode,anchor=north west,right=0.3cm of semijoin2,yshift=-0.1cm] (shreddedreldepthtwo) { \sizeshreddedtable
\begin{tabular}{ccc|cc}
    \multicolumn{5}{c}{$\phys{\depthtwo}$} \\ \hline \hline
&$x$ & $y$ & \multicolumn{2}{c}{$\{z,\{u\}\}$}\\
   &    &   & $\holshort{}$ & $w$ \\
    \hline
    \g{1}&$x_1$  & $y_1$& $1$& 2\\
    \g{2}&$x_2$  & $y_3$& $4$& 1 \\
    \g{3}&$x_3$  & $y_3$& $4$& 1 \\
\end{tabular}
};

\node [detailnode, anchor=north west,right=0.3cm of shreddedreldepthtwo,yshift=0.1cm] (shreddedreldepthtwomapping)
      {\sizeshreddedtable
\begin{tabular}{l}
$\rep{\depthtwo} =(\phys{\depthtwo}, \store{\depththree} \text{+} \store{\depthfive}, \sel{\MakeLowercase{\depthtwo}}) $
    \\
    \\
    $\sel{\MakeLowercase{\depthtwo}} = \selv{1,2,3}$\\
\end{tabular}
};

\node [detailnode,anchor=north west,right=0.3cm of unnest1,yshift=-0.1cm] (shreddedreldepthone) { \sizeshreddedtable
\begin{tabular}{cccc|cc}
    \multicolumn{6}{c}{$\phys{\depthone}$} \\ \hline \hline
&$x$ & $y$ & $z$ & \multicolumn{2}{c}{$\{u\}$}\\
   &    &     & &$\holshort{}$ & $w$ \\
    \hline
    \g{1}&$x_1$  & $y_1$& $z_1$&$2$& 2\\
    \g{2}&$x_2$  & $y_3$& $z_3$&$3$& 1 \\
    \g{3}&$x_3$  & $y_3$& $z_3$&$3$& 1 \\
\end{tabular}
};

\node [detailnode, anchor=north west,right=0.3cm of shreddedreldepthone,yshift=0.1cm] (shreddedreldepthonemapping)
      {\sizeshreddedtable
\begin{tabular}{l}
$\rep{\depthone} =(\phys{\depthone}, \store{\depthfive}, \sel{\MakeLowercase{\depthone}}) $
    \\
    \\
    $\sel{\MakeLowercase{\depthone}} = \selv{1,2,3}$\\
\end{tabular}
};

\end{tikzpicture}

   \end{center}
    \caption{Example evaluation of an NSA expression.    
      Intermediate nested relations and dictionaries are labeled $(\rootnode), (\depthone),\ldots$ Shredded processing is illustrated on the right. \label{fig:example-run} }
    \inConfVersion{\vspace{-2ex}}
\end{figure}

\smallskip\noindent\textbf{\nsa.} Our Nested Semijoin Algebra (\nsa) consists of
the standard relational operators filter ($\sigma$), projection ($\pi$),
renaming ($\rho$), bag-union ($\cup$), bag difference ($-$)---all
straightforwardly extended to operate on nested relations---and four new
operators: group-by ($\group$), nested semijoin ($\nsemijoin$), unnest
($\unnest$), and flatten ($\flatten$). We can think of $\group$, $\nsemijoin$,
and $\unnest$ as corresponding to the three separate phases of a traditional
hash-based join: hash-table building, probing, and output construction,
respectively.
We define these additional operators next and provide examples in Figure~\ref{fig:example-run}.

The \emph{group-by}
operator $\groupby{\y}{Z}$ when applied to a relation $R\colon X$ \rev{creates a dictionary  by grouping the tuples in $R$ on the attributes in
$\y$, and mapping each group-key to its group projected on $Z = X \setminus \y$.}  Formally, the
result dictionary $D\colon \dscheme{\y}{Z}$ has $\supp(\pi_{\y}(R))$ as keys, and maps each key
$t \mapsto \pi_Z(\sigma_{\y=t}(R))$.  
As an example, in Figure~\ref{fig:example-run}, 
$\groupby{\{z\}}{\{u\}}(T)$ is shown as $\depthfive$, and 
$\groupby{\{y\}}{\{z,u\}}(\depthfour)$ as $\depththree$. 

The \emph{nested semijoin} operator
$\nsemijoin$ takes two arguments, a relation $R\colon X$ and a dictionary
$D\colon \dscheme{\y}{Z}$. It is required that $X$ is \emph{compatible}
with $\dscheme{\y}{Z}$, meaning that (i) $\y \subseteq X$ and (ii) $\flatsub(Z) \cap \flatsub(X) = \emptyset$, implying that the union $X \cup \{Z\}$ is again a scheme.
Compatibility is denoted
$X \sim \dscheme{\y}{Z}$. The nested semijoin operator $R \nsemijoin D$ probes
$D$ for each tuple $t$ in $R$; if $D$ contains $\restr{t}{\y}$, then it extends
$t$ by a single nested attribute, $Z$, which contains the entire relation
associated to $\restr{t}{\y}$ by $D$, 
\begin{equation}
  \label{eq:sem-nsemijoin}
  R \nsemijoin D \defeq \bagof{t \circ \{Z \mapsto D(\restr{t}{\y})\}}{t \in R, \restr{t}{\y} \in \dom(D)}.
\end{equation}
Figure~\ref{fig:example-run} depicts the result of 
$S \nsemijoin\, \depthfive$ as $\depthfour$, and that of $R \nsemijoin\, \depththree$ 
as $\depthtwo$.

The unnest operator $\unnest_Y(R)$ unnests a nested attribute $Y \in X$ from input relation $R\colon X$ and has semantics
\begin{equation}
  \label{eq:sem-unnest}
  \unnest_Y(R) \defeq \bagof{\restr{s}{X\setminus\{Y\}} \circ t}{s \in R, t \in s(Y)}.
\end{equation}
It hence pairs each tuple $s \in R$ with all tuples in the relation $s(Y)$. 
Figure~\ref{fig:example-run} shows the result of 
$\unnest_{\{u\}}(\depthone)$ as $\rootnode$, 
and that of $\unnest_{\{z,\{u\}\}}(\depthtwo)$ as $\depthone$. 

Finally, the flatten operator $\flatten(R)$ completely flattens a nested
relation $R\colon X$, returning a flat relation with scheme
$\flatsub(X)$. Specifically, if $Y_1,\dots, Y_k$ is an
enumeration of $\sub(X)$ such that
schemes occur before their subschemes (i.e., for all $i,j$, if $Y_i \in Y_j$
then $j < i$), then $\flatten(R) \defeq \unnest_{Y_1}\dots\unnest_{Y_k}(R)$.  For
example, if $R\colon \{x,y,\{z\{u\}\}$ then
$\flatten(R) = \unnest_{\{u\}}(\unnest_{\{z,\{u\}\}}(R))$. While $\flatten$ is
hence already expressible in $\nsa$ through repeated unnests, we add $\flatten$
as a primitive operator to \nsa for reasons that will become clear in
Section~\ref{sec:representation-and-processing}.

\begin{figure}[tpb]
  \small
    \[  \inferrule{ }{R\colon \x_R} \qquad \inferrule{e\colon X \quad \y \subseteq X}{\sigma_{\theta(\y)}(e)\colon X} \qquad 
      \inferrule{e\colon X \quad Y\subseteq X }{\pi_Y(e)\colon Y}  \qquad \inferrule{e\colon X}{\rho_\varphi(e)\colon \varphi(X)}\]
  \[ \inferrule{e_1\colon X\quad e_2\colon X}{e_1 \cup e_2\colon X} \qquad \inferrule{e_1\colon \seq{x}\quad e_2\colon \seq{x}}{e_1 - e_2\colon \seq{x}} \qquad \inferrule{e\colon X \quad \y \subseteq X \quad Z = X \setminus \y}{\groupby{\seq{y}}{Z}(e)\colon \dscheme{\y}{Z}}\]
  \[
    \inferrule{e_1\colon X\quad e_2\colon \dscheme{\y}{Z} \quad X \sim \dscheme{\y}{Z}}{e_1 \nsemijoin e_2\colon X\cup \{Z\}} \quad \inferrule{e\colon X \quad Y \in X \setminus \attrs}{\unnest_Y(e)\colon X \setminus \{Y\} \cup Y}\quad
    \inferrule{e\colon X}{\flatten(e)\colon \flatsub(X)}
   \]
   \caption{\nsa type rules.\label{fig:nsa} }
   \vspace{-1ex}
  \end{figure}

Like standard relational algebra, the expressions in \nsa must be
well-typed. Figure~\ref{fig:nsa} shows the \nsa typing rules, where we write
$e\colon X$ to denote that \nra expression $e$ is well-typed and has output
scheme $X$. There, $R$ ranges over \emph{flat} input relation symbols, for which we assume
to have an associated input scheme $\x_R$. For the selection operator, $\theta(\y)$ ranges over selection predicates that concern the values in attributes in $\y$. For the renaming operator
$\rho$, the subscript $\varphi$ denotes a permutation of $\attrs$ and we denote
by $\varphi(X)$ the result of applying such a permutation recursively to scheme
$X$.

\smallskip\noindent\textbf{Complexity.}  For the complexity results that follow,
it is important to emphasize that the \nsa type rules (i) restrict to flat input
relations, and (ii) restrict all operators that involve checking tuple-equality,
like filter, difference, group-by, and nested semijoin, to check equality on
\emph{flat} tuples only. Indeed, recall that by convention $\x$ denotes a flat
scheme. Then, the type rule for group-by, for example, indicates that only flat
tuples can be group-by keys. The reason for this restriction is that tuples over
a flat scheme have a size that is constant in data complexity, whereas nested
tuples can have arbitrary size. Hence checking equality over flat tuples is
constant time, whereas it may be linear for nested tuples. We adopt the same
restriction to selection predicates $\theta$ in a selection
$\sigma_{\theta(\y)}(R)$: only predicates $\theta(\y)$ for which we can check in
constant time (in data complexity, in the RAM model of computation) that a tuple
$t\in R$ satisfies $\theta$ are allowed.

\smallskip\noindent\textbf{Relating \nsa to other operators}.  Standard
relational algebra operators such as join and flat semijoin, as well as the
lookup ($\lookup$) and expand ($\expand$) operators of BKN and the nesting
operator ($\nest$) of standard nested relational
algebra~\cite{DBLP:journals/acr/ThomasF86}\footnote{But restricted to using flat tuples as nesting keys.} are cleanly expressible in \nsa as a
composition of \nsa operators. For example, 
\allowdisplaybreaks 
\begin{align}
  R(x,y) \join S(y,z) & \equiv \unnest_{\{z\}}(R \nsemijoin \groupby{\{y\}}{\{z\}}(S)) \label{eq:join-nsa}\\
  R(x,y) \semijoin S(y,z) & \equiv \pi_{x,y}(R \nsemijoin \groupby{\{y\}}{\{z\}}(S)) \\
  R(x,y) \lookup S(y,z) & \equiv R \nsemijoin \groupby{\{y\}}{\{z\}}(S) \label{eq:lookup-nsa} \\
  \expand(R(x,y) \lookup S(y,z)) & \equiv \unnest_{\{z\}}(R \nsemijoin \groupby{\{y\}}{\{z\}}(S)) \label{eq:expand-nsa} \\
  \nest_{x,y} R(x,y,u,v) &\equiv \pi_{x,y}(R) \nsemijoin \groupby{\{x,y\}}{\{u,v\}}(R) 
\end{align}
Actually, we can take the right-hand sides in the above expressions as the \emph{definition} in \nsa of the operators on the left-hand side. As such, this provides a formalisation of \lande plans in terms of \nsa.

The advantage of our algebraic approach is that it clearly defines the underlying data model and allows free operator composition.

 \section{Shredded Processing}
\label{sec:representation-and-processing}

We next turn our attention to the efficient processing of \nsa, focusing on its
implementation in main memory column stores.
\nsa is a form of Nested Relational Algebra (\nra), and it is well-known that
one can evaluate \nra using standard flat relational algebra operators by
representing a nested relation as a collection of flat relations, and simulating
nested relational operators by flat relational operators on this
representation~\cite{DBLP:journals/tcs/Bussche01,DBLP:conf/sigmod/CheneyLW14,DBLP:journals/pvldb/SmithBNS20,DBLP:conf/pods/Wong93}.
We adapt this technique, known as \emph{query shredding}, to implement \nsa. We differ from traditional shredding in that 
some nested operators, in particular $\unnest$, are implemented by
means of a \emph{join} of flat relations. In our setting, however, we want to
use $\nsa$ as a description of physical query plans where $\group$, $\nsemijoin$
and $\unnest$ correspond to the three phases of a traditional hash join: build,
probe, and construct. Specifically, $\unnest$ must then be limited to
constructing the output tuples \emph{when the set of matching tuples have
  already previously been identified} by an earlier $\nsemijoin$ operator; its
shredded implementation hence should not require further joins.  To obtain this
behavior we modify the traditional shredded representation of a nested relation:
each nested attribute $Y$ will be encoded by a flat attribute that holds an
iterator over the elements of $Y$, instead of an abstract identifier as is
traditionally done. Additionally, to support efficient flatten ($\flatten$), we
also store the \emph{weight} of every $Y$, which is the total number of tuples
produced when flattening $Y$.
A benefit of the shredding approach 
is that it only requires modest change to existing query engines to implement:
for many \nsa operators we can
simply delegate to the implementation of existing relational algebra operators;
only $\group$, $\nsemijoin$, $\unnest$, and $\flatten$ require separate treatment.

To simplify notation in the discussion that follows, we restrict our attention
in this section to the shredded processing of nested relations $R\colon X$ for
which $\emptyset$ does not occur multiple times in $X$. So,
$X = \{x, \{y, \emptyset\}\}$ is allowed but
$X = \{x, \emptyset, \{y, \emptyset\}\}$ is not.  We silently assume throughout
this section that all considered \nsa operators consume and produce nested
relations satisfying this criterion. Our implementation does not
have this restriction.

We begin by describing how to represent nested relations in
Section~\ref{sec:representation} and describe evaluation algorithms for each
operator using this representation in Section~\ref{sec:processing}.

\subsection{The shredding representation}
\label{sec:representation}

\smallskip\noindent\textbf{Columnar layout.} We assume that we are working in
main memory, and that a flat relation $R(x_1,\dots,x_n)$ is physically
represented as a tuple $\phys{R} = (\phys{R}.x_1, \dots, \phys{R}.x_n)$ of
vectors $\phys{R}.x_i$, all of length $\card{R}$. It is understood that values
at the same offset in these vectors encode a complete tuple, i.e.,
$R = \bagof{ (\phys{R}.x_1[i],\dots,\phys{R}.x_n[i])}{1 \leq i \leq \card{R}}$.
In particular, it is possible to refer to tuples positionally, i.e., the $1$st tuple in $\phys{R}$, the second tuple in $\phys{R}$, and so on.\footnote{Note that we start our offsets at $1$, so the first tuple has offset $1$. } We will refer to $\phys{R}$ as a physical relation, and denote the number of tuples in  $\phys{R}$ by $\len{\phys{R}}$. 
If $1 \leq i \leq \len{\phys{R}}$ and
$\y = y_1,\dots,y_k$ is a subset of $\{x_1,\dots, x_n\}$, then we write
$\phys{R}[i](\y)$ for the tuple $(\phys{R}.y_1[i],\dots,\phys{R}.y_k[i])$. We denote the length of a vector $v$ by $\len{v}$. A
\emph{position vector for} $\phys{R}$ is a vector $p$ of natural numbers, all
between $1$ and $\len{\phys{R}}$. We assume an operation $\take(\phys{R}.x,p)$ that can be used to construct a new vector from an existing vector $x$ in $\phys{R}$ and a position vector $p$ on $\phys{R}$: 
$\take(\phys{R}.x, p)$ returns a new column $c$ of length $\len{p}$ such that
$c[i] = \phys{R}.x[p[i]]$ for all $i$. 
The $\take$ operation hence re-arranges the entries of $\phys{R}.x$ according to $p$, possibly repeating some entries and filtering out others.
If the entries in $p$ are strictly increasing
then $p$ is a \emph{selection vector for} $\phys{R}$. Note that this implies
$\len{p} \leq \len{\phys{R}}$, and if $\len{p} = \len{\phys{R}}$ then $p = [1,\dots, \len{\phys{R}}]$. We denote the latter vector also by $\allsel{R}$. Note that $\take(\phys{R}.x, p)$ can only filter entries when $p$ is a selection vector.

\smallskip\noindent\textbf{Weights.}  The \emph{weight} of a nested relation $R$
is the total number of tuples produced when flattening $R$, i.e.
$\weightval(R) = \card{\flatten(R)}$. Similarly, the weight of a nested tuple
$t$ is $\card{\flatten(\bag{t})}$, the total number of tuples produced when
flattening $t$.

\smallskip\noindent\textbf{Schemes shredding.}
For a scheme $X = \{y_1,\dots, y_k, Z_1, \dots, Z_\ell\}$ define the \emph{flat} schemes $\shred(X)$ and $\ishred(X)$ as
\begin{equation}
  \label{eq:shred}
  \shred(X) \defeq \{ y_1,\dots, y_k, \hol{Z_1},\dots, \hol{Z_\ell},
              \weight{Z_1}, \dots, \weight{Z_\ell}\}
\end{equation}
and $\ishred(X) = \shred(X) \cup \{\nxt\}$.  
Here, the attributes
$\hol{Z_i}$, $\weight{Z_i}$, and $\nxt$ are fresh flat attributes, all pairwise
distinct as well as distinct from the $y_j$.  Intuitively, $\hol{Z_i}$ will
store the head of a linked list that represents the contents of nested attribute
$Z_i$, whereas $\nxt$ will be used to point to the next tuple in such lists.  $\weight{Z_i}$ will store the weight of the nested $Z_i$ relations.
Observe that if $X$ is flat to begin with, then $\shred(X) = X$.

\smallskip\noindent\textbf{Relation shredding.}
The shredded representation of a nested relation $R\colon X$ is a triple $\rep{R} = (\phys{R}, \store{R}, \sel{r})$ where (i) 
$\phys{R}$ is a physical relation over $\shred(X)$;  (ii) 
$\store{R}$ is a \emph{store} over $X$: a collection of physical relations, one physical relation $\store{R}(Y)$ for every nested attribute $Y \in \sub(X)$, such that $\store{R}(Y)$ has schema $\ishred(Y)$; and (iii) $\sel{r}$ is a selection vector for $\phys{R}$.

The $\nxt$ attribute of the tuples in $\store{R}(Y)$ is used to encode a linked list of tuples: for all positions $1 \leq i \leq \len{\store{R}(Y)}$, if $\store{R}(Y).\nxt[i] = 0$ then the tuple at position $i$ in $\store{R}(Y)$ is the final tuple in the list; otherwise its successor in the list is the tuple at offset $\store{R}(Y).\nxt[i]$.
Correspondingly, each $\hol{Y}$-value of a tuple in $\phys{R}$ points to the head of the linked list in $\store{R}(Y)$ storing the nested tuples. 

\begin{example}
  To clarify the discussion that follows, we illustrate shredding by means of
  Figure~\ref{fig:shredding} which shows a nested relation $R\colon X$ with
  $X = \{x, \{y\}, \{z, \{u\}\}\}$ on the left, and its shredded
  representation $\rep{R}$ on the right. We omit the selection vector. 
  \rev{
  $\rep{R}$ contains four flat relations, or ``shreds'':
  one for the full scheme $x, \{y\}, \{u, \{v\}\}$,
  and one each for the nested schemes
  $\{u, \{v\}\}$, $\{v\}$, and $\{y\}$.
  Let us focus on the first shred $\phys{R}$.
  The first column stores values of the flat attribute $x$.
  The attribute $\{y\}$ has two columns:
  $\holshort{}$ stores pointers (offsets) to the $\store{R}(\{y\})$ shred,
  each indicating the head of the linked list of tuples nested under
  the corresponding $x$-value.
  The $w$ column stores the weight, which is the size of the nested
  relation if we were to flatten it.
  In each nested shred, the $\nxt$ column link together tuples
  that belong to the same nested segment.
  For example, the first tuple in $\phys{R}$
  has $x = a_1$, $\{y\}.\holshort{} = 2$, $\{y\}.w = 2$;
  this points to the head of the linked list in $\store{R}(\{y\})$
  at offset $2$, whose $\nxt$ column points to tuple above it.
  Similarly, the $\{u, \{v\}\}$ attribute of the first tuple
  has $\holshort{} = 2$ and $w = 3$,
  pointing to the 2nd tuple in $\store{R}(\{u, \{v\}\})$.
  However, note that although $w = 3$,
  the linked list has length 2, becuase the shred over $\store{R}(\{u, \{v\}\})$
  further nests the shred over $\{v\}$.}
\end{example}

The shredded representation of $R$ works as follows: every tuple $t \in R$ is
represented by exactly one tuple in $\phys{R}$. Let $i$ be the index of the
tuple in $\phys{R}$ representing $t$. Then $t(x) = \phys{R}.x[i]$ for every flat
$x \in X$. For every nested attribute $Y \in X$ we have that $\phys{R}.\weight{Y}[i] = \weightval(t(Y))$. Furthermore, 
$\phys{R}.\hol{Y}[i] = j$ for some $1 \leq j \leq \len{\store{R}(Y)}$, which is the head index of the linked
list of tuples in $\store{R}(Y)$ that together represent the tuples occurring in
$t(Y)$.
Note that the tuples in
$t(Y)$ may themselves contain further nested relations, and the shredding hence
proceeds recursively.

Every tuple in $R$ will be represented in the above sense in $\phys{R}$. To
allow efficient implementation of repeated semijoins of the form
$(S \nsemijoin e_1) \nsemijoin e_2$, we do allow that in the shredding
$\rep{R}$ for $R = S \nsemijoin e_2$, the physical relation $\phys{R}$ contains tuples that have already been filtered out by
the nested semijoin,
i.e., we allow that $\len{\phys{R}} \geq \card{R}$.  
 In that case, the
selection vector $\sel{r}$ of $\sel{R}$ contains the offsets of the \emph{valid}
tuples in $\phys{R}$, i.e., those that actually represent tuples in $R$ (having
passed previous $\nsemijoin$). So, we always have $\len{\sel{r}}$ equal to the
cardinality of $R$.  Additionally, if $X$ is a flat scheme, then $\phys{R}$ is
not allowed to contain redundant tuples, i.e., $\len{\phys{R}} = \len{\sel{r}}$,
and $\sel{r} = \allsel{R}$.  It is important to observe that if $X$ is a flat
scheme, then $\store{R}$ is empty; the shredding $\rep{R}$ of $R$ is then simply
$\rep{R} = (\phys{R}, \emptyset, \allsel{R})$.

\smallskip\noindent\textbf{Dictionary shredding.} The shredded representation of
a dictionary $D\colon \dscheme{\y}{Z}$ is similarly defined as the shredding of
a nested relation, except that it has a hash-map as first component and does not
have a selection vector. Concretely, the shredding of $D$ is a pair
$\rep{D} = (\phys{h}, \store{D})$ where $\phys{h}$ is a hash-map, mapping
$\y$-tuples to pairs $(j,w)$, and $\store{D}$ is a store over $\{Z\}$. For every
$\y$-tuple $t$, if $\phys{h}(t) = (j,w)$ then $j$ is the head index in
$\store{D}(Z)$ of the linked list of tuples that together
represent the nested relation $D(t)$, and $w = \weightval(D(t))$.
See node (\depthfive) in Figure~\ref{fig:example-run} for an example.

\begin{figure*}[tbp]
  \small
\begin{varwidth}[t]{0.33\textwidth}
          \tt
		\begin{algorithmic}[1]
                  \Function{\opgroup}{\phys{R},\store{R},\sel{r},$X$,$\y$,$Z$}\algorithmicdo
                  \State w = multiply\_weights(R, X)
                  \State $\nxt$ = [0] * $\len{\phys{R}}$
                  \State h = \{\} \Comment{maps keys -> (pos, weight)}
                  \For{i in \sel{r}}
                  \State key = $\phys{R}[i](\y)$
                  \If{h.contains(key)}
                  \State (j, prev\_w) = h[key]
                  \State nxt[i] = j
                  \State h[key] = (i, prev\_w + w[i])
                  \Else
                  \State{h[key] = (i, w[i])}
                  \EndIf
                  \EndFor
                  \State \store{R}(Z) = \newphys()
                  \State \store{R}(Z).nxt = nxt 
                  \State \textbf{for} $u$ in $\shred(Z)$ \{ \store{R}(Z).$u$ = \phys{R}.$u$ \}
                  \State \Return (h, $\store{R}$)
                  \EndFunction
                \end{algorithmic}
         \end{varwidth}\qquad
\begin{varwidth}[t]{0.33\textwidth}
          \tt
		\begin{algorithmic}[1]
                  \Function{\opsemijoin}{\phys{R},\store{R},\sel{r},\phys{h},\store{D},$X$,$\y$,$Z$}\algorithmicdo
                  \State sel = [] \label{alg:nsemijoin:probestart}
                  \State \phys{R}.\hol{$Z$} = [0] * \len{\phys{R}} 
                  \State \phys{R}.\weight{$Z$} = [0] * \len{\phys{R}}
                  \For{i in \sel{r}}
                  \State key = $\phys{R}[i](\y)$ 
                  \If{h.contains(key)}
                  \State sel.append(i)
                  \State (\phys{R}.\hol{$Z$}[i], \phys{R}.\weight{$Z$}[i]) = h[key]
                  \EndIf
                  \EndFor \label{alg:nsemijoin:probestop}
                  \State \Return (\phys{R},\store{R} + \store{D}, sel)  \label{alg:nsemijoin:output}
                  \EndFunction
                  \State{}
                  \State{\Comment{iterator over linked  list at row}}
                  \Function{itr}{\phys{R},\store{R},$Y$,row}\algorithmicdo
                  \State curr = \phys{R}.\hol{$Y$}[row]
                  \While{curr != 0}
                  \State \textbf{yield} curr
                  \State curr = \store{R}($Y$).nxt[curr]
                  \EndWhile
                  \EndFunction
                \end{algorithmic}
         \end{varwidth}\qquad
\begin{varwidth}[t]{0.33\textwidth}
          \tt
		\begin{algorithmic}[1]
                  \Function{unnest}{\phys{R},\store{R},\sel{r},$X$,$Y$}\algorithmicdo
                  \State pos\_$\phys{R}$ = []; pos\_$Y$ = [] 
                  \For{i in \sel{r}}
                    \For{j in itr(\phys{R}, \store{R}, $Y$, i)}\label{alg:unnest:itr}
                      \State{pos\_$\phys{R}$.append(i)}
                      \State{pos\_$\phys{Y}$.append(j)}
                      \EndFor
                  \EndFor
                  \State \phys{O} = \newphys() \label{alg:unnest:outputstart}
                  \For{$u$ in $\shred(X) \setminus \{\hol{Y}\}$} 
                  \State $\phys{O}.u$ = take($\phys{R}.u$, pos\_$\phys{R}$)
                  \EndFor
                  \For{$u$ in $\shred(Y) \setminus \{\nxt\}$} 
                  \State $\phys{O}.u$ = take($\store{R}(Y).u$, pos\_$Y$)
                  \EndFor
                  \State del(\store{R}, Y)\label{alg:unnest:outputstop}
                  \State \Return (O, \store{R}, \allsel{O})  
                  \EndFunction
                \end{algorithmic}
         \end{varwidth}\caption{Physical implementation of group-by, nested semijoin, and unnest.  \label{fig:physical-operators}}
\end{figure*}

\begin{figure}[tbp]
  \small
	\begin{varwidth}[t]{0.9\textwidth}
          \tt
		\begin{algorithmic}[1]
                 \Function{flatten}{\phys{R},\sel{r},\store{R},$X$}\algorithmicdo
                  \State \phys{O} = \newphys()
                  \State rep = [1] * \len{\sel{r}}
                  \State \opflatteninner\!\!(\phys{R},\sel{r},rep,\store{R},$X$,\phys{O})
                  \State \Return $(\phys{O}$, $\emptyset$, $\allsel{O})$
                  \EndFunction
                  \State
\Function{\opflatteninner}{\phys{R},pos,rep,\store{R},$X$,\phys{O}}\algorithmicdo
                  \State w = multiply\_weights(\phys{R}, $X$) \label{op:flatten:inner:calc-weights}
                  \State \Comment{Generate output columns for all flat attrs in \phys{R}} \label{op:flatten:inner:generate}
                  \State \opgenerate(\phys{R},pos,rep,$X$,\phys{O})
         \State \Comment{Recursively generate columns for nested attrs in \phys{R}}         
                  \For{$Y$ in $X \setminus \attrs$} \label{op:flatten:inner:start-nested-attrs}
                  \State npos = []; nrep = [] \label{op:flatten:inner:start-npos}
                  \State \Comment{w = total weight of all remaining nested attrs}
                  \State w = div(w, \phys{R}.\weight{Y})
                  \For{(row, i) in enumerate(pos)} 
                  \For{k = 1 to rep[row]}
                  \For{j in itr(\phys{R},\store{R},$Y$,i)}
                  \State npos.append(j); nrep.append(w[i])
                  \EndFor
                  \State \Comment{already update rep for the next $Y$ }
                  \State rep[row] *= \phys{R}.weight{Y}[i]
                  \EndFor
                  \EndFor \label{op:flatten:inner:stop-npos}

                  \State{\opflatteninner(\store{R}(Y),npos,nrep,\store{R},$Y$, \phys{O})} \label{op:flatten:inner:recursive-call}
\EndFor \label{op:flatten:inner:end-nested-attrs}
                  \EndFunction
                  \State

\Function{\opgenerate}{\phys{R},pos,rep,w,$X$,\phys{O}}\algorithmicdo
                  \State{rwpos = []}
                  \For{(i,r) in zip(pos, rep), j in 1..r*w[i]\hspace{-1ex}}
                  \State{rwpos.append(i)}
                  \EndFor
                  \For{u in $X \cap \attrs$}
                  \State{\phys{O}.u = take(\phys{R}.u, rwpos)}
                  \EndFor
                  \EndFunction

                \end{algorithmic}
        \end{varwidth}\caption{Physical implementation of flatten. \label{fig:flatten-physical-operators}}
        \vspace{-2ex}
\end{figure}

\subsection{Processing}
\label{sec:processing}

We implement \nsa by defining a physical operator $\phys{f}$ for every \nsa
operator $f$. Physical operators consume and produce shredded representations:
if $\rep{R}$ is the shredding of $R$ then $\phys{f}(\rep{R})$ is the shredding
of $f(R)$. For the \nsa operators $f \in \{ \sigma, \pi, \rho, \cup, - \}$ that
also exist in flat \ra, the physical operator $\phys{f}$ simply consists of
applying the corresponding flat physical \ra operator to one or more physical
relations in $\rep{R}$, possibly with a slight variation. For example, consider
$f = \pi_Y$ and $R\colon X$. To get a representation of $\pi_Y(R)$ from
$\rep{R} = (\phys{R}, \store{R}, \sel{r})$, it suffices to simply return
$(\pi_{\shred(Y)}(\phys{R}), \store{R}', \sel{r})$ where $\store{R}'$ is
obtained from $\store{R}$ by removing all entries in $\sub(X \setminus Y)$. This
works because $\pi_{\shred(Y)}(\phys{R})$ retains only those columns in
$\phys{R}$ required to represent $\pi_Y(R)$. Because the nested attributes in
$\sub(X \setminus Y)$ are removed from $R$ we can also remove them from
$\store{R}$.  The other standard operators in $\{ \sigma, \pi, \rho, \cup, -\}$
are similarly implemented by calling flat \ra physical operators.
\inConfVersion{Because of space constraints, we 
  omit their definition. 
  The interested reviewer may find them in 
  ~\cite{anon-full-version}.} \inFullVersion{We refer to the
  Appendix for their description.}

\begin{toappendix}
In this section we give the full definition of the physical operators for $f \in \{ \sigma, \pi, \rho, \cup, -\}$. Their implementation is shown in Figure~\ref{fig:other-phyiscal-operators}.

Let $R\colon X$. The implementation of selection $\sigma_{\theta(\x)}(R)$ takes as argument the shredding $(\phys{R}, \store{R}, \sel{r})$ of $R$ as well as the scheme $X$, predicate $\theta$ and its scheme $\x$. It simply builds a new selection vector $\sel{s}$ by  iterating over all tuples in the selection vector $\sel{r}$, and checking if the tuple satisfies $\theta$.

 The implementation of projection $\pi_Y(R)$ in Figure~\ref{fig:op-project} was already discussed. 

For the implementation of renaming $\rho_\varphi(R)$ we may assume w.l.o.g. that the $\varphi$ only renames the flat attributes occurring in $X$, and leaves the flat $\hol{Z}$, $\weight{}$, and $\nxt$ attributes that we have invented for shredding untouched. The implementation first extends $\varphi$ to also map $\hol{Z} \mapsto \hol{\varphi(Z)}$ and then applies flat RA renaming with the extended $\varphi$ to all physical relations in $(\phys{R}, \store)$.

For the implementation of difference $R - S$ we note that due to the typing rules, $R$ and $S$ must have the same flat scheme $\x$. The implementation takes the shreddings $(\phys{R}, \store{R}, \store{r})$ and $\phys{S}, \store{S}, \store{s})$ as input. Because the scheme of $R$ and $S$ is flat, $\store{R}$ and $\store{S}$ are empty, $\sel{r} = \allsel{R}$, and $\sel{s} = \allsel{S}$. It hence suffices to simply take flat RA difference of the top-level physical represenations $\phys{R}$ and $\phys{S}$.

For the implementation of union $R \cup S$ we note that due to the typing rules,
$R$ and $S$ must have the same scheme $X$. The implementation takes the
shreddings $(\phys{R}, \store{R}, \sel{r})$ and $\phys{S}, \store{S}, \sel{s})$ as input, as well
as their scheme $X$. For every nested attribute occurring (directly or
recursively in $X$), we will take the point-wise union of $\store{R}(Y)$ and
$\store{S}(Y)$ to compute the store of the output shredding. We have to take
care to maintain the linked-lists encoded in the $\nxt$ column, however as well
as the head-of-list positions encoded in the various $\hol{Z}$ for
$Z \in \sub(Y)$: when we take the pointwise union the positions in
$\store{S}(Y)$ need to be shifted by the number of entries in $\store{R}(Y)$.
We do so by means of the auxiliary recusive function $\texttt{fix}$: given a
(mutable reference to) physical relation $\phys{T}$, it modifies $T$ by
modifying its $\hol{Z}$ columns as well as its $\nxt$ column (if it has it). The
modification is done through the function $\texttt{offset}$ (definition not shown) which takes a
(mutable reference to) a vector and a natural number $n$, and increments each
entry with a non-zero value by $n$.

Having computed the output store, it then remains to also take the flat RA union of the top-level physical represenations $\phys{R}$ and $\phys{S}$, also offsetting the $\hol{Y}$ entries in $\phys{S}$.

\begin{figure*}[tbp]
  \small
   \subcaptionbox{\label{fig:op-select}}{\begin{varwidth}[t]{0.33\textwidth}
          \tt
		\begin{algorithmic}[1]
                  \Function{select}{$\phys{R}, \store{R},\sel{r} \theta(\x)$}\algorithmicdo
                  \State s = []
                  \For{i in \sel{r}}
                  \State key = $\phys{R}[i](\x)$
                  \If{key $\models \theta$}
                  \State s.append(i)
                  \EndIf
                  \EndFor
                  \If{$X$ is flat and \len{\sel{s}} < \len{\phys{R}}} 
\For{$u$ in $\shred(X)$}
                  \State \phys{R}.$u$ = take(\phys{R}.$u$, \sel{s}) 
                  \EndFor \label{alg:group:restrict-end}
                  \State s = \allsel{R}
                  \EndIf
                  \State \Return (\phys{R}, \store{R}, s)
                  \EndFunction
                \end{algorithmic}\end{varwidth}}
         \qquad
   \subcaptionbox{\label{fig:op-project}}{\begin{varwidth}[t]{0.33\textwidth}
          \tt
		\begin{algorithmic}[1]
                  \Function{project}{$\phys{R}, \store{R}, \sel{r}, X,  Y$}\algorithmicdo
                  \For{$Z \in (X \setminus Y) \setminus \attrs$}
                  \State drop(\store{R}, $Z$)
                  \EndFor
                  \State \Return $(\pi_{\shred(Y)}(\phys{R}), \store{R})$
                  \EndFunction
                \end{algorithmic}\end{varwidth}}
   \qquad
      \subcaptionbox{\label{fig:op-difference}}{\begin{varwidth}[t]{0.33\textwidth}
          \tt
        	\begin{algorithmic}[1]
                  \Function{difference}{$\phys{R}, \store{R}, \sel{r}, \phys{S}, \store{S}, \sel{s}
                    $}\algorithmicdo
                  \State \Comment{Well-typedness ensures that $R$ and $S$ have equal flat scheme $\x$}
                  \State \Comment{$\store{R}$ and $\store{S}$ are hence empty}
                  \State \Comment{Selection vectors are hence \allsel{R} and \allsel{S}, respectively}
                  \State \Return ($\phys{R} - \phys{S}, \store{R}, \allsel{\phys{R} - \phys{S}}$)
                  \EndFunction
                \end{algorithmic}
         \end{varwidth}}
\\[2ex]
   \subcaptionbox{\label{fig:op-rename}}{\begin{varwidth}[t]{0.33\textwidth}
          \tt
        	\begin{algorithmic}[1]
                  \Function{rename}{$\phys{R}, \store{R}, \sel{r}, \varphi$}\algorithmicdo
                  \State  \Comment{assume $R\colon X$}
                  \State \Comment{assume $\varphi$ renames only attrs in $X$,}
                  \State \Comment{not the  new $\hol{Z}, \weight{Z}$ attributes }

                  \For{$Y$ in $\sub(X)$}
                  \State $\varphi(\hol{Z}) = \hol{\varphi(Z)}$
                  \State $\varphi(\weight{Z}) = \w{\varphi(Z)}$
                  \EndFor
                  \State $\store{}$ = empty store
                  \For{$Y$ in $\sub(X)$}
                  \State $\store{}(\varphi(Y))$ = $\rho_{\varphi}(\store{R}(Y))$
                  \EndFor
                  \State \Return $(\rho_{\varphi}(R), \store{}, \sel{r})$
                  \EndFunction
                \end{algorithmic}\end{varwidth}}
       \quad
   \subcaptionbox{\label{fig:op-union}}{\begin{varwidth}[t]{0.5\textwidth}
          \tt
        	\begin{algorithmic}[1]
                  \Function{union}{$\phys{R}, \store{R}, \sel{r}, \phys{S}, \store{S}, \sel{s}, X$}\algorithmicdo
                  \State \Comment{scheme of $R$ = scheme of $S$ = $X$}
                  \If{\len{\sel{r}} < \len{\phys{R}}} \Comment{\sel{r} != \allsel{R}} 
                  \For{$u$ in shred($X$)}
                  \State \phys{R}.$u$ = take(\phys{R}.$u$, \sel{r}) 
                  \EndFor
                  \EndIf
                  \If{\len{\sel{s}} < \len{\phys{S}}} \Comment{\sel{s} != \allsel{S}} 
                  \For{$u$ in shred($X$)}
                  \State \phys{S}.$u$ = take(\phys{S}.$u$, \sel{s}) 
                  \EndFor
                  \EndIf
                  \State $\store{}$ = empty store
                  \For{$Y$ in $X \setminus \attrs$}
                  \State \texttt{fix}($\store{S}(Y), \store{R}, \store{S}, Y$)
                  \EndFor
                  \For{$Y$ in $\sub(X)$}
                  \State $\store{}(Y)$ = $\store{R}(Y)\ \cup \store{S}(Y)$
                  \EndFor
                  \State \texttt{fix}($\phys{S}, \store{R}, \store{S}, X$)
                   \State \Return ($\phys{R} \cup \phys{S}, \store{}, \allsel{\phys{R} \cup \phys{S}}$)
                  \EndFunction
                \end{algorithmic}
         \end{varwidth}}
   \quad
      \subcaptionbox{\label{fig:op-fix}}{\begin{varwidth}[t]{0.5\textwidth}
          \tt
        	\begin{algorithmic}[1]
                  \Function{fix}{$\phys{T}, \store{R}, \store{S}, Y$}\algorithmicdo
                  \State \Comment{scheme of $T$ = $Y$}
                  \For{$Z \in Y \setminus \attrs$}
                  \State n = \len{\store{R}(Z)}
                  \State \texttt{offset}($\phys{T}.\hol{Y}$, n)
                  \State \texttt{fix}($\store{S}(Z), \store{R}, \store{S}, Z$)
                  \EndFor
                  \If{\nxt\xspace is an attribute of $\phys{T}$}
                  \State n = \len{\store{R}(Y)}
                  \State \texttt{offset}($\phys{T}.\nxt$, n)
                  \EndIf
                  \EndFunction
                \end{algorithmic}
         \end{varwidth}}
  \caption{Physical implementation of the ``standard'' NSA operators, using traditional flat \ra operators.}
  \label{fig:other-physical-operators}
\end{figure*}

 \end{toappendix}

The implementation of $\group, \nsemijoin$, and $\unnest$ is defined in
Figure~\ref{fig:physical-operators} and illustrated on an example in Figure~\ref{fig:example-run}. Group-by $\groupby{\y}{Z}(R)$ with
$R\colon X$ follows conventional hash-build. Given shredding
$\rep{R} = (\phys{R}, \store{R}, \sel{r})$ of nested relation $R$, as well as
schemes $\y$ and $Z = X \setminus \y$, we compute the weight of each tuple in $\phys{R}$, using a function
\texttt{multiply\_weights} (definition not shown) that returns a vector $w$ with
$w[i]$ equal to the product over all nested attributes $Y \in X$ of
$\phys{R}.\weight{Y}[i]$. Then vector $\nxt$ of length $\len{\phys{R}}$ is
created, initialized to $0$ so that initially all tuples terminate the linked
lists encoded in $\nxt$. We further initialize \texttt{h} to the empty
hash-map. We then iterate over the group-by keys mentioned in $\sel{r}$, adding
them to \texttt{h}, and storing the position of the most recent $\phys{R}$-tuple
with the current key as well as the total weight of the key. If we have
previously already encountered the same key, the current tuple's $\nxt$ value is
set to point to the position in $\phys{R}$ of the previous tuple with the same key, and the
weight is updated.
 Finally, we create the store entry for $Z$ by selecting the
columns in $Z$ from $\phys{R}$, and adding $\nxt$.

The implementation of nested semijoin $R \nsemijoin D$ takes as argument the
shredding $\rep{R} = (\phys{R}, \store{R}, \sel{r})$ of nested relation
$R\colon X$ and the shredding $\rep{D} = (\phys{h}, \store{D})$ of dictionary
$D\colon \dscheme{\y}{Z}$, as well as the schemes $X$, $\y$, and $Z$. It simply
executes as a conventional hash join probe. We collect in selection vector
\texttt{sel} the positions of the valid tuples in $\phys{R}$ whose keys can be
successfully probed in $D$. We add new vectors $\hol{Z}$ and $\weight{Z}$ to
$\phys{R}$, in which we store the matching positions in $\store{D}(Z)$ according
to $\phys{h}$, as well as their weights.  In line~\ref{alg:nsemijoin:output},
$\store{R} \texttt{+} \store{D}$ denotes the disjoint union of the two stores
$\store{R}$ and $\store{D}$.\footnote{This is disjoint because the type rules
  for $\nsemijoin$ require $X$ compatible with $\y \to Z$. In particular,
  $X \cup \{Z\}$ is a scheme; therefore the domains of $\store{R}$ and
  $\store{D}$ must be disjoint.}

Unnesting $\unnest_Y(R)$ takes as argument the shredding
$(\phys{R}, \store{R}, \sel{r})$ of $R$ as well as $Y$. It first creates two
position vectors, \texttt{pos\_}$\phys{R}$ and \texttt{pos\_}$\phys{Y}$ that are
populated with valid positions in $\phys{R}$ and $\store{R}(Y)$,
respectively. Specifically, for every tuple $t \in R$ that is represented at
position $i$ in $\phys{R}$ we append the positions of all the tuples in
$\store{R}(Y)$ that encode the elements of $t(Y)$ to \texttt{pos\_}${\phys{Y}}$;
and we add $i$ to \texttt{pos\_}$\phys{R}$ as many times as $\card{t(Y)}$. In
line~\ref{alg:unnest:itr}, \texttt{itr(\phys{R},\store{R},$Y$,i)} returns an
iterator over the positions in $\store{R}(Y)$ that encode the elements of
$t(Y)$. We use the position vectors to index into $\phys{R}$
resp. $\store{R}(Y)$ to create the physical representation $\phys{O}$ of the
output in
lines~\ref{alg:unnest:outputstart}--\ref{alg:unnest:outputstop}. Finally, we
remove the entry for $Y$ from $\store{R}$ as this is no longer required.

\smallskip\noindent\textbf{Flatten.} When multiple unnest operations are applied
in sequence, there is an overhead in the number of \take operations
applied. To illustrate, consider $\unnest_{\{u\}} \unnest_{\{z,\{u\}\}}(\depthtwo)$ from
Figure~\ref{fig:example-run}, where
$\depthtwo \colon \{x, y, \allowbreak \{z, \{u\}\}\}$. The first unnest,
$\unnest_{\{z, \{u\}\}}(\depthtwo)$, will already perform a \take on $x$, $y$,
and $z$ (among others) to produce the result with scheme $\{x, y, z,
\{u\}\}$. The second unnest performs a \take again on $x, y, z$ to
produce the final result. While this overhead is modest in
Figure~\ref{fig:example-run}, it grows linearly in the number of $\unnest$
applied sequentially. For $\unnest_{Y_1}\unnest_{Y_2}\dots\unnest_{Y_k}(R)$, the
outer-most flat attributes would be copied and re-arranged $k$ times by means of
\take before producing the final, flat relation. It is for this reason
that we have included flatten ($\flatten$) as a primitive operator in \nsa, and
provide the dedicated physical implementation \opflatten shown in
Figure~\ref{fig:flatten-physical-operators}. It performs only a single
\take operation per flat attribute.

Flatten is implemented by calling the auxiliary function $\opflatteninner$,
which takes a physical relation $\phys{R}$ as argument, a position vector
\posvec for $\phys{R}$, and a numerical vector \repvec of the same length as
\posvec containing only strictly positive numbers, called the \emph{repetition
  vector}.\footnote{Additionally, \opflatteninner takes the associated store
  $\store{R}$ as input, as well as the physical relation $\phys{O}$ in which the
  output is to be constructed. We ignore these in our explanation.} Initially,
\posvec is the selection vector of $\rep{R}$ and \repvec contains all $1$s, but
this changes when we call $\opflatteninner$ recursively.  Intuitively, for each
tuple $i$ specified in \posvec and matching repetition number $r$ specified in \repvec, \opflatteninner will completely flatten the
$i$-th tuple of $\phys{R}$, but additionally \emph{repeat} each produced
flattened tuple $r$ times. To be precise, let $s_i$ denote the nested
tuple represented at offset $i$ in $\phys{R}$. Let us write $\flatten(s_i)$ for
$\flatten(\bag{s_i})$. When we implement $\flatten(s_i)$, it will produce tuples
in a certain order; say it produces the flattened tuples $t_1,\dots, t_n$. Then, let 
 $\mu^*(s_i,r)$ be this sequence with every $t_j$
repeated $r$ times as follows
\begin{equation}
  \label{eq:flatten-repeat}
  \flatten(s_i,r) \defeq \underbrace{t_1,\dots,t_1}_{r \text{ times}}, \dots,
  \underbrace{t_n, \dots, t_n}_{r \text{ times}}.
\end{equation}
Assuming $\posvec = [i_1, \dots, i_p]$ and $\repvec = [r_1,\dots, r_p]$, the
call to $\opflatteninner(\phys{R}, \posvec, \repvec)$ will produce a physical
relation that represents the sequence of tuples
$\flatten(s_{i_1},r_1), \dots, \flatten(s_{i_p}, r_p)$, in this order.

To understand how \opflatteninner works consider the flattening of a single
tuple $s_i$ having flat attributes $\x$ and nested attributes
$Y_1,\dots,Y_k$.  By definition,
\[ \flatten(s_i) = \bag{s_i[\x]} \times \flatten(s_i(Y_1)) \times \dots \times
  \flatten(s_i(Y_k)). \] All tuples in $\flatten(s_i)$ hence have the same
$\x$-values, which is combined with the cartesian product of
flattening $Y_1,\dots,Y_k$. As such, for each $u \in \x$ we can easily produce
the entire $u$-column of $\flatten(s_i)$ by taking $s_i(u)$ and repeating this
value
$r_i * \weightval(s_i(Y_1)) \times \dots \times \weightval(s_i(Y_k))$
times. 
This is exactly what \opflatteninner does in
lines~\ref{op:flatten:inner:calc-weights} and ~\ref{op:flatten:inner:generate}
by first calculating the total weight for each tuple, and subsequently calling
\opgenerate to produce each column.

Next, 
lines~\ref{op:flatten:inner:start-nested-attrs}--\ref{op:flatten:inner:end-nested-attrs} produce $\flatten(s_i(Y_1)) \times \dots \times \flatten(s_i(Y_k))$
in a column-wise fashion, so that (i) the recursive $\flatten$ calls can
independently produce the columns for their respective flat attributes and (ii)
these independent calls produce the flattened tuples in an order so that all
columns together give a physical representation for the entire cartesian product. For
the recursive call on $Y_1$ this is trivial: flatten each tuple in $s_i(Y_1)$
and repeat it $\weightval(s_i(Y_2)) \times \dots \times \weightval(s_i(Y_k))$
times to account for the cartesian products that follow. For the recursive call on $Y_\ell$ with $2 \leq \ell \leq k$ 
this becomes more involved. Assume that $s_i(Y_\ell)$ contains the tuples
represented at offsets $[j_1,\dots,j_m]$ in $\store{R}(Y_\ell)$. Then, letting
$r' = r_i \times \weightval(s_1(Y_1)) \times \dots \weightval(s_1(Y_{\ell-1})$, the recursive call to \opflatteninner in line~\ref{op:flatten:inner:recursive-call} will flatten
$Y_\ell$ with the position vector containing
\begin{equation*}
  \label{eq:2}
  \underbrace{[j_1,\dots, j_m] \texttt{+} \dots \texttt{+} [j_1,\dots,j_m]}_{r' \text{ times}} 
\end{equation*}
This ensures that every tuple already produced in the recursive calls for $s(Y_1), \dots, s(Y_{\ell-1})$ get paired with every tuple of $s(Y_\ell)$. To ensure that the they also get paired with the recursive calls that follow,  the repetition vector for $Y_{\ell}$ specifies that the flattened result of each $j_q$ is to be repeated $\weightval(s_i(Y_{\ell+1})) \times \dots \times \weightval(s_i(Y_k))$ times. Lines~\ref{op:flatten:inner:start-npos}--\ref{op:flatten:inner:stop-npos} construct the correct position and repetition vector in this respect.

 \section{Instance-optimal \nsa expressions.}
\label{sec:nsa-instance-optimal}

In this section we study the asymptotic complexity of shredded processing and
identify a class of instance-optimal \nsa expressions. We focus on the RAM model
of computation with unit cost model and assume that hashing is $\bigo(1)$ per
tuple, both for hash map building as well as probing. We focus on data
complexity, i.e., the \nsa operators to be executed as well as the input and
scheme of each operator is
fixed. 

\begin{toappendix}
  This appendix contains the proof of Theorem~\ref{thm:two-phase-nsa-linear}. We require the following auxiliary definitions and results.

Recall that a store on scheme $X$ is a collection $\store{}$ of physical
relations, one physical relation $\store{}(Y)$ for every $Y \in \sub(X)$.  If
$\store{}$ is a store for a scheme $X$, then we define the size of $\store{}$,
denoted $\size(\store{})$, as $\sum_{Y \in \sub(X)} \len{\store{}(Y)}$, the sum
of cardinalities of all physical relations in $\store{}$.

\newcommand{\dsize}{\textit{dsize}}

Define the \emph{detailed size of shredding}
$\rep{R} = (\phys{R}, \store{R}, \sel{r})$ of nested relation $R\colon X$,
denoted $\dsize(\rep{R})$, to be the pair $(\len{\phys{R}},
\size(\store{R}))$. Note that, since $\sel{r}$ is a selection vector on
$\phys{R}$, we necessarily have $\len{\sel{r}} \leq \len{\phys{R}}$, which is
why we do not include it in the detailed size as we are working under data
complexity.

Similarly, define the \emph{detailed size of} shredding $\rep{D} = (\phys{h}, \store{D})$ of dictionary $D$ to be the pair $(\len{\phys{h}}, \size(\store{D}))$. Recall that $\len{\phys{h}}$ is the number of keys in hash-map $\phys{h}$.

\begin{definition}
  \label{def:strongly-linear}
  Let $f$ be an \nsa operator and let $\phys{f}$ be its physical implementation.
  Call $f$ \emph{strongly linear} if the following holds.
  \begin{itemize}
  \item If $f$ is unary, then for every legal input $I$ to $f$ (which may be a
    relation, or a dictionary), and for every shredded representation $\rep{I}$
    of $I$ with detailed size $(N, M)$ it holds that the output representation
    $\phys{f}(\rep{I})$ is computed in time $\bigo(N + M)$ under data
    complexity\footnote{Recall that under data complexity we consider all schema information to be of constant size.} and, moreover, the detailed size of the output shredded
    representation is $(\bigo(N), \bigo(M))$.\footnote{I.e., if the detailed
      size of output $\phys{f}(\phys{I})$ is $(N',M')$ then $N'= \bigo(N)$ and
      $M' = \bigo(M)$.}
  \item If $f$ is binary, then for all legal pairs of inputs $I_1$ and $I_2$ to $f$ (where $I_1$ will be a relation and $I_2$ may be a relation or a dictionary), and for all shredded representations $\rep{I}_1$ and $\rep{I}_2$ of $I_1$ and $I_2$, respectively, with detailed sizes $(N_1,M_1)$ and $(N_2, M_2)$, it holds that the output representation $\phys{f}(\rep{I}_1, \rep{I}_2)$ is computed in time $\bigo(N_1 + M_1 + N_2 + M_2)$ and, moreover, the detailed size of this representation is $(\bigo(N_1 + N_2), \bigo(M_1 + M_2))$.
  \end{itemize}
\end{definition}

From the definition of the physical operators given in
Figures~\ref{fig:physical-operators} and
Figure~\ref{fig:other-physical-operators} it is straightforward to obtain the
following.
\begin{proposition}
  \label{prop:nsa-operators-strongly-linear}
  All \nsa operators except $\groupby{}{}$, $\unnest$ and $\flatten$ have
  shredded implementations that are strongly linear.
\end{proposition}

We note that the \opgroup implementation of $\groupby{}{}$ is almost strongly linear, but does not satisfy the detailed output size requirement. Instead it is straightforward to see:
\begin{proposition}
  \label{prop:groupby-detailed-complexity}
  Let $R\colon X$ be an input to $\groupby{\y}{}$ and let $\rep{R}$ be a
  shredded representation of $R$ of detailed size $(N,M)$. On input $\rep{R}$,
  the shredded implementation \opgroup computes in time $\bigo(N + M)$ a
  shredded representation of $\groupby{\y}{}(R)$ that has detailed size
  $(\bigo(N), \bigo(N+M))$.
\end{proposition}

We next analyze the detailed complexity of $\unnest$ and $\flatten$, using the
following notion.

\begin{definition}
  \label{def:strongly-io-linear}
  Let $f$ be a unary \nsa operator and let $\phys{f}$ be its physical
  implementation.  Call $f$ \emph{strongly input-output linear} (strongly IO
  linear) if for every legal input $I$ to $f$ and every shredded representation
  $\rep{I}$ of $I$ with detailed size $(N_I, M_I)$, $\phys{f}$ computes a
  shredded representation $\rep{J}$ of $J = f(I)$ of detailed size $(N_J, M_J)$
  such that (i) $M_J = \bigo(M_I)$ and (ii) this shredded representation is
  computed in time $\bigo(N_I + M_I + N_J + M_J)$.
\end{definition}

\begin{proposition}
  \label{prop:unnest-flatten-strongly-output-linear}
  The shredded implementations of both $\unnest$ and $\flatten$ are strongly
  input-output-linear.
\end{proposition}
\begin{proof}
  For $\unnest$ and its physical operator \opunnest given in
  Figure~\ref{fig:physical-operators}, observe that the store of the output
  representation is the store of the input representation where some entries
  have been dropped. The size of the output store is hence at most that of the
  input store. This hence establishes the condition on the detailed output
  representation size. The time complexity follows straightforwardly from the
  definition of \opunnest: the vectors \texttt{pos\_R} and \texttt{pos\_Y} will
  be populated to be of the same length as $N_J$, the size of physical relation
  in the output representation $\phys{J}$ computed by \opunnest; hence the time
  spent constructing these vectors in lines 2--6 is $\bigo(N_J)$. The \take
  operations in lines 8--11 are linear in \texttt{pos\_R} and \texttt{pos\_Y},
  respectively, hence $\bigo(N_J)$.

  For $\flatten$ and its physical operator \opflatten given in
  Figure~\ref{fig:flatten-physical-operators}, observe that the store of the
  output representation is empty. It hence trivially satisfies the condition on
  the detailed output representation size. The time complexity follows
  straightforwardly from the definition of \opflatten: in every recursive call
  of \opflatteninner, the position vector \posvec is bounded in length by the
  length of $N_J$, the size of physical relation in the output representation
  $\phys{J}$ computed by \opflatten. To be more precise, it is possible to show
  using an inductive argument that in every such call also
  $\sum_{1 \leq \text{row}\leq \len{\posvec}} \repvec[\text{row}] \times
  \len{\posvec}$ and
  $\sum_{1 \leq \text{row} \leq \len{\posvec}} \repvec[\text{row}] \times
  w[\text{row}]$ are bounded by $N_J$, where $w$ is the weight vector computed
  in Line 8. Hence, computing the new vectors \texttt{npos} and \texttt{nrep} in
  lines 13--21, which provide the arguments to the recursive call and therefore
  have the same bound, runs in time $\bigo(N_J)$. The computation of \texttt{w}
  in Line 8 is linear in the size of argument physical relation $\phys{R}$,
  which is either the physical relation of \opflatten's shredded input
  $\rep{I}$, or a physical relation of the store in $\rep{I}$. Hence, this is
  certainly $\bigo(N_I + M_I)$. The call to generate in line 10 is also
  $\bigo(N_J)$ since
  $N_J \geq \sum_{1 \leq \text{row} \leq \len{\posvec}} \repvec[\text{row}]
  \times w[\text{row}]$.
\end{proof}

\begin{proposition}
  \label{prop:shrinking-non-shrinking}
  The \nsa operators $\pi, \rho, \cup, \unnest$ and $\flatten$ are
  non-shrinking, whereas $\sigma, -, \nsemijoin$ and $\groupby{}{}$ are shrinking.
\end{proposition}
\begin{proof}
  Projection $\pi$ is non-shrinking because projection is bag-based; hence it
  has exactly the same output cardinality as the input. For $\rho$ and $\cup$,
  non-shrinking is trivial by definition. Unnest and flatten themselves are
  non-shrinking because inner nested relations cannot be empty.

  All of $\sigma$, $-$ and $\nsemijoin$ may return a nested relation whose
  output cardinality is smaller than the cardinality of the largest input nested
  relation. Indeed, $\nsemijoin$ takes as input a nested relation $R$ and a
  dictionary $D$; the produced output relation may have cardinality that is
  smaller than $\card{R}$.

  Also $\groupby{}{}$ is shrinking: it returns a dictionary whose cardinality can be smaller than the cardinality of the input. This happens in particular when the input has multiple copies of the same key, which will only occur once as a key in the dictionary. Therefore the dictionary's cardinality may be smaller than the cardinality of the input relation.
\end{proof}

 \end{toappendix}

Define the \emph{size of shredding} $\rep{R} = (\phys{R}, \store{R}, \sel{r})$ of
relation $R\colon X$, to be the sum of cardinalities of all physical relations in
$\rep{R}$, i.e.  $\len{\phys{R}} + \sum_{Y \in \sub(X)}
\len{\store{R}(Y)}$. Note that $\card{R}$ equals the size of
$\rep{R}$ for flat relations.  Similarly, the size of shredding $\rep{D} = (\phys{h},
\store{D})$ of dictionary $D\colon \dscheme{\y}{Z}$ is
$\len{\phys{h}}$ plus $\sum_{Y \in \sub(Z)}
\len{\store{D}(Y)}$ where $\len{\phys{h}}$ is the number of keys in
$\phys{h}$. By analysis of the physical operators proposed in
Section~\ref{sec:processing} we readily obtain:
\begin{proposition}
  \label{nsa-operators-linearly-implementable}
  For every \nsa operator except $\unnest$ and $\flatten$, shredded processing
  runs in time $\bigo(\insize)$ while $\unnest$ and $\flatten$ run in
  $\bigo(\insize + \outsize)$ where $\insize$ and $\outsize$ are the sizes of
  the operator's shredded input, and output, respectively.
\end{proposition}

General \nsa expressions may suffer from the diamond problem. Indeed, 
every binary join plan is a valid \nsa expression by means of the equivalence
\eqref{eq:join-nsa}. Hence, the \nsa expression in Figure~\ref{fig:example-nsa-plans}b, which is the equivalent of binary join plan
$R(x,y) \join (S(y,z) \join T(z,u))$ exhibits the diamond problem when run on
instances like $\db_2$ from Figure~\ref{fig:threeq-inputs:bad} (see also
Example~\ref{ex:threeq-binaryjoin}). The utility of \nsa for avoiding the
diamond problem is as follows: all operators except $\unnest,\flatten$ are
\emph{linear} and hence produce shredded outputs whose size is at most linear in
that of the input.  In contrast to the standard join, this is true in particular
for the nested semijoin $R \nsemijoin D$. Indeed, every tuple in $R$ can produce
at most one tuple in $R \nsemijoin D$.  As such, like the classic flat semijoin,
the output of $R \nsemijoin D$ cannot increase in size. This is also the reason
why we call $\nsemijoin$ a \emph{nested semijoin}.  By contrast, $\unnest$ and
$\flatten$, because they pair each tuple $t$ in input $R$ with the tuples in an
inner relation of $t$, can produce outputs whose cardinality is not linear in
the input size.  Observe that $\unnest,\flatten$ are hence the \emph{only}
operators that can cause ``dangling tuples'' to be created. This happens when
they generate a more-than-linear subresult while another operator applied later
removes tuples from this subresult. If no such later operator is applied, all
tuples produced by $\unnest, \flatten$ will appear in the output, and none will be
dangling. This motivates the following definition.

\begin{definition}
  \label{def:2-phase}
  An \nsa expression is \emph{non-shrinking} if it always produces an output
  (relation or dictionary) whose cardinality is at least as large as the
  cardinality of its largest input.  An \nsa expression is \emph{2-phase} if,
  when viewed as a syntax tree, every $\unnest$ and $\flatten$ operator has only
  non-shrinking operators as ancestors.
\end{definition}

In other words, the output of a 2-phase expression $e$ is computed in two
phases: a first phase where subexpressions generate linear-sized subresults
(possibly filtering tuples from their input), and a second phase (delimited by
the first $\unnest$ or $\flatten$) where subexpressions may create subresults of
larger-than-linear cardinality but where the tuples in these subresults, once
created, can afterwards never be eliminated from the final output.  It is
important to stress that non-shrinking is a requirement on the output
\emph{cardinality} produced by an operator, not its size. In particular, the
store in the operator's output may be smaller than that of either input.  

The following theorem shows that all 2-phase \nsa expressions avoid the diamond
problem. \inConfVersion{We refer to \cite{anon-full-version} for the proof.}\inFullVersion{The proof is in the Appendix.}
\begin{theoremrep}
  \label{thm:two-phase-nsa-linear}
  Every 2-phase \nsa expression that maps flat input relations to flat output
  relations is evaluated in time $\bigo(\insize + \outsize)$ by shredded
  processing, where $\insize$ is the sum of the cardinalities of the
  expression's flat input relations, and $\outsize$ is the output cardinality. 
\end{theoremrep}
\begin{proof}
  Let $e\colon X$ be a 2-phase \nsa expression such that $X$ is flat. Let $\db$
  be an input database and let
  $\insize = \sum_{R(\x) \text{ atom in } e} \card{\db(R)}$ be the sum of the
  cardinalities of the input relations in $\db$.

  First, note that because $X$ is flat, no ancestor of an $\unnest$ or
  $\flatten$ operator in $e$ can be $\groupby{}{}$. This is because
  $\groupby{}{}$ produces a dictionary which is not a relation, and which itself
  hence cannot be the output of $e$. Therefore, if an ancestor of $\unnest$ or
  $\flatten$ is $\groupby{}{}$ then the dictionary produced by $\groupby{}{}$
  has to be ``consumed'' by a later operator before we are at $e$'s root. The
  only operation that we can do with dictionaries, is using it in a $\nsemijoin$
  operator. But since $\nsemijoin$ is a shrinking operator by
  Proposition~\ref{prop:shrinking-non-shrinking}, it cannot occur as an ancestor
  of $\unnest$ or $\flatten$.

  This implies that our expression $e$ is a well-typed expression generated by
  the following grammar
  \begin{align*}
    e  ::=\, & f \mid \unnest_Y(e) \mid \flatten(e) \mid \pi_Y(e) \mid \rho_{\varphi}(e) \mid e_1 \cup e_2 \\
    f  ::=\, & R(\x) \mid \sigma_{\theta(\y)}(f) \mid \pi_Y(f) \mid \rho_{\varphi}(f) \mid f_1 \cup f_2 \\
    & \mid f_1 - f_2 \mid \groupby{\y}{}(f) \mid f_1 \nsemijoin f_2
  \end{align*}

  We now note the following. Recall that for flat input relations $R$, the
  shredded representation is of the form
  $\rep{R} = (\phys{R}, \emptyset, \allsel{R})$ and has detailed size
  $(\card{R}, 0)$. Therefore, if $f$ is any \nsa expression in which neither
  $\unnest$ nor $\flatten$ occurs (cf the grammar above), we know from
  Propositions~ \ref{prop:nsa-operators-strongly-linear} and
  \ref{prop:groupby-detailed-complexity} that given any input database $\db$,
  shredded processing will compute a shredded representation of $f(\db)$ in time
  $\bigo(\insize)$ and this shredded representation has detailed size
  $\big(\bigo(\insize), \bigo(\insize)\big)$.

  We next show by induction on $e'$ that for any expression $e'$ generated by
  the above grammar, computing the representation of $e'(\db)$ runs in time
  $\bigo(\insize + \card{e'(\db)})$ and has detailed size
  $\big(\bigo(\insize + \card{e'(\db)}), \bigo(\insize)\big)$, from which the
  Theorem follows by taking $e'= e$ and observing that
  $\outsize=\card{e'(\db)}$.
  \begin{itemize}
  \item Case $e'$ is of the form $f$. We have already established that computing
    a representation of $f(\db)$ by means of shredded processing is done in time
    $\bigo(\insize)$ and has detailed size $(\bigo(\insize), \bigo(\insize))$, from which the claim clearly follows.

  \item Case $e' = \unnest_Y(e'')$. By induction hypothesis, a representation of
    $e''(\db)$ can be computed in time $\bigo(\insize + \card{e''(\db)})$ and has
    detailed size $\big(\bigo(\insize + \card{e''(\db)}),
    \bigo(\insize)\big)$. Let $(N'', M'')$ be the concrete detailed size of this
    representation. Because $\unnest$ is strongly IO-linear by Proposition
    ~\ref{def:strongly-io-linear}, shredded processing computes a representation
    of $\unnest_Y(e''(\db))$ with concrete detailed size $(N', M')$ in time
    $\bigo(N' + M' + N'' + M'')$ such that $M' = \bigo(M'')$. By definition of
    $\opunnest$, the output representation
    $\rep{R} = (\phys{R}, \store{R}, \sel{r})$ that it produces satisfies
    $\sel{r} = \allsel{R}$. Hence
    $N' = \len{\phys{R}} = \len{\allsel{R}} =\len{\sel{r}}$. Because always
    $\len{\sel{r}}$ is the cardinality of the represented nested relation, we
    hence have $N' = \card{e'(\db)}$. Therefore,
    $(N', M') = (\bigo(\card{e'(\db)}), M'') = \big(\bigo(\insize + \card{e'(\db)}),
    \bigo(\insize)\big)$, as desired. It also follows that the computation time is
    \begin{align*}
      \bigo(N' & + M' + N'' + M'') \\
      & = \bigo(N' + M'' + N'' + M'') \\
      & = \bigo(N' + N'' + M'') \\
      &= \bigo(\card{e'(\db)}) + \bigo(\insize + \card{e''}(\db)) + \bigo(\insize)\\
      &= \bigo(\insize + \card{e'(\db)})
    \end{align*}
    where in the last step we use the fact that
    $\card{e'(\db)} \geq \card{e''(\db)}$ as $\unnest$ is non-shrinking.

  \item Case $e' = \flatten(e'')$. Completely analogous to the previous case.

  \item Case $e' = \pi_Y(e'')$. By induction hypothesis, a representation of
    $e''(\db)$ can be computed in time $\bigo(\insize + \card{e''(\db)})$ and has
    detailed size $\big(\bigo(\insize + \card{e''(\db)}), \bigo(\insize)\big)$.  Because $\pi$ is strongly linear, shredded processing on this representation  computes a representation for $e'(\db)$ in time $\bigo\big(\insize + \card{e''(\db)}\big) + \bigo(\insize)$ that has detailed size $\big(\bigo(\insize + \card{e'(\db)}), \bigo(\insize)\big)$. Since $\pi$ is non-shrinking, $\card{e'(\db)} \geq \card{e''(\db)}$. Therefore, the computation time is
\begin{align*}
\bigo\big(\insize + \card{e''(\db)} + \insize\big) = \bigo(\insize + \card{e'(\db)}
\end{align*}
and the detailed representation size is   $\big(\bigo(\insize + \card{e'(\db)}), \bigo(\insize)\big)$, as desired.
  
\item Case $e = \rho_{\varphi}(e')$ and $e = e_1 \cup e_2$ are completely analogous to the previous case. \qedhere
\end{itemize}
\end{proof}

A similar result was observed in~\cite{robust-diamond-hardened-joins} for
expressions with only $\lookup$ and $\expand$. Here, we generalize it to include
all other \nsa operators. 

It is straightforward to verify that  $\pi$, $\rho$, $\cup$, $\unnest$
and $\flatten$ are the only non-shrinking operators in \nsa.  For $\pi$ this holds because projection is bag-based; hence it has exactly the same output cardinality as the input. For $\rho$ and $\cup$ this
is trivial. Unnest and flatten themselves are non-shrinking because inner nested relations cannot be empty.

\begin{figure}[tbp]
\begin{minipage}[t]{0.48\columnwidth}
  \begin{tikzpicture}[font=\small,
grow=down, level distance=0.7cm, sibling distance=1.3cm, level 2/.style={sibling distance=1.8cm}, level 3/.style={sibling distance=1cm} 
    ]
    
    \node {$\flatten$}
child { node (sj1) {$\nsemijoin$}
          child [xshift=-0.4cm] {node {$R(x,y)$}}
          child {node {$\groupby{\{y\}}{\{z,\{u\}\}}$}
            child { node (sj2) {$\nsemijoin$}
              child [xshift=-0.1cm]{node {$S(y,z)$}}
              child {node {$\groupby{\{z\}}{\{u\}}$}
                child {node {$T(z,u)$}}
              }
            }
          }
        };

  \node[right=0.01cm of sj1] {\scriptsize $\{x,y,\{z,\{u\}\}\}$};
  \node[right=0.01cm of sj2] {\scriptsize $\{y,z,\{u\}\}$};  
  \node[right=-2cm of sj2]{{\bf (a)}};    
  \end{tikzpicture}
  
 \end{minipage}
\begin{minipage}[t]{0.48\columnwidth}
  \begin{tikzpicture}[font=\small,
grow=down, level distance=0.7cm, sibling distance=1.3cm, level 2/.style={sibling distance=1cm}, level 3/.style={sibling distance=1cm} 
    ]
    
    \node {$\unnest_{\{z,u\}}$}
      child { node (sja) {$\nsemijoin$}
          child [xshift=-0.2cm] {node {$R(x,y)$}}
          child {node {$\groupby{\{y\}}{\{z,u\}}$}
            child {node {$\unnest_{\{u\}}$}
              child { node (sjb) {$\nsemijoin$}
                child [xshift=-0.1cm]{node {$S(y,z)$}}
                child {node {$\groupby{\{z\}}{\{u\}}$}
                  child {node {$T(z,u)$}}
                }
              }
            }          
          }
      };

  \node[right=0.01cm of sja] {\scriptsize $\{x,y,\{z,u\}\}$};
  \node[right=0.01cm of sjb] {\scriptsize $\{y,z,\{u\}\}$};  
  \node[right=1cm of sjb,yshift=1cm]{{\bf (b)}};     
  \end{tikzpicture} \end{minipage}
  \caption{(a) A 2-phase \nsa
   plan. (b) A non 2-phase \nsa
    plan. \label{fig:example-nsa-plans}} 
\end{figure}

\begin{example}
  Figure~\ref{fig:example-nsa-plans}a is a two-phase \nsa
  expression. Figure~\ref{fig:example-nsa-plans}b is not two-phase since
  $\unnest_{\{u\}}$ has $\nsemijoin$ as ancestor. Note that this plan is
  equivalent to $R(x,y) \join (S(y,z) \join T(z,u))$, which exhibits the diamond
  problem.
\end{example}

We should hence prefer 2-phase \nsa expressions as physical query plans since
these are the only expressions guaranteed to avoid the diamond problem. This
begs the question of when a 2-phase \nsa expression exists for a given
query. The following theorem answers this question for join queries.  Call an
\nsa expression a \emph{join plan} if it uses only the operators $\group$,
$\nsemijoin$, $\unnest$, and $\flatten$.

\begin{theoremrep}
  \label{thm:2-phase-join-acyclic}
  A join query $Q$ can be evaluated by means of a 2-phase \nsa join plan if and
  only if $Q$ is acyclic.
\end{theoremrep}
\begin{proofsketch}
  \rev{ BKN show that every acyclic join query can be evaluated by means of a
    2-phase \lande plan. Since every 2-phase \lande plan is also a 2-phase \nsa
    join plan, this hence establishes the ``if'' direction.}
  \rev{For the converse direction, let $e$ be a 2-phase \nsa join plan for
    $Q$. Using the fact that $e$ is 2-phase, we first establish that $e$ must be of the form $m_1 \dots m_k (f)$ with every $m_i$ either of the form $\unnest_{Y_i}$ or $\flatten$, and $f$ containing only $\nsemijoin$ and $\groupby{}{}$ as operators. We then convert $f$ into a rooted join tree by induction on $f$: if $f$ is an input relation the join tree consists only of the input relation, which is the root. Otherwise, $f = f_1 \nsemijoin \groupby{\y}{} f_2$ and we attach the root of the join tree created for $f_2$ as the child of the root of the join tree created for $f_1$. The  connectedness property required to be a valid join tree is ensured by  the \nsa typing rules.}
\end{proofsketch}
\begin{proof}
  BKN show that every acyclic join query can be evaluated by means of a 2-phase
  \lande plan. Every 2-phase \lande plan is also a 2-phase \nsa join plan by
  \eqref{eq:lookup-nsa} and \eqref{eq:expand-nsa}. This hence proves the ``if''
  direction.

  For the converse direction, let us write $\atoms(Q)$ for the bag (i.e.,
  multiset) of all atoms in $Q$. Let $Q = R_1(\x_1) \join \dots \join R_k(\x_k)$
  be a join query and let $e$ be a 2-phase \nsa join plan that expresses
  $Q$. Then every atom in $Q$ must be occur as an input relation in $e$, and it
  must occur as many times as it appears in $Q$ or the cardinality of the flat
  output relation that it computes will not be correct. For a subexpression $e'$
  of $e$, let $\atoms(e')$ be the sub-bag of $\atoms(Q)$ that appear in $e$.

  Let $f$ be the largest subexpression of $e$ that does not include $\unnest$ or
  $\flatten$. Because $e$ is two-phase, because $\nest$ and $\unnest$ can only have
  non-shrinking operators as ancestors, and because $\nsemijoin$ is the only
  binary operator in $\{\unnest, \flatten, \groupby{}{}, \nsemijoin\}$ but is
  shrinking (Proposition~\ref{prop:shrinking-non-shrinking}), it necessarily
  follows that $\atoms(f) = \atoms(e)$.  We next show how to construct from $f$
  a join tree for $\atoms(f) = \atoms(e) = \atoms(Q)$, which hence shows that
  $Q$ is acyclic.

  For our construction it will actually be simpler to construct a \emph{width-1
    generalized hypertree decomposition (GHD)} for $\atoms(f)$ instead of a
  regular join tree. It is well-known that a width-1 GHD tree exists for a
  multiset of atoms if and only if a join tree exists.

  A \emph{width-1 GHD} is a rooted tree $\jtree$ such that (1) all of $\jtree$'s
  nodes are either atoms, or flat schemes; and (2) for every flat attribute
  appearing somewhere in $\jtree$, all nodes mentioning $x$ are connected in
  $\jtree$.

  We claim that for every subexpression $g\colon X$ of $f$ that produces a
  nested relation, we can create a width-1 GHD $\jtree_{g}$ for $\atoms(g)$ such
  that the root of $\jtree_{g}$ is exactly the scheme $X \cap \attrs$. And, for
  every dictionary subexpression $g\colon \y \to Z$ similarly we can create
  $\jtree_g$ for $\atoms(g)$ with root the scheme $\y$.  The result follows from
  this claim since in particular $\jtree_f$ will be a width-1 GHD for
  $\atoms(f) = \atoms(Q)$.

  The claim is proved by induction on $g$.
  \begin{itemize}
  \item If $g = R_i(\x_i)$ for some $i$ then $\jtree_g$ consists of two nodes:
    the leaf node $g$ itself, and the scheme $\x_i$ which is the parent of $g$
    and forms the root of $\jtree_g$.
  \item If $g= e_1 \nsemijoin e_2$ then $e_1\colon X$ and $e_2\colon \y \to Z$
    with $\y \subseteq X$, for some $X, \y$, and $Z$. By induction hypothesis we
    have width-1 GHDs $J_{e_1}$ and $J_{e_2}$ for $\atoms(e_1)$
    resp. $\atoms(e_2)$ with as roots the schemes $X \cap \attrs$ and $\y$,
    respectively. Then create the tree $J_{g}$ by taking the union of the two
    trees $\jtree_{e_1}$ and $\jtree_{e_2}$ where we make the root of
    $\jtree_{e_2}$ a child of the root of $\jtree_{e_1}$. In particular, the
    root of $\jtree_{e_1}$ is the root of $\jtree_{g}$.  It is readily verified
    that the result is a join tree (i.e., has the connectedness property) and
    its root has scheme $X \cap \attrs$.
  \item If $g = \groupby{\y}{Z}(g')$ then $g'\colon X$ for some $X$ with
    $\x \subseteq Z$ and $Z = X \setminus \y$. By induction hypothesis, there
    exists $\jtree_{g'}$ for $\atoms(g') = \atoms(g)$ whose root has scheme
    $X \cap \attrs$. Then let $\jtree_g$ be the width-1 GHD obtained by adding $\y$ as new root on top of $\jtree_{g'}$. \qedhere
  \end{itemize}

\end{proof}
\inFullVersion{The full proof is in the Appendix.}
\inFullVersion{We note that BKN~\cite{robust-diamond-hardened-joins} have already illustrated the ``if''
direction of Theorem~\ref{thm:2-phase-join-acyclic}; here we generalize it to a
characterisation of the acyclic joins. }

 \section{Comparing binary join plans to 2-phase \nsa plans}
\label{sec:left-deep}

For parsimony, let us refer to binary join plans simply as ``binary plans'' and
to 2-phase \nsa plans as ``\twonsa plans'' in what follows.\footnote{Recall that
  \twonsa plans are 2-phase \nsa expressions using only
  $\group, \nsemijoin,\unnest$, and $\flatten$.} In
Section~\ref{sec:nsa-instance-optimal} we have shown that \twonsa plans are
\emph{robust} physical plans for acyclic join queries, as we can evaluate such
plans instance-optimally in $\bigo(\insize + \outsize)$ time.
Instance-optimality, however, is only concerned with asymptotic complexity and
may hide an important constant factor. Indeed, recall from
Section~\ref{sec:prelim} that \ya is also instance-optimal but often slower in
practice because the required semijoin reduction implies building extra hash
tables and doing extra probes compared to binary plans.  In this section we therefore move
from asymptotic complexity to analyzing plans based on a more detailed cost
model that accounts for the sizes of the hashmaps built, the number of probes
done in them, and the number of input data accesses.

\smallskip\noindent\textbf{Cost model.} We adopt three abstract cost functions,
\begin{align*}
  \cbuild&\colon \nats \to \posreals & \cprobe&\colon \nats^2 \to \posreals& \cgen&\colon \nats \to \posreals,
\end{align*}
so that $\cbuild(N)$ represents the runtime cost of a building a hash map on a
relation with $N$ tuples; $\cprobe(N,M)$ represents the cost of probing $N$ keys
in a hash map of $M$ entries; and $\cgen(N)$ represents the cost of generating a
single column (vector) of length $N$, whose content is populated by doing $N$
accesses in an already existing column.  In other words, $\cgen(N)$ is the cost
of \texttt{take(u,\posvec)} when $\len{pos} = N$.  We assume monotonicity: if
$N \leq N'$ and $M \leq M'$ then $\cprobe(N, M) \leq \cprobe(N',M')$ and
similarly for $\cbuild$ and $\cgen$.

Let us analyze binary join and the \nsa plan operators in this cost model. Let
$\numcols{R}$ denote the number of attributes (flat or nested) in the scheme of
$R$, i.e., if $R\colon X$ then $\numcols{R} = \card{X}$. Consider a traditional
binary join $R \Join S$ of flat relations $R$ and $S$ on join keys $\y$. It will
build on $S$, yielding a hash map with $\card{\groupby{\y}{}(S)}$ keys. It
probes into this hashmap from $R$, and needs to construct all columns in
$R\Join S$. Its total cost hence is
\begin{equation*}
 \cost{R\!\Join\!S}{} = \cbuild(\card{S}) + 
\cprobe(\card{R},\card{\groupby{\y}{}(S)})\  + \numcols{(R\!\Join\! S)} \times \cgen(\card{R\!\Join\! S}).
\end{equation*}
Furthermore, by inspecting the physical operators given in Figures~\ref{fig:physical-operators} and~\ref{fig:flatten-physical-operators} we obtain, for nested relations $R$ and $S$ and dictionary $D$
\begin{align*}
  \cost{\groupby{\y}{}(S)}{} & =     \cbuild(\card{S}) \\
  \cost{R \nsemijoin D}{} & = \cprobe(\card{R},\card{D}) \\
  \cost{\unnest_{Y}(R)}{} & = \numcols{(\unnest_{Y}(R))}  \times \cgen(\card{\unnest_Y(R)}) \\
  \cost{\flatten(R)}{} & = \numcols{(\flatten(R))}  \times \cgen(\card{\flatten_Y(R)})
\end{align*}
The cost of an entire plan (binary or \nsa) is then the sum of costs of each
individual operator, given the true cardinalities of the relations produced by
the operator's subexpressions.

\begin{example}
  \label{ex:cost-comparison}
  It is instructive to compare the cost of right-deep binary plan $P = R(x,y) \join (S(y,z) \join T(z,u))$ with that of the \twonsa plan $e$ shown Figure~\ref{fig:example-nsa-plans}a. Let $k = \numcols{(S \join T)}$ and $\ell = \numcols(P)$.  Then\begin{multline*}
    \cost{P}{} = \cbuild(\card{T}) + \cprobe(\card{S},\card{\groupby{\{z\}}{}T}) + k
    \cgen(\card{S\Join T})\, +\\ \cbuild(\card{S\!\join\! T}) + \cprobe(\card{R},
    \card{\groupby{\{y\}}{}(S\! \Join\! T)}) + \ell \cgen(\card{R\! \join\! S\! \join\! T}).
  \end{multline*}
  We note that this is exactly the cost of the (non-two-phase) \nsa plan in Figure~\ref{fig:example-nsa-plans}(b), which is obtained by applying equivalence \eqref{eq:join-nsa} to $P$. The embedding of binary plans in \nsa hence preserves cost.

  To compute the cost of the \twonsa plan $e$ of
  Figure~\ref{fig:example-nsa-plans}a, we first note that the subexpression
  $S(y,z) \nsemijoin \groupby{\{z\}}{} T(z,u)$ produces a nested relation whose
  cardinality is exactly $\card{S \semijoin T}$ (the flat semijoin between $S$
  and $T$). Its parent operator $\groupby{\{y\}}{}$ therefore builds a hashmap
  on $\card{S \semijoin T}$ tuples. Continuing this reasoning yields
  \begin{multline*}
    \cost{e}{} = \cbuild(\card{T}) + \cprobe(\card{S}, \card{\groupby{\{z\}}{}T})\, + \cbuild(\card{S \semijoin T})
 + \\ + \cprobe(\card{R}, \card{\groupby{\{y\}}{}(S \semijoin T)}) + \ell \cgen(\card{R \join S \join T}).
\end{multline*}
Since $\card{S \semijoin T} \leq \card{S \join T}$, this is at most $\cost{P}{}$
due to monotonicity.
\end{example}

The crucial reason why in Example~\ref{ex:cost-comparison} \twonsa plan
$e$ has at most the cost of binary plan $P$ is that $e$ can be obtained from $P$
by first turning $P$ into an \nsa plan using equivalence \eqref{eq:join-nsa}
(which yields the plan of Fig.~\ref{fig:example-nsa-plans}b) and then rewriting
the latter into a \twonsa plan by pulling to the top all $\unnest$ operations, and combining them into a single $\flatten$. We next show that we can generalize this rewriting to arbitrary binary plans as long as they are \emph{well-behaved}.

\smallskip\noindent\textbf{Well-behaved plans.} Denote by $\lleaf(P)$ the
\emph{left-most leaf atom} of binary plan $P$, when viewing $P$ as a tree. For
example, if $P = (R \Join (S \join T)) \join U$ then $\lleaf(P) = R$. For plans
that consist of a single atom, $\lleaf(P)$ is the atom itself. Denote by
$\la(P)$ the set of attributes of $\lleaf(P)$ and by $\ja(P)$ the set of join
attributes of $P$'s root join node. For example,
$\ja\left(R(x,y) \join (S(y,z) \join T(z,u))\right) = \{y\}$. If $P$ is a leaf relation
then $\ja(P) = \emptyset$.  A binary join plan $P$ is \emph{well-behaved} if for
every subplan $P'= P_1 \join P_2$ in $P$ (including $P$ itself) we have
$\ja(P') \subseteq \la(P_1)$ and $\ja(P') \subseteq \la(P_2)$.

To illustrate, both the right-deep $R(x,y) \join \big(S(y,z) \join T(z,u)\big)$ and the bushy $[ R(x,y) \join \big(S(y,z) \join T(z,u)\big) ] \join U(x,v))$ are well-behaved while $R(x,y) \join \big(T(z,u) \join S(y,z)\big)$ is not: there $\ja(P) = \{y\} \not \subseteq \{z\} =\la(T(z,u) \join S(y,z))$.

If $P$ is well-behaved, then let $\tonsemijoin{P}$ be the \twonsa expression obtained by recursively replacing every $\join$ by means of $\nsemijoin \groupby{\ja(P)}{}$:
\[ \tonsemijoin{R} = R \qquad \tonsemijoin{(P_1 \join P_2)} = \tonsemijoin{P_1} \nsemijoin \groupby{\ja(P_1 \join P_2)}{\attrs(P_2)\setminus\ja(P)} \tonsemijoin{P_2}.\]

\begin{toappendix}
  To prove Theorem~\ref{thm:well-behaved-same-cost}, we first observe the
following equivalences. To avoid confusion in what follows, we explicitly add
the database on which an \nsa expression is executed in our cost
formulas. Hence, $\cost{e}{\db}$ denotes the cost of executing expression $e$ on
database $\db$. We denote by $e(\db)$ the result of executing \nsa expression $e$ on database $\db$.

\begin{lemma}
  \label{lem:lift-unnest-left}
  Let $e_1\colon X$ and $e_2\colon \dscheme{\y}{Z}$ be \nsa expressions with
  $X \sim \dscheme{\y}{Z}$. Then
  \begin{align*}
    \flatten\left(\flatten(e_1) \nsemijoin e_2\right) & \equiv \flatten(e_1 \nsemijoin e_2) \text{ and}\\
\cost{\flatten\left(\flatten(e_1) \nsemijoin e_2\right)}{\db} & \geq \cost{\flatten(e_1\nsemijoin e_2)}{\db}    
  \end{align*}
for every database $\db$.
\end{lemma}
\begin{proof}
  Abbreviate
  \begin{equation*}
    f \defeq \flatten\left(\flatten(e_1) \nsemijoin e_2\right) \qquad \text{ and}  \qquad
    g  \defeq \flatten(e_1\nsemijoin e_2).
  \end{equation*}
  The scheme of $\flatten(e_1)$ is $\attrs(X)$.   Since $X \sim \dscheme{\y}{Z}$ we know in particular that
  $\y \subseteq X \subseteq \attrs(X)$. Hence, when computing subexpression
  $\flatten(e_1) \nsemijoin e_2$ of $f$, all flat attributes in the scheme
  $\attrs(X)$ of $\flatten(e_1)$ that are required to do a nested semijoin with
  $e_2$ are actually already in $X \cap \attrs$. Therefore,
  $\flatten(\flatten(e_1) \nsemijoin e_2)$ is equivalent to instead computing
  $e_1 \nsemijoin e_2$, and flattening the result. Hence, $f \equiv g$.

  To see the claim concerning the cost, fix an arbitrary database $\db$. Let
  $k = \numcols{\flatten(e_1)}$ and
  $\ell = \numcols{f} = \numcols{g}$. Then
  \begin{align*}
    \cost{f}{\db}&  = \cost{\flatten(e_1)}{\db} + \cost{e_2}{\db} \\
    &\phantom{=} \quad + \cprobe(\card{\flatten(e_1)(\db)}, \card{e_2(\db)}) \\
    &\phantom{=} \quad  + \ell \cgen(\card{f(\db)}) \\
    & = \cost{e_1}{\db} + k \cgen(\card{\flatten(e_1)(\db)}) + \cost{e_2}{\db} \\ 
    &\phantom{=} \quad  + \cprobe(\card{\flatten(e_1)(\db)}, \card{e_2(\db)}) \\
    &\phantom{=} \quad  + \ell \cgen(\card{f(\db)}) \\
    & \geq \cost{e_1}{\db} + \cost{e_2}{\db} \\ 
    &\phantom{=} \quad  + \cprobe(\card{e_1(\db)}, \card{e_2(\db)}) \\
    &\phantom{=} \quad  + \ell \cgen(\card{g(\db)}) \\
    & = \cost{g}{\db}
  \end{align*}
Here, the inequality uses the fact that $\card{\flatten(e_1)(\db)} \geq \card{e_1(\db)}$ and the fact that $f(\db) = g(\db)$.
\end{proof}

\begin{lemma}
  \label{lem:lift-unnest-right}
  Let $e_1\colon X$ and $e_2\colon Y$ be \nsa expressions  and
  let $\y = \attrs(X) \cap \attrs(Y)$. If $\y \subseteq X$ and $\y \subseteq Y$,
  then
  \begin{align*}
    \flatten\left(e_1 \nsemijoin \groupby{\y}{}\flatten(e_2)\right) & \equiv \flatten(e_1 \nsemijoin \groupby{\y}{} e_2) \text{ and}\\
\cost{\flatten\left(e_1 \nsemijoin \groupby{\y}{}\flatten(e_2)\right)}{\db} & \geq \cost{\flatten(e_1 \nsemijoin \groupby{\y}{} e_2)}{\db}    
  \end{align*}
for every database $\db$.
\end{lemma}
\begin{proof}
    Abbreviate
  \begin{equation*}
    f \defeq \flatten\left(e_1 \nsemijoin \groupby{\y}{}\flatten(e_2)\right) \qquad \text{ and}  \qquad
    g  \defeq \flatten(e_1 \nsemijoin \groupby{\y}{} e_2).
  \end{equation*}
  The scheme of $\flatten(e_2)$ is $\attrs(Y)$, that of
  $\groupby{\y}{}\flatten(e_2)$ is $\dscheme{\y}{\attrs(Y) \setminus \y}$.  Because $\y = \attrs(X) \cap \attrs(Y)$, $\y \subseteq X$ and $\y \subseteq Y$ we know that $X \sim \dscheme{\y}{\attrs(Y) \setminus \y}$. Hence, $f$ is well-typed.

  The scheme of $e_2$ is $Y$, that of $\groupby{\y}{}e_2$ is
  $\dscheme{\y}{Y \setminus \y}$. Since $\y \subseteq X$ and
  $\attrs(X) \cap \attrs(Y\setminus \y) = \emptyset$,
  $X \sim \dscheme{\y}{Y \setminus \y}$. Hence, also $g$ is well-typed.

  Since  $\y \subseteq Y$ we know that when computing subexpression
  $\groupby{\y}{} \flatten(e_2)$ of $f$, all flat attributes $\y$ in the scheme
  $\attrs(Y)$ of $\flatten(e_2)$ that are required for the
  grouping and that serve as keys for the later nested semijoin with $e_1$, are
  actually already in $Y$, the scheme of $e_2$. Therefore,
  $\flatten\left(e_1 \nsemijoin \groupby{\y}{}\flatten(e_2)\right)$ is
  equivalent to instead computing $e_1 \nsemijoin \groupby{\y}{} e_2$, and
  flattening the result. Hence, $f \equiv g$.

  To see the claim concerning the cost, fix an arbitrary database $\db$. Let
  $k = \numcols{\flatten(e_2)}$ and
  $\ell = \numcols{f} = \numcols{g}$. Then
  \begin{align*}
    \cost{f}{\db}&  = \cost{e_1}{\db} + \cost{\flatten(e_2)}{\db} \\
    &\phantom{=} \quad + \cbuild(\card{\flatten(e_2)(\db)}) \\
    &\phantom{=} \quad + \cprobe(\card{e_1(\db)}, \card{\groupby{\y}{}\flatten( e_2) (\db)}) \\
    &\phantom{=} \quad  + \ell \cgen(\card{f(\db)}) \\
    & = \cost{e_1}{\db} + \cost{e_2}{\db} + k \cgen(\card{\flatten(e_2)(\db)})  \\ 
    &\phantom{=} \quad + \cbuild(\card{\flatten(e_2)(\db)}) \\
    &\phantom{=} \quad  + \cprobe(\card{e_1(\db)}, \card{\groupby{\y}{}\flatten(e_2)(\db)}) \\
    &\phantom{=} \quad  + \ell \cgen(\card{f(\db)}) \\
    & \geq \cost{e_1}{\db} + \cost{e_2}{\db} \\ 
    &\phantom{=} \quad + \cbuild(\card{e_2(\db)}) \\
    &\phantom{=} \quad  + \cprobe(\card{e_1(\db)}, \card{\groupby{\y}{}e_2(\db)}) \\
    &\phantom{=} \quad  + \ell \cgen(\card{g(\db)}) \\
    & = \cost{g}{\db}
  \end{align*}
Here, the inequality uses the fact that $\card{\flatten(e_2)(\db)} \geq \card{e_2(\db)}$, the fact that  $\card{\groupby{\y}{}\flatten(e_2)(db)} \geq \card{\groupby{\y}{} e_2(\db)}$ because every $\y$-key of $\flatten(e_2)(\db)$ is also in $e_2(\db)$, and the fact that $f(\db) = g(\db)$.
\end{proof}

\begin{lemma}
  \label{lem:tonsemijoin-is-welltyped}
  If binary plan $P$ is well-behaved, then $\tonsemijoin{P}$ is well-typed with scheme
  $X$. The flat attributes in $X$ are exactly $\la(P)$, i.e.,
  $X \cap \attrs = \la(P)$, and the set of all attributes contained in $X$
  (directly or recursively) is exactly $\attr(P)$, the set of all flat
  attributes mentioned in $\attr(P)$, i.e., $\attrs(X) = \attr(P)$.
\end{lemma}
\begin{proof}
  The proof is by induction on $P$.
  \begin{itemize}
  \item If $P$ is an atom $R(\x)$ then $\tonsemijoin{P} = R(\x)$, and the result
    trivially holds.
  \item If $P = P_1 \join P_2$, then $\tonsemijoin{P} = \tonsemijoin{P_1} \nsemijoin \groupby{\ja(P)}{} \tonsemijoin{P_2}$. By induction hypothesis, the lemma holds for $\tonsemijoin{P_1}$ and $\tonsemijoin{P_2}$. Let $X_1$ be the scheme of $\tonsemijoin{P_1}$ and $X_2$ be the scheme of $\tonsemijoin{P_2}$.
Then, because $P$ is well-behaved, $\ja(P) \subseteq \la(P_1) \subseteq X_1$ and $\ja(P) \subseteq \la(P_2) \subseteq X_2$. Hence, the expression $\groupby{\ja(P)}{} \tonsemijoin{P_2}$ is well-typed, and has dictionary scheme $\dscheme{\ja(P)}{Z_2}$ where $Z_2 = X_2 \setminus \ja(P)$. This dictionary scheme is compatible with $X_1$ since $\ja(P) \subseteq X_1$ and 
\begin{align*}
  &\attrs(X_1) \cap \attrs(Z_2) \\
  & = \attrs(X_1) \cap \attrs(X_2 \setminus \ja(P)) \\
  & = \attrs(X_1) \cap (\attrs(X_2) \setminus \ja(P)) \\
  & = \attr(P_1) \cap (\attr(P_2) \setminus \ja(P)) \\
  & = \attr(P_1) \cap (\attr(P_2) \setminus (\attr(P_1) \cap \attr(P_2)))\\
& = \emptyset.
\end{align*}
Hence, $\tonsemijoin{P}$ is well-typed and has scheme $X = X_1 \cup \{ Z_2\}$. The flat attributes of $X$ are exactly the flat attributes of $X_1$, which by induction hypothesis equals $\la(P_1) = \la(P)$. Moreover,
\begin{align*}
  \attrs(X) & = \attrs(X_1) \cup \attrs(Z_2) \\
  & = \attrs(X_1) \cup \attrs(X_2 \setminus \ja(P)) \\
  & = \attr(P_1) \cup (\attr(P_2) \setminus \ja(P)) \\
  & = \attr(P_1) \cup (\attr(P_2) \setminus (\attr(P_1) \cap \attr(P_2)))\\
& = \attr(P_1) \cup \attr(P_2) \qedhere
\end{align*}
  \end{itemize}
\end{proof}

\begin{corollary}
  \label{cor:totwonsa-is-welltyped}
  If binary plan $P$ is well-behaved, then $\totwonsa{P}$ is well-typed and has
  scheme $\attr(P)$. 
\end{corollary}
\begin{proof}
  By Lemma~\ref{lem:tonsemijoin-is-welltyped}, $\tonsemijoin{P}$ is well-typed with scheme $X$ such that $\attrs(X) = \attr(P)$. Therefore, $\totwonsa{P}$ is also well-typed, with scheme $\attrs(X) = \attr(P)$.
\end{proof}

 \end{toappendix}

\begin{theoremrep}
  \label{thm:well-behaved-same-cost}
  $\tonsemijoin{P}$ is a well-typed \twonsa expression for every well-behaved binary plan $P$. Moreover, $\totwonsa{P} \equiv P$ and the cost of $\totwonsa{P}$ is at most that of $P$, on every database.
\end{theoremrep}
\begin{proof}
  Assume $P$ is well-behaved. Then $\totwonsa{P}$ is well-typed by
  Corollary~\ref{cor:totwonsa-is-welltyped}. Since $\tonsemijoin{P}$ does not
  contain any $\unnest$ or $\flatten$, expression $\totwonsa{P}$ is trivially
  two-phase.  It remains to show equivalence and the cost bound, which we do by
  induction on $P$.

  If $P = R(\x)$ the statement holds trivially. So, assume $P = P_1 \join P_2$
  and assume that the induction hypothesis holds for $P_1$ and $P_2$. We
  distinguish four cases.

  \smallskip
  (1) Both $P_1$ and $P_2$ are atoms, say $P_1 = R(\x)$ and $P_2 = S(\y)$. In
  that case $P = R(\x) \join S(\y)$ and
  $\totwonsa{P} = \flatten \big(R \nsemijoin \groupby{\z}{} S\big)$ where
  $\z = \ja(P) = \x \cap \y$. These two are clearly equivalent (use 
  equivalence \eqref{eq:join-nsa}, replacing $\unnest$ by $\flatten$). Moreover, the binary join plan
  has cost
  \[ \cost{P}{} = \cbuild(\card{S}) + \cprobe(\card{R}, \card{\groupby{\z}{} S})
    + \ell \cgen(\card{R \join S}), \] where $\ell = \card{\x \cup \y}$ denotes
  the total number of attributes occurring in the join result. The \twonsa plan
  has cost
  \[ \cost{\totwonsa{P}}{} = \cbuild(\card{S}) + \cprobe(\card{R},
    \card{\groupby{\z}{} S}) + \numcols{\totwonsa{P}}
    \cgen(\card{\totwonsa{P}}) \] Due to equivalence of $P$ and $\totwonsa{P}$,
  we necessarily have $\numcols{\totwonsa{P}} = \ell$ and
  $\card{\totwonsa{P}} = \card{R \join S}$. Hence, $P$ and $\totwonsa{P}$ have
  equal cost.

  \smallskip (2) Neither $P_1$ nor $P_2$ are atoms. By induction hypothesis,
  $P_1$ is equivalent to $\totwonsa{P_1}$ and $P_2$ is equivalent to
  $\totwonsa{P_2}$; both \twonsa expressions have a cost that is no worse than
  that of $P_1$ resp. $P_2$.  Let $P(\db)$ denote the result of evaluating $P$
  on $\db$, and similarly for $P_1, P_2$. Let $\ell = \card{\attr(P)}$. Then
  \begin{align*}
    \cost{P}{\db} & =\ \cost{P_1}{\db} +  \cost{P_2}{\db} \\
                  &\phantom{=} + \cbuild(\card{P_2(\db)}) \\
                  &\phantom{=} + \cprobe(\card{P_1(\db)}, \card{\groupby{\ja(P)}{} P_2(\db)})\\
                  &\phantom{=} + \ell \cgen(\card{P(\db)}).
  \end{align*}
  Let $e$ be the \nsa expression that simulates $P$ by executing
  $\totwonsa{P_i}$ instead of $P_i$ for $i = 1,2$, and joining the results,
  \[ e = \flatten\left(\totwonsa{P_1} \nsemijoin \groupby{\ja(P)}{}
      \totwonsa{P_2}\right).\] 
Note that while $e$ itself is not two-phase, it
  is clearly equivalent to $P$. Moreover,
  \begin{align*}
    \cost{e}{\db} & =\ \cost{\totwonsa{P_1}}{\db} +  \cost{\totwonsa{P_2}}{\db} \\
                  &\phantom{=} + \cbuild(\card{\totwonsa{P_2}(\db)}) \\
                  &\phantom{=}+ \cprobe(\card{\totwonsa{P_1}(\db)}, \card{\groupby{\ja(P)}{} \totwonsa{P_2}(\db)})\\
                  &\phantom{=} + \numcols{e(\db)} \cgen(\card{e(\db)}).
  \end{align*}
  Because $e$ is equivalent to $P$, we have $\numcols{e(\db)} = \ell$ and
  $\card{e(\db)} = \card{P(\db)}$. Moreover, because $\totwonsa{P_i}$ is
  equivalent to $P_i$ we have $\card{\totwonsa{P_i}(\db)} =
  \card{P_i(\db)}$. Finally, by induction hypothesis we have
  $\cost{\totwonsa{P_i}}{(\db)} \leq \cost{P}{\db}$. Combining all of these, we conclude
  \begin{align*}
    \cost{e}{\db} & \leq\ \cost{P}{\db}.
  \end{align*}
  We next prove the theorem by showing that $\totwonsa{P}$ is equivalent to $e$,
  and has a cost that is no worse than $e$. We will do this by using Lemmas
  ~\ref{lem:lift-unnest-left} and \ref{lem:lift-unnest-right}. 

  To that end, first observe that subexpression
  $\groupby{\ja(P)}{} \totwonsa{P_2}$ of $e$ has dictionary scheme
  $\dscheme{\ja(P)}{\attr(P_2) \setminus \ja(P)}$ since $\totwonsa{P_2}$ has
  scheme $\attr(P_2)$ by Corollary~\ref{cor:totwonsa-is-welltyped}. By
  Lemma~\ref{lem:tonsemijoin-is-welltyped}, subexpression $\tonsemijoin{P_1}$ of
  $e$ has a scheme $X$ such that $\attrs(X) = \attr(P_1)$ and
  $\ja(P)\subseteq \la(P_1) \subseteq X$. Because $\attr(P_1) $ and $\attr(P_2)$ have exactly
  $\ja(P)$ in common, it follows that the scheme $X$ of $\tonsemijoin{P_1}$ is
  compatible with the dictionary scheme
  $\dscheme{\ja(P)}{\attr(P_2) \setminus \ja(P)}$ of $\totwonsa{P_2}$. Hence,
  the preconditions of Lemma~\ref{lem:lift-unnest-left} apply when taking
  $e_1 = \tonsemijoin{P_1}$ and $e_2 = \groupby{\ja(P)}{} \totwonsa{P_2}$.  Let
  $e'$ be the following expression:
  \[ e' = \flatten\left(\tonsemijoin{P_1} \nsemijoin \groupby{\ja(P)}{}
      \totwonsa{P_2}\right). \]
  By Lemma~\ref{lem:lift-unnest-left}, $e' \equiv e$ and has a cost that is no worse then $e$.

  We next apply Lemma~\ref{lem:lift-unnest-right} to $e'$. To that end, observe
  that by Lemma~\ref{cor:totwonsa-is-welltyped}, subexpression
  $\tonsemijoin{P_2}$ has a scheme $Y$ such that $\attrs(Y) = \attr(P_2)$ and
  $\ja(P) \subseteq \la(P_2) \subseteq Y$. Since, by definition
  $\ja(P) = \attr(P_1) \cap \attr(P_2)$ it follows that
  $\attrs(X) \cap \attrs(Y) = \ja(P)$ and $\ja(P) \subseteq X$ and
  $\ja(P) \subseteq Y$. Hence, the preconditions of
  Lemma~\ref{lem:lift-unnest-left} apply when taking $e_1 = \tonsemijoin{P_1}$
  and $e_2 = \tonsemijoin{P_2}$. Then, let $e''$ be the following expression:
  \[ e''= \flatten\left(\tonsemijoin{P_1} \nsemijoin \groupby{\ja(P)}{}
      \totwonsa{P_2}\right) \]
  This expression is equivalent to $e'$ by Lemma~\ref{lem:lift-unnest-right} and has a cost that is no worse than $e'$. Note that $e'' = \totwonsa{P}$. Hence
  \[ \totwonsa{P} = e'' \equiv e' \equiv e \equiv P\]
  and 
  \[ \cost{\totwonsa{P}}{\db} = \cost{e''}{\db} \leq \cost{e'}{\db} \leq
    \cost{e}{\db} \leq \cost{P},\] i.e., its cost is at most that of $P$.

  \smallskip 
  (3) and (4) The cases where one of $P_1, P_2$ are atoms is entirely similar to the case (2).   
\end{proof}

Theorem~\ref{thm:well-behaved-same-cost} identifies a large class of binary
plans for which we can find equivalent \twonsa plans without
regret. \inConfVersion{Ill-behaved plans may incur additional cost when
  converted to \twonsa plans, as we illustrate in the full
  paper~\cite{anon-full-version} .}  \inFullVersion{We illustrate in
  Example~\ref{ex:not-well-behaved-not-same-cost} in the Appendix that this is
  not possible for ill-behaved plans: equivalent \twonsa plans may incur
  additional cost.}
\begin{toappendix}
  \smallskip\noindent\textbf{Discussion regarding ill-behaved plans.}
  Theorem~\ref{thm:well-behaved-same-cost} identifies a large class of binary
  plans for which we can find equivalent \twonsa plans without regret. The
  following Example illustrates that
  this is not possible for ill-behaved plans: equivalent \twonsa plans may incur
  additional cost.
\begin{example}
  \label{ex:not-well-behaved-not-same-cost}
  Consider $P_1= R(x,y) \join (T(z,u) \join (S(y,z))$ which, as previously discussed, is not well-behaved. It builds on $S$ and probes from $T$. Subsequently, it builds on $T \join S$ and probes from $R$. We cannot guarantee the same costs in \twonsa: if we convert subplan $O =T(z,u) \join (S(y,z)$ to $\tonsemijoin{O}$ then we also build on $S$ and probe from $T$, but we cannot construct a legal nested semijoin of $R$ with $\tonsemijoin{O}$ since $O$ does not have the join attributes between $R$ and $O$ as flat attributes. The only legal equivalent \twonsa expressions will hence either have to build on $S$ instead of $T$, or build on $R$ and probe from $O$.
As another example, consider the left-deep plan
  $P_2 = (R(x,y) \join S(y,z)) \join T(z,u)$, which is also not well-behaved. It builds on $S$ and $T$, and probes from $R$ into $S$, and from $R \join S$ into $T$. If we convert subplan $N = R(x,y) \join S(y,z)$ to $\tonsemijoin{N}$ then again we cannot legally perform nested-semijoin with (the group-by on) $T$ since the join attributes of $P$ are not flat attributes of $\tonsemijoin{N}$. We can of course construct the plan from Figure~\ref{fig:two-phase-plan} instead, which builds on $S \semijoin T$ and $T$, and hence is no worse than $P$ in terms of build cost. However, note that in this \nsa plan the number of probes into $T$ is $\card{S}$, while in $P$ this is $\card{R \join S}$, which may be lower.
\end{example}
\end{toappendix}

\smallskip\noindent\textbf{Making binary plans well-behaved.} We propose the
following strategy for generating \twonsa plans that may performance-wise
compete with the binary plans generated by existing query optimizers, while
additionally being provably instance-optimal. If the optimizer already outputs a
well-behaved plan, we simply execute $\totwonsa{P}$, which is guaranteed to
match the cost. \inConfVersion{\rev{Otherwise, we apply a dynamic programming algorithm
to ``repair'' the ill-behaved plan, making it well-behaved while minimizing the additional cost, and execute the obtained well-behaved plan.}
\rev{This repair algorithm is described in detail in the full paper~\cite{anon-full-version}.}}
\inFullVersion{Otherwise, we apply the following dynamic programming algorithm to ``repair'' ill-behaved plans $P$ while minimizing additional cost.

Our ``repair'' algorithm actually constructs a join tree given an ill-behaved
binary plan. This suffices, 
since given a join tree $J$, we can generate a well-behaved plan by
induction: if $A$ is the root of $J$, having child trees $J_1,\dots, J_n$ (in
this order) then construct the plan $((A \join P_1) \join \dots ) \join P_n$
where $P_i$ is the plan recursively constructed for $J_i$. In the Appendix we also show the converse direction: given a well-behaved plan one can construct a join tree. Well-behaved plans hence correspond one-to-one to join trees.
}

\begin{toappendix}
  \smallskip\noindent\textbf{Correspondence between well-behaved plans and join
  trees.}  To transform a well-behaved $P$ into a join tree $J$, iteratively
replace every join node in $P$ with its left child. Figure~\ref{fig:correspondence-well-behaved-plan-jointree} illustrates
the construction. Well-behavedness of $P$ guarantees that the result satisfies
the connectedness property, and is hence a valid join tree. Remark that the
conversion of a join tree into a well-behaved plan described in Section~\ref{sec:left-deep} is exactly the inverse operation. Well-behaved plans and join trees are hence in one-to-one correspondence.

\begin{figure}[tpb]
  \begin{tikzpicture}[sibling distance=0.9cm, level distance=0.5cm]
    \node {$\join$}
      [sibling distance=1.5cm]
       child { node {$\join$}
         [sibling distance=1.2cm]
         child { node {$\join$}
           [sibling distance=0.5cm]
           child { node{$A$} }
           child { node{$B$} }
         }
         child { node{$\join$}
           [sibling distance=1cm]
           child { node{$\join$}
             [sibling distance=0.5cm]
             child { node{$C$} }
             child { node{$D$} }
           }
           child { node{$\join$}
             [sibling distance=0.5cm]
             child { node{$E$} }
             child { node{$F$} }
           }
        }
    }
    child { node {$\join$}
      [sibling distance=0.5cm]
       child {node {$G$}}
       child {node {$H$}}
    }
    ;
  \end{tikzpicture}
  \qquad
  \begin{tikzpicture}[sibling distance=0.75cm, level distance=0.6cm]
    \node {$A$}
    child { node {$B$} }
    child { node {$C$}
      child { node {$D$} }
      child { node {$E$}
        child { node {$F$} }
      }
    }
    child { node {$G$}
      child {node {$H$} }
    };
\end{tikzpicture}

  \caption{Correspondence between a well-behaved plan $P$ (left) and a join
    tree (right). Letters $A$,$B$,\dots denote atoms.\label{fig:correspondence-well-behaved-plan-jointree}}
\end{figure}

 \end{toappendix}

\inFullVersion{We define the repair algorithm by means of the set of mutually recursive
functions shown in Figure~\ref{fig:dp}.  For every subplan $P$ and atom
$A \in P$ such that $\jvar(P) \subseteq A$, $\tau_A(P)$ outputs the ``optimal''
join tree that is rooted by $A$ and that covers all atoms of $P$. Here,
``optimal'' means that it incurs minimal cost penalty compared to $P$. This
penalty is computed using $\delta$. In the base case where $P=A$, $\tau_A(A)$
simply returns $A$.  In the recursive case where $P = P_1 \join P_2$, we first
compute, for every atom $A$, the optimal $A$-rooted tree for the subplan that
contains $A$. Then, for the other subplan $P'$ we first use the function
$\beta_{P'}(P)$ to compute the atom $B$ of $P'$ that contains $\jvar(P)$ and for
which $\tau_B(P')$ has minimal penalty among all candidates $B$. Subsequently we
concatenate the trees with $t_1 \concat t_2$, attaching $t_2$ as the right-most
child of the top-most atom (node) in $t_1$ containing $\jvar(P_1 \join P_2)$.
The penalty $\delta_A(P)$ of the optimal tree rooted at $A$ has a similar
recursive structure.  
In the base case of a single atom the penalty is zero.  In the recursive case,
picking a root on the left incurs no additional cost, while picking a root $B$
on the right incurs the extra cost of building on a relation of size $\card{B}$.
Finally, the optimal tree for the entire plan $P$ is the optimal tree for each
choice of roots.

We note that the algorithm above assumes that the binary plan $P$ satisfies the following property:
for every node $P_1 \join P_2 \in P$, there exist atoms $A \in P_1$ and $B \in P_2$ such that
$\jvar(P_1 \join P_2) \subseteq \attr(A) \cap \attr(B)$. This ensures in particular that $P$ is acyclic. All binary plans encountered in our experiments satisfy this property. 

\begin{figure}
  \small
\begin{align*}
\tau_A(A) &= A \\
\tau_A(P_1 \join P_2) &= \begin{cases}
  \tau_A(P_1) \concat \tau_{\beta_{P_1}(P_2)}(P_2) & \text{if } A \in P_1 \\
  \tau_A(P_2) \concat \tau_{\beta_{P_2}(P_1)}(P_1) & \text{if } A \in P_2 
\end{cases} \\
\delta_A(A) &= 0 \\
\delta_A(P_1 \join P_2) &= \begin{cases}
\delta_A(P_1) + \delta_{\beta_{P_1}(P_2)}(P_2) & \text{if } A \in P_1 \\
\delta_A(P_2) + \delta_{\beta_{P_2}(P_1)}(P_1) + \card{\beta_{P_2}(P_1)} & \text{if } A \in P_2
\end{cases}\\
\beta_{P'}(P) &= \argmin_{B \in P \mid \jvar(P' \join P) \subseteq \attr(B)} \delta_B(P)\\
\tau(P) &= \tau_\alpha(P) \text{ where } \alpha = \argmin_{A \in P \mid \jvar(P) \subseteq \attr(A)} \delta_A(P)
\end{align*}
\caption{Dynamic programming algorithm to convert a binary plan into a well-behaved plan.}
\label{fig:dp}
\end{figure}
}

\section{Experimental Evaluation}
\label{sec:experiments}

\inFullVersion{We conduct an empirical evaluation of \algnamefull by comparing it to binary hash join on a comprehensive set of queries from well-established benchmarks.}

\smallskip
\noindent {\bf Implementation.} 
Leveraging the shredding approach introduced in
Section~\ref{sec:representation-and-processing}, we implemented \twonsa plans
inside Apache Datafusion~\cite{DBLP:conf/sigmod/LambSHCKHS24} (v.34), a
high-performance columnar query engine written in Rust that uses Apache Arrow as
its \inFullVersion{in-memory} data representation.  Since \inConfVersion{Datafusion} \inFullVersion{Datafusion’s query planner} lacks a
join order optimizer, we use the columnar engine DuckDB~\cite{DBLP:conf/sigmod/RaasveldtM19}
(v1.0.0) to generate optimized plans for all considered queries. DuckDB’s
optimizer may introduce projections and filters in-between hash
joins. To ensure that the resulting plans are strictly binary, we remove these
intermediate filters and projections in the Datafusion binary plans, but keep
filters and projections on input relations. 
\inFullVersion{Corresponding \twonsa plans are obtained from the Datafusion binary plans through the algorithm of Figure~\ref{fig:dp}.}
\inConfVersion{Datafusion binary plans are subsequently transformed into  \twonsa plans  \rev{as discussed in Section~\ref{sec:left-deep}.}}

\smallskip
\noindent{\bf Setup.} 
\rev{
We first focus on interpreted columnar engines.  There, we consider three ways of executing queries: DuckDB, using its original binary-join plans (\duckdbplan); Datafusion executing the stripped binary-join plans (\binplan); and our \twonsa implementation in Datafusion running the \twonsa plans (\algname). Subsequently, we extend this comparison to the compiled engine Umbra using the reproducability package of~\cite{robust-diamond-hardened-joins}, both with and without \lande plans.
To ensure fair comparison, we focus on plan execution time reporting the median of 10 runs. This excludes query optimization time but includes hash join time as well as reading input from disk, filters, projections and aggregations if applicable.
\inConfVersion{We have also compared systems when considering only the hash join times. Due to space reasons, we defer these results, which show even more pronounced speedups, to the full paper version~\cite{anon-full-version}.}
\inFullVersion{We have also compared systems when considering only the hash join times, these results can be found in the Appendix.}}
All experiments are conducted on a Ubuntu 22.04.4 LTS machine configured to use a single thread with an Intel Core i7-11800 CPU and 32GB of RAM.

\smallskip
\noindent
{\bf Benchmarks.} We employ three established benchmarks: the Join Order Benchmark (JOB)~\cite{DBLP:journals/pvldb/LeisGMBK015}, STATS-CEB~\cite{DBLP:journals/pvldb/HanWWZYTZCQPQZL21}, and the cardinality estimation (CE) graph benchmark~\cite{DBLP:journals/pvldb/ChenHWSS22}. 
Both the JOB and STATS-CEB benchmarks consist of acyclic queries with only base table filters and equijoins, followed by a single aggregation. We excluded query 7c from JOB due to an offset overflow error that prevented its execution in Datafusion. Additionally, we removed three queries from STATS-CEB with an output cardinality exceeding $10^{10}$, resulting in a final set of 112 queries for JOB and 143 for STATS-CEB.
The CE benchmark contains both cyclic and acyclic queries. After discarding the cyclic queries and the acyclic queries that ran out of memory, 1,594 queries remained. In summary, we employ a suite of \nrofqueries\ queries for our experiments.

\begin{figure*}

  \newcommand{\interfiguredistance}{5cm}
  \newcommand{\dinterfiguredistance}{10cm}

  \pgfplotsset{width=5.2cm}

  \begin{tikzpicture}
\begin{scope}
        \begin{loglogaxis}[
          xlabel={{\bf (a)} \duckdbplan},
          ylabel={\binplan},
          xtick pos=left,
          ytick pos=both,
          ylabel style={yshift=-0.4cm},
          xmin=1e-4, xmax=4e2,
          ymin=1e-4, ymax=1e2,
          ytick={0.0001, 0.001, 0.01,0.1, 1, 10,100,1000},
          xtick={0.0001, 0.001, 0.01, 0.1, 1, 10,100,1000},
          every x tick label/.style={font=\tiny},
          every y tick label/.style={font=\tiny},
legend style={
            at={(0.5,1.05)}, anchor=south,
            legend columns=-1, draw=none, column sep=0.3em, /tikz/mark size=5pt, }
      ]
      \addplot [
        only marks,
        every mark/.append style={gray, mark=*, scale=0.3},
      ]
table[x index=1, y index=2] {./data-revision/ce_executiontime_duckdbbin_vs_dfbin.dat};
      \addplot [
        only marks,
        every mark/.append style={red, mark=*, scale=0.3},
      ]
table[x index=1, y index=2] {./data-revision/job_executiontime_duckdbbin_vs_dfbin.dat};
      \addplot [
          only marks,
          every mark/.append style={blue, mark=*, scale=0.3},
      ]
table[x index=1, y index=2] {./data-revision/statsceb_executiontime_duckdbbin_vs_dfbin.dat};
      \draw [gray] (axis cs:0.0001,0.0001) -- (axis cs:1000,1000);
      \legend{CE,JOB,STATS-CEB}
      \end{loglogaxis}
    
    \end{scope}

\begin{scope}[xshift=\interfiguredistance]
      \begin{loglogaxis}[
        xlabel={{\bf (b)} \binplan},
        ylabel={\algname},
        ylabel style={yshift=-0.4cm},
        xtick pos=left,
        ytick pos=both,
        xmin=1e-4, xmax=1e2,
        ymin=1e-4, ymax=1e2,
        ytick={0.0001, 0.001, 0.01,0.1, 1, 10,100,1000},
        xtick={0.0001, 0.001, 0.01, 0.1, 1, 10,100,1000},
        every x tick label/.style={font=\tiny},
        every y tick label/.style={font=\tiny},
legend style={
          at={(0.5,1.05)}, anchor=south,
          legend columns=-1, draw=none, column sep=0.3em, },
        scatter/classes={
            True={mark=*,black,scale=.5},
            False={mark=x,gray,scale=1}   },
    ]
    \addplot [
            scatter,only marks,
            scatter src=explicit symbolic,
        ]
    table[x index=1, y index=2, meta=well_behaved]
    {./data-revision/job_executiontime_dfbin_vs_sya.dat};
    \addplot [
      scatter,only marks,
      scatter src=explicit symbolic,
  ]
  table[x index=1, y index=2, meta=well_behaved]
  {./data-revision/statsceb_executiontime_dfbin_vs_sya.dat};
  \addplot [
            scatter,only marks,
            scatter src=explicit symbolic,
        ]
      table[x index=1, y index=2, meta=well_behaved]
      {./data-revision/ce_executiontime_dfbin_vs_sya.dat};
    \draw [gray] (axis cs:0.0001,0.0001) -- (axis cs:100,100);

    \legend{Well-behaved,Ill-behaved}
    \end{loglogaxis}
    \end{scope}

    \begin{scope}[xshift=\dinterfiguredistance]
    \begin{semilogxaxis}[ boxplot,
        width=7.5cm,
        height=4.5cm,
        boxplot/draw direction=x,
        ytick={1,2,3,4,5,6},
        yticklabels={CE, , STATS, , JOB, },
        xtick={.0001, 0.001, 0.01, 0.1, 1, 10, 100},
        every x tick label/.style={font=\tiny},
        every y tick label/.style={font=\footnotesize, rotate=90},  
        yticklabel style={xshift=.23cm, yshift=.1cm},
        legend entries = {\algname, \binplan},
        legend columns=-1, legend to name={legend},
        name=border,
        legend style={
draw=none, column sep=0.3em, /tikz/mark size=10pt, },
        boxplot/box extend=0.55,
xlabel={{\bf (c)}},
        ]
\addplot [red, mark=*, mark size=.8pt] table [y index=2] {./data-revision/ce_executiontime_dfbin_vs_sya.dat};
\addplot [blue, mark=x, mark size=1.5pt]  table [y index=1] {./data-revision/ce_executiontime_dfbin_vs_sya.dat};

\addplot [red, mark=*, mark size=.8pt]  table [y index=2] {./data-revision/statsceb_executiontime_dfbin_vs_sya.dat};
\addplot [blue, mark=x, mark size=1.5pt]  table [y index=1] {./data-revision/statsceb_executiontime_dfbin_vs_sya.dat};

\addplot [red, mark=*, mark size=.8pt]  table [y index=2] {./data-revision/job_executiontime_dfbin_vs_sya.dat};
\addplot [blue, mark=x, mark size=1.5pt]  table [y index=1] {./data-revision/job_executiontime_dfbin_vs_sya.dat};

        \draw [gray] (axis cs:0.0001,2.5) -- (axis cs:500,2.5);
        \draw [gray] (axis cs:0.0001,4.5) -- (axis cs:500,4.5);
    
      \end{semilogxaxis}
      \node[above] at (border.north) {\ref{legend}};
  \end{scope}
  \end{tikzpicture}
  \caption{\rev{\label{fig:loglogplots}Comparison of runtime performance in seconds: (a) \binplan vs.\ \duckdbplan; (b-c) \algname vs.\ \binplan.}}
\end{figure*}

\smallskip
\noindent
{\bf DuckDB \& Datafusion.}
We use log-log scatter plots where each point corresponds to the runtime of a specific query, allowing us to compare the performance of two approaches. The diagonal line represents equal runtimes for both approaches. Points that lie above (below) this diagonal indicate cases where the runtime of the approach on the Y-axis is slower (faster).

Figure \ref{fig:loglogplots}a compares \duckdbplan with \binplan on the complete set of queries. \rev{\binplan is faster than \duckdbplan in 43\% of the cases when considering total plan runtime, and in 60.4\% of the cases when considering solely join time (not shown). Moreover, the average query runtime in DuckDB is 0.828 seconds, while Datafusion executes the same queries in 0.518 seconds on average.} We conclude that \binplan is a robust baseline to use for further comparison against \algname, and focus on this comparison next.

\begin{table}[]
  \resizebox{\columnwidth}{!}{\begin{tabular}{ll|lll}
                           &         & SYA                  & SYA              & \umbrale \\
                           &         & absolute speedup (s) & relative speedup & relative speedup        \\ \hline
\multirow{4}{*}{JOB}       & min     & -0.09                & 0.76             & \textbf{0.77}           \\
                           & average & 0.11                 & \textbf{1.21}    & 1.08                    \\
                           & median  & 0.01                 & 1.01             & \textbf{1.03}           \\
                           & max     & 2.21                 & \textbf{6.42}    & 1.85                    \\ \hline
\multirow{4}{*}{STATS-CEB} & min     & -0.01                & 0.87             & \textbf{1.22}           \\
                           & average & 0.97                 & \textbf{3.13}    & 2.87                    \\
                           & median  & 0.01                 & 2.07             & \textbf{2.62}           \\
                           & max     & 37.43                & \textbf{23.97}   & 19.09                   \\ \hline
\multirow{4}{*}{CE}        & min     & -0.70                & \textbf{0.40}    & 0.33                    \\
                           & average & 0.20                 & \textbf{1.81}    & 1.65                    \\
                           & median  & 0.01                 & 1.21             & \textbf{1.44}           \\
                           & max     & 14.96                & \textbf{62.54}   & 16.82                  
\end{tabular}}
  \caption{\rev{Speedups of \algname over \binplan and \umbrale over \umbra. Relative speedups smaller than 1 and negative absolute speedups are slowdowns. Highest relative speedups are in bold.}}
  \label{tab:speedups}
  \inConfVersion{\vspace{-2ex}}
  \end{table}

\rev{ Figure \ref{fig:loglogplots}b compares \binplan with \algname. To be precise, \algname   matches or outperforms \binplan on  $94.6$\% of the queries (JOB), $94.4$\% (STATS-CEB), and $83.8$\% (CE). 
The improved robustness of \algname over \binplan is illustrated in Figure~\ref{fig:loglogplots}c, which shows a box-plot comparison of the absolute runtimes (log-scale) per dataset.
The maximum speedups and slowdowns, both relative and absolute, are shown in Table~\ref{tab:speedups}. 
Slowdowns are small, as the maximum relative/absolute slowdowns are limited to 1.3x/89ms (JOB), 1.15x/5ms (STATS-CEB), and 2.5x/0.70s (CE). The maximum relative/absolute speedups are 6.4x/2.2s (JOB), 24x/37.4s (STATS-CEB), and 62.5x/15s (CE). We conclude that \algname is always competitive with binary join plans, and often much better, while at the same time guaranteeing robustness.} 

\rev{
Recall from  Section~\ref{sec:left-deep}, that well-behaved plans can theoretically be translated into \twonsa plans without increasing their execution cost. We observe that 849 (46\%) of the binary plans are well-behaved.
777 of these (92\%) are indeed evaluated at least as fast by \algname as by \binplan in our experiments. 
For the remaining 72 queries where this is not the case, the absolute slowdown is at most 77ms.}
We conclude that our cost model, while an abstraction of reality, accurately predicts performance in the vast majority of cases.
54\% of the binary plans are not well-behaved. For such plans the rewriting into a \twonsa plan is not guaranteed to be cost-preserving. \rev{However, in our experiments, for 80\% of them, \algname is at least as fast as \binplan.} This demonstrates that the benefit obtained by avoiding the diamond problem often outweighs the additional build and/or probe cost introduced by converting binary into \twonsa plans.

\rev{
We next discuss some queries qualitatively.
The query with the highest relative speedup ($62.5$x), is \texttt{yago\_acyclic\_star\_6\_39} from the CE benchmark. This well-behaved binary plan clearly suffers from the diamond problem, explaining the observed speedup: the query produces an intermediate of $1.5 \times 10^7$ tuples, while the inputs are not larger than $2.1 \times 10^5$ and the output cardinality is $7.1 \times 10^5$. }
\rev{
The query with the highest slowdown (2.5x)  is query \texttt{yago\_acyclic\_chain\_12\_39} from the same benchmark. We observe that the binary plan is already well-optimized and does not suffer from the diamond problem. Moreover, the binary plan is not well-behaved, leading to a higher build cost in the \twonsa plan. This increased build cost, combined with the absence of the diamond problem in the binary plan, accounts for the slowdown observed with \algname.
}

\begin{toappendix}
   
\rev{
  In Section~\ref{sec:experiments}, all reported runtimes include file scans, filters, projections, aggregations and joins. 
  This appendix further examines the speedups of \algname by focusing exclusively on join time. 
  Table~\ref{fig:jointime-speedup} shows a comparison of \algname's relative speedups over \binplan, considering both total query runtime (left) and join time alone (right). The differences in relative speedups and slowdowns become more pronounced when isolating join time. For completeness, we also provide the same plots as in Figure~\ref{fig:loglogplots} but now focusing on join time only (Figure~\ref{fig:loglogplots_jointime}).      
  }

    \begin{table}[ht]
  \begin{tabular}{ll|ll}
        &         & \algname speedup & \algname speedup       \\ 
        &         & (total runtime) & (join time)      \\ \hline
        \multirow{4}{*}{JOB}       & min     & \textbf{0.76} & 0.31           \\
        & average & 1.21          & \textbf{1.51}  \\
        & median  & 1.01          & \textbf{1.11}  \\
        & max     & 6.42          & \textbf{13.94} \\ \hline
  \multirow{4}{*}{STATS-CEB} & min     & \textbf{0.87} & 0.73           \\
        & average & 3.13          & \textbf{3.66}  \\
        & median  & 2.07          & \textbf{2.46}  \\
        & max     & 23.97         & \textbf{29.01} \\ \hline
  \multirow{4}{*}{CE}        & min     & \textbf{0.40} & 0.23           \\
        & average & 1.81          & \textbf{2.10}  \\
        & median  & 1.21          & \textbf{1.33}  \\
        & max     & 62.54         & \textbf{90.42}
      \end{tabular}\caption{Relative speedups of \algname over \binplan when considering total query runtime (left) and join time only (right). Highest number in each row is in bold.}
      \label{fig:jointime-speedup}
  \end{table}
  
  \begin{figure*}
  
        \newcommand{\interfiguredistance}{5cm}
        \newcommand{\dinterfiguredistance}{10cm}
      
        \pgfplotsset{width=5.2cm}
      
        \begin{tikzpicture}
  \begin{scope}
              \begin{loglogaxis}[
                xlabel={{\bf (a)} \duckdbplan},
                ylabel={\binplan},
                xtick pos=left,
                ytick pos=both,
                ylabel style={yshift=-0.4cm},
                xmin=1e-4, xmax=4e2,
                ymin=1e-4, ymax=1e2,
                ytick={0.0001, 0.001, 0.01,0.1, 1, 10,100,1000},
                xtick={0.0001, 0.001, 0.01, 0.1, 1, 10,100,1000},
                every x tick label/.style={font=\tiny},
                every y tick label/.style={font=\tiny},
  legend style={
                  at={(0.5,1.05)}, anchor=south,
                  legend columns=-1, draw=none, column sep=0.3em, /tikz/mark size=5pt, }
            ]
            \addplot [
              only marks,
              every mark/.append style={gray, mark=*, scale=0.3},
            ]
  table[x index=1, y index=2] {./data-revision/ce_jointime_duckdbbin_vs_dfbin.dat};
  \addplot [
              only marks,
              every mark/.append style={red, mark=*, scale=0.3},
            ]
  table[x index=1, y index=2] {./data-revision/job_jointime_duckdbbin_vs_dfbin.dat};
  \addplot [
                only marks,
                every mark/.append style={blue, mark=*, scale=0.3},
            ]
  table[x index=1, y index=2] {./data-revision/statsceb_jointime_duckdbbin_vs_dfbin.dat};
  \draw [gray] (axis cs:0.0001,0.0001) -- (axis cs:1000,1000);
            \legend{CE,JOB,STATS-CEB}
            \end{loglogaxis}
          
          \end{scope}
      
  \begin{scope}[xshift=\interfiguredistance]
            \begin{loglogaxis}[
              xlabel={{\bf (b)} \binplan},
              ylabel={\algname},
              ylabel style={yshift=-0.4cm},
              xtick pos=left,
              ytick pos=both,
              xmin=1e-4, xmax=1e2,
              ymin=1e-4, ymax=1e2,
              ytick={0.0001, 0.001, 0.01,0.1, 1, 10,100,1000},
              xtick={0.0001, 0.001, 0.01, 0.1, 1, 10,100,1000},
              every x tick label/.style={font=\tiny},
              every y tick label/.style={font=\tiny},
  legend style={
                at={(0.5,1.05)}, anchor=south,
                legend columns=-1, draw=none, column sep=0.3em, },
              scatter/classes={
                  True={mark=*,black,scale=.5},
                  False={mark=x,gray,scale=1}   },
          ]
          \addplot [
                  scatter,only marks,
                  scatter src=explicit symbolic,
              ]
          table[x index=1, y index=2, meta=well_behaved]
          {./data-revision/job_jointime_dfbin_vs_sya.dat};
          \addplot [
            scatter,only marks,
            scatter src=explicit symbolic,
        ]
        table[x index=1, y index=2, meta=well_behaved]
        {./data-revision/statsceb_jointime_dfbin_vs_sya.dat};
        \addplot [
                  scatter,only marks,
                  scatter src=explicit symbolic,
              ]
            table[x index=1, y index=2, meta=well_behaved]
            {./data-revision/ce_jointime_dfbin_vs_sya.dat};
          \draw [gray] (axis cs:0.0001,0.0001) -- (axis cs:100,100);
      
          \legend{Well-behaved,Ill-behaved}
          \end{loglogaxis}
          \end{scope}
      
          \begin{scope}[xshift=\dinterfiguredistance]
          \begin{semilogxaxis}[ boxplot,
              width=7.5cm,
              height=4.5cm,
              boxplot/draw direction=x,
              ytick={1,2,3,4,5,6},
              yticklabels={CE, , STATS, , JOB, },
              xtick={.0001, 0.001, 0.01, 0.1, 1, 10, 100},
              every x tick label/.style={font=\tiny},
              every y tick label/.style={font=\footnotesize, rotate=90},  
              yticklabel style={xshift=.23cm, yshift=.1cm},
              legend entries = {\algname, \binplan},
              legend columns=-1, legend to name={legend},
              name=border,
              legend style={
  draw=none, column sep=0.3em, /tikz/mark size=10pt, },
              boxplot/box extend=0.55,
  xlabel={{\bf (c)}},
              ]
  \addplot [red, mark=*, mark size=.8pt] table [y index=2] {./data-revision/ce_jointime_dfbin_vs_sya.dat};
  \addplot [blue, mark=x, mark size=1.5pt]  table [y index=1] {./data-revision/ce_jointime_dfbin_vs_sya.dat};
      
  \addplot [red, mark=*, mark size=.8pt]  table [y index=2] {./data-revision/statsceb_jointime_dfbin_vs_sya.dat};
  \addplot [blue, mark=x, mark size=1.5pt]  table [y index=1] {./data-revision/statsceb_jointime_dfbin_vs_sya.dat};

  \addplot [red, mark=*, mark size=.8pt]  table [y index=2] {./data-revision/job_jointime_dfbin_vs_sya.dat};
  \addplot [blue, mark=x, mark size=1.5pt]  table [y index=1] {./data-revision/job_jointime_dfbin_vs_sya.dat};

              \draw [gray] (axis cs:0.0001,2.5) -- (axis cs:500,2.5);
              \draw [gray] (axis cs:0.0001,4.5) -- (axis cs:500,4.5);
          
            \end{semilogxaxis}
            \node[above] at (border.north) {\ref{legend}};
        \end{scope}
        \end{tikzpicture}
        \caption{\rev{\label{fig:loglogplots_jointime}Comparison of runtime performance in seconds when considering join time only: (a) \binplan vs.\ \duckdbplan; (b-c) \algname vs.\ \binplan.}}
      \end{figure*} 
    \end{toappendix}

\smallskip
  \noindent
  \textbf{Umbra.}
  \rev{
Absolutely speaking,  binary join plans in Umbra (\umbra) execute on average one order of magnitude faster than \duckdbplan and \binplan. The same holds for \lande plans in Umbra (\umbrale) compared to \algname (which is roughly Datafusion + \lande). We attribute this significant difference in absolute runtime to 
the fundamental 
differences between compiled query engines and interpreted column stores as well as other differences between the systems including the query planner and various low-level optimizations. When we compare the relative speedups obtained by  \algname over \binplan to the speedups obtained by \umbrale over \umbra, however, as shown in  Table~\ref{tab:speedups} we see that these speedups are comparable for all benchmarks. We can conclude that the benefits of \lande, originally implemented in a compiled query engine, can be successfully translated to column stores.}

\section{Conclusion}
\label{sec:discussion}

We have shown how to implement the idea of \lande decomposition
inside column stores using nested relations and \nsa as the logical  model, and query shredding as physical model. We have used this approach to illustrate the feasibility of implementing
Yannakakis-style instance-optimal join processing inside a conventional
main-memory columnar query engine \emph{without regret}: fast on every acyclic
join, and not only asymptotically.  We hope that this perspective can help
system engineers to better understand \ya, and pave the way for its adoption
into existing systems. 
 
\begin{acks}
  This work was initiated during the research program on Logic and Algorithms in
  Database Theory and AI at the Simons Institute for the Theory of Computing.
  This research was supported by Hasselt University Bijzonder Onderzoeksfonds
  (BOF) Grants No. BOF22DOC07 and BOF20ZAP02, and the Research Foundation
  Flanders (FWO) under Grant No. G0B9623N.
\end{acks}

\bibliographystyle{ACM-Reference-Format}
\bibliography{references}

\end{document}